# Economic consequences of the spatial and temporal variability of climate change

*Francisco Estrada[1,2,3*], Richard S.J. Tol[4,2,5,6,7,8,9], Wouter Botzen[2,10]*

[1]Centro de Ciencias de la Atmosfera, Universidad Nacional Autónoma de México, CDMX, Mexico; [2]Institute for Environmental Studies, VU Amsterdam, Amsterdam, the Netherlands; [3]Programa de Investigación en Cambio Climático, Universidad Nacional Autónoma de México, CDMX, Mexico; [4]Department of Economics, University of Sussex, Falmer, UK; [5]Department of Spatial Economics, Vrije Universiteit, Amsterdam, The Netherlands; [6]Tinbergen Institute, Amsterdam, The Netherlands; [7]CESifo, Munich, Germany; [8]Payne Institute for Public Policy, Colorado School of Mines, Golden, CO, USA; [9]College of Business, Abu Dhabi University, UAE; [10]Utrecht University School of Economics (U.S.E.), Utrecht University, Utrecht, Netherlands.

**Damage functions in integrated assessment models (IAMs) map changes in climate to economic impacts and form the basis for most of estimates of the social cost of carbon. Implicit in these functions lies an unwarranted assumption that restricts the spatial variation (Svar) and temporal variability (Tvar) of changes in climate to be null. This could bias damage estimates and the climate policy advice from IAMs. While the effects of Tvar have been studied in the literature, those of Svar and their interactions with Tvar have not. Here we present estimates of the economic costs of climate change that account for both Tvar and Svar, as well as for the seasonality of damages across sectors. Contrary to the results of recent studies which show little effect that of Tvar on expected losses, we reveal that ignoring Svar produces large downward biases, as warming is highly heterogeneous over space. Using a conservative calibration for the damage function, we show that previous estimates are biased downwards by about 23-36%, which represents additional losses of about US$1,400-US$2,300 billion by 2050 and US$17-US$28 trillion by the end of the century, under a high emissions scenario. The present value of losses during the period 2020-2100 would be larger than reported in previous studies by $47-$66 trillion or about ½ to ¾ of annual global GDP in 2020. Our results imply that using global mean temperature change in IAMs as a summary measure of warming is not adequate for estimating the costs of climate change. Instead, IAMs should include a more complete description of climate conditions.**

Key words: climate change; climate impacts; spatial variation; temporal variability; social cost of carbon

JEL Classification: Q54

## I. Introduction

Climate is commonly defined as the long-term weather conditions over a specific region. These long-term conditions are usually approximated by the statistical description in terms of central tendency and variability of weather over a long period of time (i.e., 30 years)[1]. In particular, the average of weather variables over such a period is referred to as climatology or climate normal and, in practice, it is frequently used as a shorthand description of climate. However, this description is incomplete as it ignores variability and omits information that may be relevant for a variety of users/applications. The same is true for using only changes in the mean to describe climate change and to estimate its impacts, which are known to be a function of both changes in central tendency measures and variability[2–4].

Integrated assessment models of the climate and the economy usually translate changes in global or regional mean annual surface temperatures to economic damages through stylized mathematical representations, called damage functions[5,6]. Temperature change in IAMs is assumed to be gradual and smooth, and excludes any variability in the time and spatial domains[7,8].

This omission has two important implications. First, intra- and interannual variability are assumed to have no effect on the warming rate. All months and seasons warm by the same amount and variability modes, such as the Atlantic Multidecadal Oscillation (AMO) and the Pacific Decadal Oscillation (PDO), cannot amplify nor dampen the warming trend, nor is there any interplay between variability and climate change that affect the projected damages. These are oversimplifications that are useful for keeping IAMs as simple and transparent as possible, but do not fare well with what is known about climate change: warming rates do vary between months and seasons[9,10]; low-frequency oscillations modulate the warming rate; there are important interactions and synergies between natural variability and the response of the climate system to changes in external forcing, and natural variability affects the magnitude of damages[8,11].

Second, in IAMs warming is incorrectly assumed to be spatially homogeneous over the globe and/or world regions. Observations, as well as climate models' projections, show that warming rates and changes in other climate variables (e.g., precipitation) vary widely over space[12,13]. In IAMs, global (or, in some cases, regional) annual temperatures act as an index of changes in the relevant climate variables over space (and time) which drive the physical and economic losses that damage functions aim to represent. While there is considerable evidence in favor of spatial stationarity of changes in climate, this does not imply that the spatial variance of changes remains constant for different levels of warming[2,14]. On the contrary, differential warming around the globe is clear from all global circulation climate models and implies that the variance of changes in temperature increases with global temperature rise. As an illustration, consider the case of the pattern scaling technique which has been extensively evaluated and is consider to adequately approximate the output from complex physical climate models, particularly in the case of mean surface temperatures[14–16]. This technique assumes spatial stationarity of changes in variables such as temperatures and precipitation. Under such assumption, a matrix of fixed weights $P^v = p^v_{i,j}$, which represents the spatial pattern of change for variable $v$ and for each pair of geographical coordinates $i, j$, can be obtained from climate models' output to approximate what the magnitude of change at the grid cell would be for variable $v$ and for a given value of change in global annual temperature $\Delta T^g$. Even in this case of spatial stationarity, the variance of the pattern $P^v$ would increase as warming rises due to the properties of the variance: $var(v) = var(\Delta T^g P^v) = (\Delta T^g)^2 var(P^v)$. As is shown in the next sections, similar results are obtained using warming output from global physical models. Having spatial climate change projections that are consistent with physics is important for generating economic impacts of climate change in recently developed local IAMS in which the economic exposure to climate impacts and vulnerability varies across space[17].

Few studies have focused on incorporating the effects of climate variability or stochasticity into commonly used climate damage functions. These modified functions aim to represent the impacts of climate change in terms of mean and variability, which includes intra-annual to decadal, centennial, and even longer frequency oscillations. However, these few studies have been restricted to include variability in the time domain, ignore spatial variation, and do not account for the seasonality of climate impacts. Here we show that these limitations can have significant effects on their estimates and conclusions. Estrada and Tol[7] proposed recalibrating existing damage functions in such a way that their parameters account for the stochasticity in climate time series projections. They focus on global average temperatures and conclude that the mean monetary impacts produced by the stochastic damage function is similar those of the deterministic function, but that the welfare impacts are larger[7]. A recent study[18] suggests that accounting for interannual variability in global mean temperature time series, the risk of economic damages increases by trillions of dollars even if variability is assumed to be moderate. However, the expected losses are not biased and are the same as when temporal variability is not included. They use a zero-dimensional stochastic climate model and the Weitzman damage function, one of the damage functions with higher response to temperature change, to project the distribution of losses[18]. The authors calculate the risk premium of the effects of including internal variability in the time domain and combined with the uncertainty of the equilibrium climate sensitivity. They find that while there is no considerable effect on the estimated expected losses, the risk premium can be up to $46 trillion dollars (or about 58% of global output). Of this risk premium about US$32 trillion areis due to aleatory uncertainty and US$14 trillion to the interaction of aleatory and epistemic uncertainty. These studies provide relevant information about the effects of internal variability in the time domain on economic impacts and conclude that omitting this factor can have important effects on some aspects of the estimation of the consequences of climate change. Another study looking at past impacts showed that there are interaction effects between natural variability and external forcing that could bias the estimated costs of climate change[8]. That is, variability may affect not only risk premium but also the estimated level of economic losses.

In the present study, we show analytically and numerically that the way internal variability has been handled in current studies of the economic costs of climate change leads to important downward biases. We expand previous studies by considering temporal inter- and intra-annual variability and spatial variation, as well as by developing new annual, seasonal and monthly damage functions designed to include climate variability and take advantage of sectoral estimates of damages available in the literature. Our results show that considering internal climate variability, both in the time and space domains, significantly increases the projected economic damages, not only the risk premium a social planner would pay to avoid stochasticity. Therefore, the current estimates importantly underestimate the economic costs of climate change. Furthermore, we provide an analytic solution to include climate variability that does not require conducting simulation experiments and that is embedded in the most common currently used damage functions. This extension can be easily adopted in most IAMs currently available, as we illustrate by extending the RICE damage functions[19,20] to calculate the regional social cost of carbon.

To include temperature change variability in time and space we incorporate three extensions to the commonly used climate-economy IAM framework (see Methods). First, coupled ocean-atmosphere general circulation models' output is used to create the monthly and spatially explicit temperature scenarios. Two approaches are used for this extension, one based on monthly scaling patterns (SP)[14,16,21] calculated by means of linear regressions (Methods) to produce gridded temperature projections from the output of MAGICC7[22], a reduced complexity climate model. The second approach is based on creating centered 30-year rolling climatologies (RM) to approximate the climate signal at each grid cell[23].

A considerable fraction of the available damage functions in IAMs are derived from the function proposed by Nordhaus in the DICE model, with only small changes implemented over the years[24–28]. A general representation of this damage function is $I_t = a\left(\frac{\overline{\Delta T_t}}{C}\right)^2 = \alpha(\overline{\Delta T_t})^2$, where $\overline{\Delta T_t}$ is global mean annual surface temperature increase, $I_t$ is the projected economic damages, $a$ represents the losses in percent of global GDP for an increase of $C$ in global temperature, and $\alpha = \left(\frac{a}{C^2}\right)$. For the estimates presented here, parameter values $a = 2.01\%$ and $C = 2.5$ºC were obtained from a review of published estimates[24,29,30] (see Table S1).

Note that $\overline{\Delta T_t} = E[\Delta T_t]$ is the expected value of temperature change over the globe and a sufficiently long period of time (e.g., 30 years). Using the definition of the variance (i.e., $var(X) = E[X^2] - E[X]^2$, for any variable *X*), the original damage function $I_t$ can be expressed as (Methods):

$$I_t^* = \alpha(\overline{\Delta T_t})^2 + \alpha[var(\Delta T_t)] = \alpha\overline{\Delta T_t^2} \qquad (1)$$

$I_t^* = \alpha\overline{\Delta T_t^2}$ offers a computationally efficient way of incorporating spatial and/or temporal variability into commonly used damage functions in IAMs. A relevant feature of $I_t^*$ is that $I_t = \alpha(\overline{\Delta T_t})^2$ arises as a particular case when $var(\Delta T_t)$ is restricted to be zero. Note that since the variance is a positive number, damages are necessarily larger if this restriction is relaxed. As shown in Methods, this extension is easily implemented in the RICE model's regional damage functions.

Moreover, damage functions are usually calibrated for producing annual loss estimates, implicitly assuming that there are no differences in intra-annual warming and that for the different sectors seasonality plays a limited role in their production. We use sectoral estimates of the costs of climate change that are available in the literature[29] and propose weights that reflect the seasonality in production by sector to develop monthly damage functions for 15 sectors (Methods; Table S1 [RICHARD, ARE THERE SOME REFERENCES TO SUPPORT OUR SELECTION OF SEASONAL WEIGHTS?]). These monthly damage functions allow to account for spatial and temporal variability in warming, as well as for the seasonality in expected damages (Methods).

**Results**

*Effects on the estimates of economic losses of accounting for temporal and spatial variation in warming*

From equation (1) follows that accounting for climate variability has a direct effect on the level of damages, and not only through the risk premium[7,18]. Here we quantify the losses attributable to changes in the mean of global temperature, to spatial variation (Svar) and temporal variability (Tvar) and to those that are produced by the interaction of both factors. The main estimates presented here correspond to the SP approach of the ensemble mean (one member per model) of the Coupled Model Intercomparison Project Phase 6 (CMIP6), while the Supplementary Information contains a sensitivity analysis that considers both the SP and RM approaches from six climate models with different ensemble sizes (Methods). The differences in the number of realizations per model are used to illustrate the effects of different levels of noise around the climate signal.

Figure 1 compares the economic impacts with and without climate variability. Moreover, it decomposes these results into spatial and temporal effects. This figure underlines the important role of Svar in the change in temperature for projecting the economic damages of climate change. Allowing for both spatial variation and temporal variability (S&Tvar) in $\Delta T_t$ leads at the end of this century to damages of about 9.65% of global GDP, which are about 2.0% of global GDP higher than when no variability is included in the calculations (7.66% of global GDP). The largest change in the projected damages can be attributed to Svar, with damages raising to 9.11% of global GDP. The effect of incorporating Tvar by itself is quite small (0.01% of global GDP) and similar to what has been reported previously in the literature[7]. When variability in both spatial and temporal domains is allowed, the interaction effects of these two domains makes losses noticeably larger, with an increase in damages of about 0.54% of global GDP. The differences in damages increase with warming: according to the SP estimates, by 2050 if S&Tvar is ignored, damages would be underestimated by about US$1.6 trillion (US$1.4-US$2.3 trillion; Table S3) and about US$20 trillion by 2100 (US$17-US$28 trillion; Table S3). These results are like those obtained with the RM approach, except for those obtained with the CanESM5 which projects about 2ºC larger than the CMIP6 ensemble mean by 2085. Also, since the RM estimates are based on 30-year rolling statistics, the last year that can be calculated is 2085 and not 2100. In consequence, the magnitudes of these estimates are typically lower than those of the SP for 2100.

Figure 2 shows the present value of the losses accumulated during the period 2020-2100 for each of the annual damage functions, using a 4% discount rate. Tables S3-S4 presents a sensitivity analysis using other climate models and both the SP and RM approaches. The present value of losses obtained with the damage function that excludes temperature change variability amounts to US183,841 billion, while the present value including S&Tvar is US$231,363 billion (US$225,679-US$249,750; Table S4), which represents a difference of about 55% of global GDP in 2020 (US$47 trillion with a range of US$42-US$66 trillion; about 26% higher than the original damage function). The present value of losses for the Svar damage function is US$218,489 billion (US$210,443-US$224,870), and US$184,086 (US$183,880-US$184,683) for Tvar which is much closer to that of the original damage function. The results in Table S3 show that not only the differences in the level of warming by various models (RM) strongly influence the projected losses, but that even when the differences in warming are controlled for (i.e., using the pattern scaling method; SP), the variability in space and time of each model increases damages between 23%-36%. Decomposing these additional losses (Table S4) reveals that about 65% (62%-75%) of them can be attributed to spatial variation and about 1% (-3%-2%) to temporal variability. The interaction between spatial variation and temporal variability has a large effect, accounting for about 33% (24%-45%) of the increases in losses. The effects of incorporating variability in the space and time domains are also important when comparing the costs associated to different reference and policy emissions trajectories. As shown in Table S5, when S&Tvar is included, the estimated damages are considerably larger for any emissions scenario. Moreover, using the SSP585 as an illustration, Table S5b-c shows that the benefits (in terms of avoided damages) are also larger for all levels of mitigation effort deemed possible under the SSP5 socioeconomic scenario[31].

When using the monthly damage function with all sectors aggregated by month, the calculated damages are very similar to those obtained with the annual damage function (Tables S3-S5). For instance, by 2100 using the S&Tvar damage function, the present value of losses is about 0.03% of global GDP smaller compared with that of the annual damage function. These differences are due to the heterogeneity in warming during the year. The differences become clearer when monthly damages are analyzed per sector (Tables S6-S11). While including spatial variation always leads to higher damages, temporal variability can lead to lower damages in some cases when compared with projections from the annual damage function. These are sectors that are strongly influenced by seasonal variation, such as agriculture, forestry, water, and air pollution (Figure S3, S4). This occurs because the average annual warming is higher than the increase in temperatures in the months that are most relevant for the sector. In contrast, for some sectors temporal variability increases impacts significantly, as the warming that occurs in the relevant months is larger than the annual mean. These sectors are amenity, energy, health and time use, a sector for which warming and seasonal variability produces benefits. This illustrates the importance of relaxing the assumption of zero variability that is implicitly made in most of the current damage functions (equation 1). As suggested by Figure S4, damages on health, amenity and water would occur earlier in this century, reaching monthly losses of exceeding 0.1% of global GDP during the 2050s and doubling in the 2070s. If mitigation efforts that lead to an emissions trajectory like that of the SSP245 are achieved, exceeding 0.1% of GDP in monthly damages in sectors such as health could be delayed to the end of the 2080s, and losses larger than 0.2% would not be attained during this century.

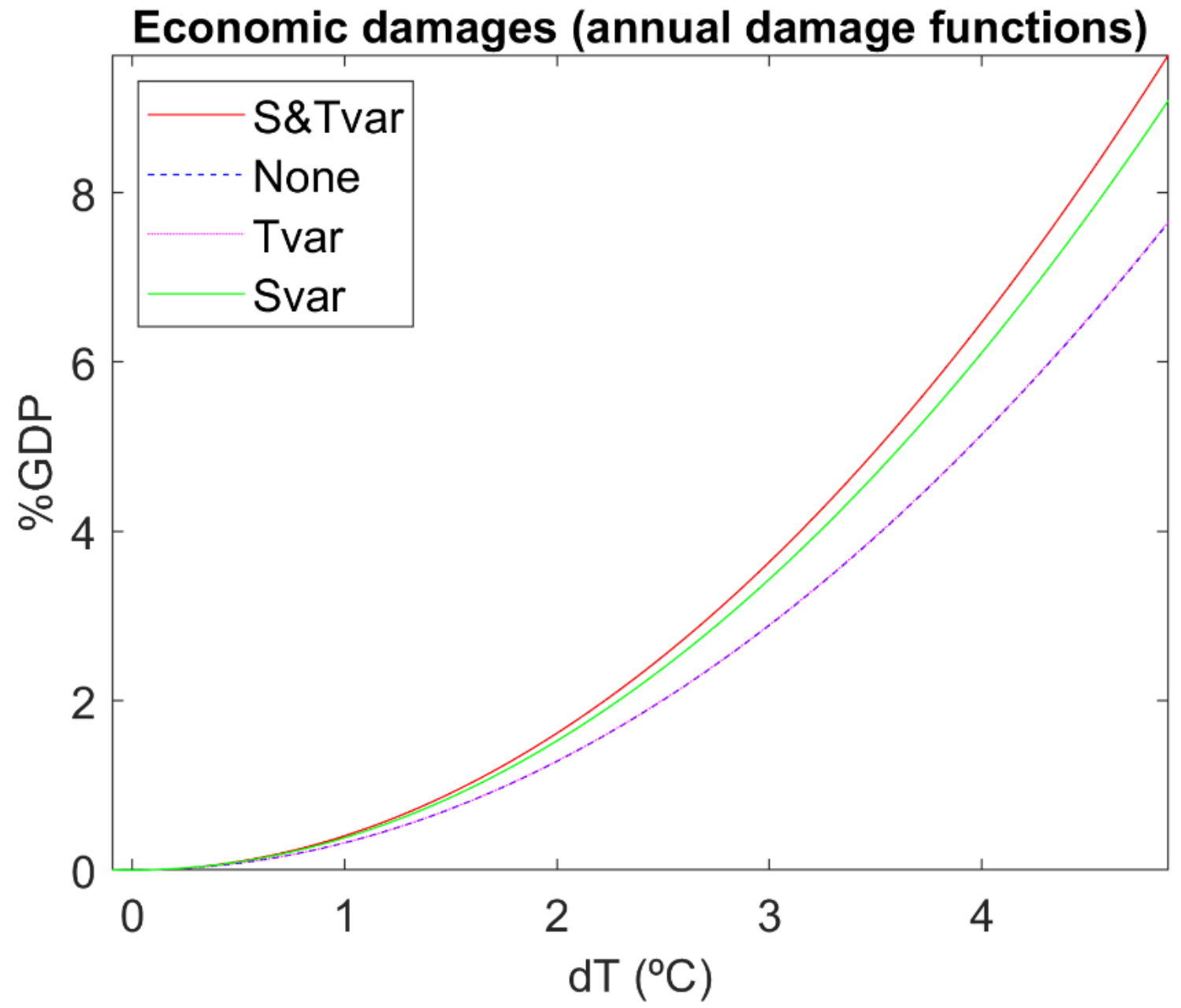

Figure 1. Projected economic damages (% GDP) as a function of changes in global temperature according to the aggregated annual damage functions and the SSP585 scenario. S&Tvar (red), Tvar (dotted red) and Svar (green) denote the use of damage functions that include both spatial variation and temporal variability, and temporal variability and spatial variation alone, respectively. The slashed blue line labeled None shows the results produced with the damage function that does not include any type of variability in temperature change.

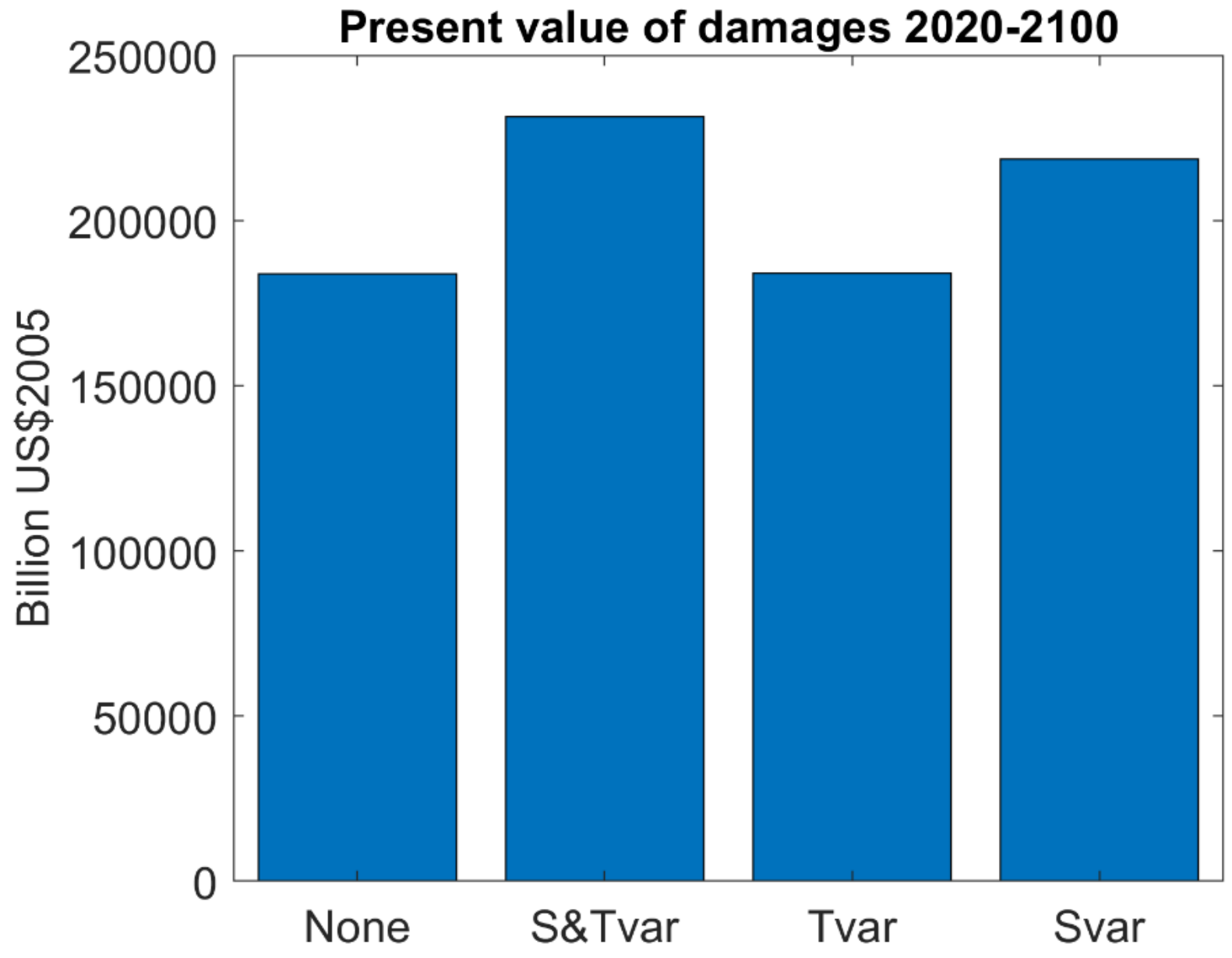

Figure 2. Present value of losses accumulated over 2020-2100 for the SSP585 scenario. S&Tvar, Tvar and Svar denote the use of damage functions that include both spatial variation and temporal variability, and temporal variability and spatial variation alone, respectively. The bar labeled None shows the results produced with the damage function that does not include any type of variability in temperature change. Calculations are based on a 4% discount rate.

Figure 3 shows the present values calculated per sector under the SSP5-8.5, using monthly damage functions and different assumptions about temperature change variability (see Tables S6-S11 for other SSP scenarios and climate models). Except for *Time use*, climate change has negative effects in all sectors over this century and their magnitudes are highly dependent on the assumptions made about the temporal variability and spatial variation of warming. The largest losses correspond to the S&Tvar damage functions. Compared with the losses obtained under the no-variability assumption, the aggregated present value is about US$47 trillion larger and the sectors with larger differences are *Health* (US$12 trillion), *Amenity* (US$10.5 trillion), *Catastrophe* (US$6.3 trillion) and *Time use* for which climate change represents considerable benefits (US$9.1 trillion). In the case of the Svar damage function, the aggregated present value is about $35 trillion larger when compared with the damage function that assumes no variability. Although the magnitudes are smaller, the sectors with larger losses are the same: *Health* ($6.6 trillion), *Amenity* ($5.8 trillion), *Catastrophe* ($4.6 trillion) and *Time use* (-$5.0 trillion). The differences in present values between Tvar and the original no-variability damage functions are much smaller, about $813 billion. Tvar has a more complex effect on how damages compare with the no-variability damage function. Due to the variability in

warming, and for some sectors that have strong seasonal dependence, the present values are smaller than those obtained using the no-variability damage function (*Agriculture, Water, Air pollution, Forestry*). This highlights the importance not only of incorporating temperature change variability, but also of developing intra-annual, sectoral damage functions.

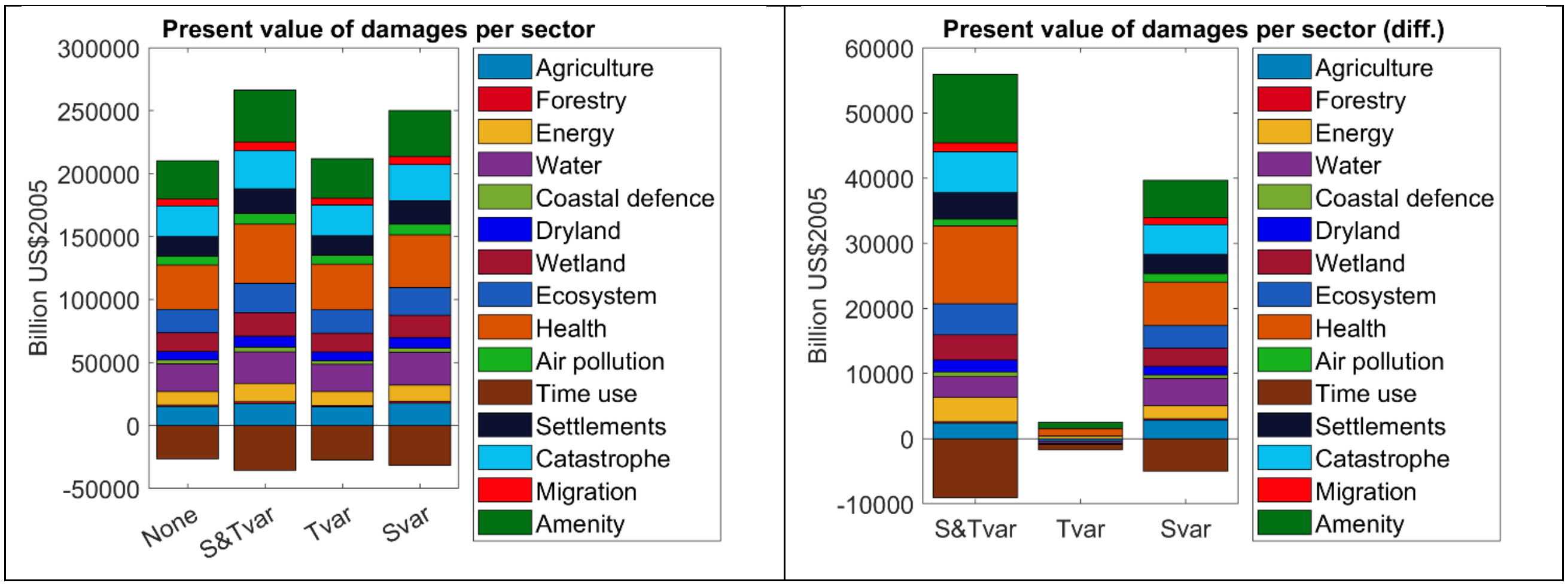

Figure 3. Present value of the economic losses over 2020-2100 per sector. Panel a) shows the present value of economic losses per sector for different assumptions about temperature change variability. Panel b) depicts the present value of differences between the projections using the damage function in equation (1) and from those that allow for different types of variability. S&Tvar, Tvar and Svar denote the use of damage functions that include both spatial variation and temporal variability, and temporal variability and spatial variation alone, respectively. The bar labeled None shows the results produced with the damage function that does not include any type of variability in temperature change. Calculations are based on a 4% discount rate.

*Social cost of carbon under different temperature change variability assumptions and SSP scenarios.*

The social cost of carbon (SCC) is a commonly used metric to inform decision-making about the impacts of adding an extra metric ton of $CO_2$ into the atmosphere, or put in a different way, the benefits society can have by mitigating a ton of $CO_2$[24,32,33]. The SCC can be used in cost-benefit analysis of greenhouse gas mitigation projects, and its level can guide setting the optimal (Pigouvian) carbon tax[29]. The SCC was estimated using a simple and physically robust response function of the marginal global mean near-surface temperatures to a one-time pulse of 1 GtC[34,35] released in 2020 (see Methods). This response was added to a set of baseline SSP global temperature scenarios. These scenarios represent a variety of narratives that encompass very high emissions and very high economic growth scenarios (SSP585), one consistent with current trends (SSP370), a middle of the road scenario (SSP245), and two high economic growth scenarios with the successful implementation of stringent mitigation policies (SSP126 and SSP119).

Figure 4a shows the estimated SCC values per ton of $CO_2$ for each of the damage functions discussed in this paper and for the SSP585, SSP370, SSP126 and SSP119 scenarios. These estimates are based on a 4% discount rate, but a sensitivity analysis based on 3% and 1.5% discount rate is also included Table S12. In all cases the results show that including spatial variation and temporal variability (S&Tvar) leads to considerably higher SCC values, followed by the S variability damage function, and that temporal variability alone has a marginal effect on SCC. As expected, the largest values of SCC occur under the SSP585 reaching about US$62.24/t$CO_2$ in the case of the no-variability damage function and up to US$78.34/t$CO_2$ with the S&T variability damage function. The S&T estimate for a 3% discount rate is US$120.10/t$CO_2$ (Table S12), which is more than twice that of the US government under the Biden administration for this discount rate[36], and similar to the price of the carbon tax in Sweden (US$130/t$CO_2$) which is one of the highest in the world. These figures provide further support substantial climate taxes as high as those in countries with the most stringent climate policy[29].

Despite the large changes in climate this trajectory implies, the SCC values for the SSP370 are the lowest due to the much slower economic growth that is associated with such socioeconomic scenario. Using the no-variability damage functions, the SCC value reaches US$25.08/t$CO_2$ compared with US$31.57/t$CO_2$ using the S&T variability damage functions.

The SSP2-4.5 and the SSP1-2.6 and SSP1-1.9 imply considerably different levels of warming: while the SSP2-4.5 leads to about 2.72ºC increase in 2100 with respect to preindustrial values, the warming under the SSP1-2.6 and SSP1-1.9 is about 1.77ºC and 1.43ºC, respectively. At the end of this century, the GDP levels under the SSP2 and SSP1 are very similar, but economic growth is faster during midcentury in the SSP1. Due to the combinations of warming and growth rates, the SCC value is slightly higher under the SSP1-2.6 scenario reaching about US$38.76/t$CO_2$ (US$30.79/t$CO_2$) using the S&Tvar (None) damage functions; for the SSP2-4.5, the SCC values are US$38.76/t$CO_2$ (US$30.42/t$CO_2$) with the S&Tvar (None) damage functions. Slightly lower SCC values are obtained with the SSP1-1.9 of about US$35.30/t$CO_2$ (US$28.05/t$CO_2$) dollars using the S&Tvar (None) damage functions.

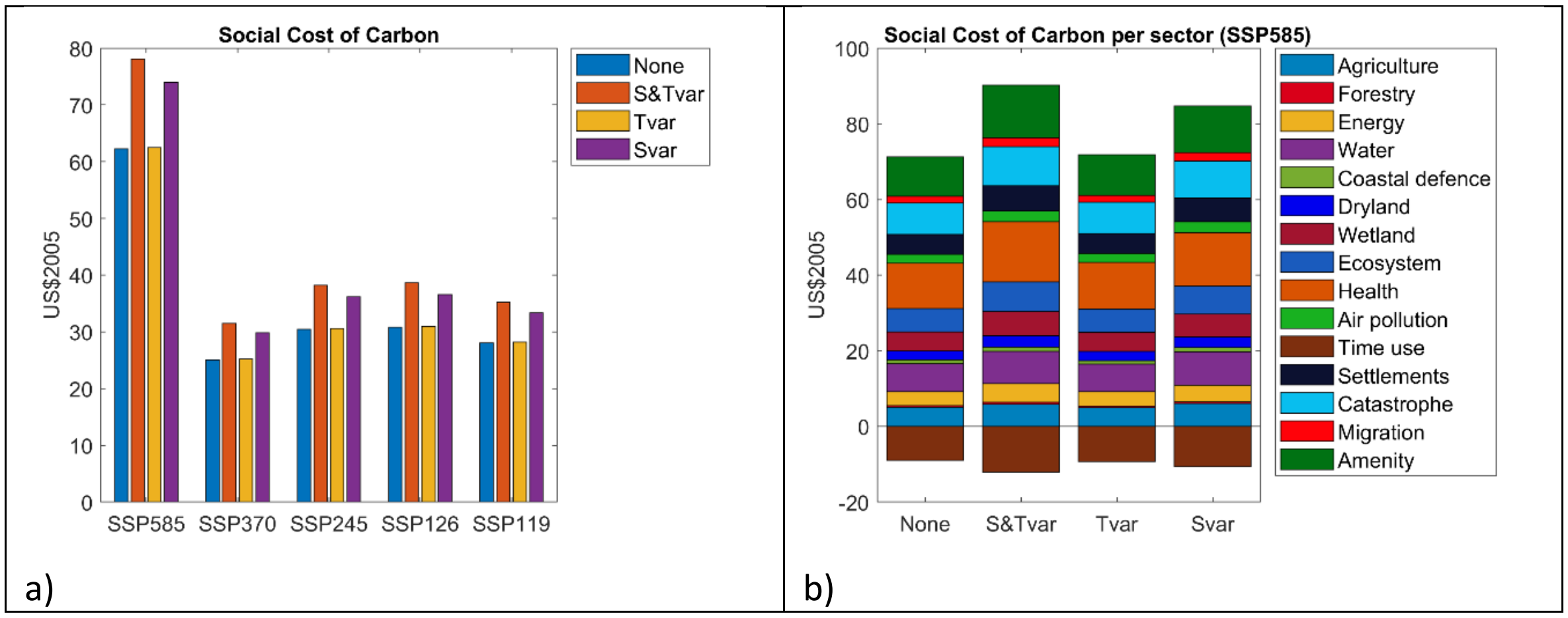

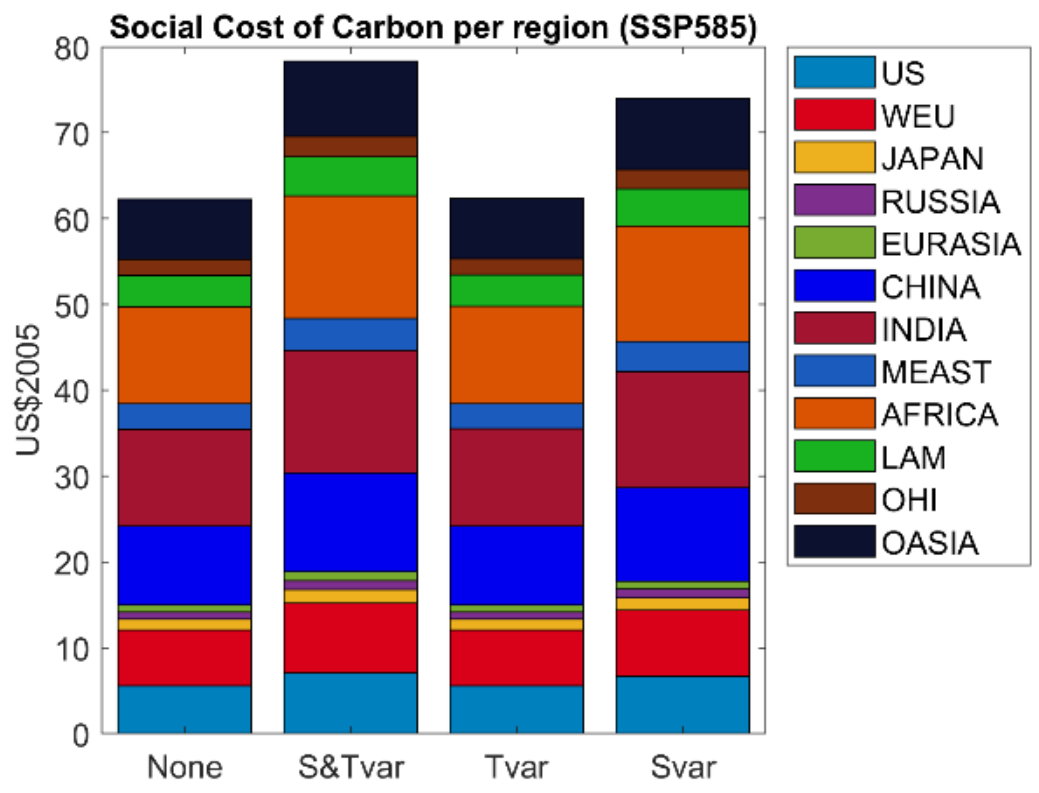

c)

Figure 4. Estimates of the SCC ($tCO_2$) for five SSP scenarios, 15 sectors and 12 world regions. Panel a) shows the SCC values for five SSPs and four damage functions with different assumptions about temperature change variability. Panel b) shows the SCC values for 15 sectors and four damage functions with different assumptions about temperature change variability. Panel c) shows the SCSC values for 12 world regions and four damage functions with different assumptions about temperature change variability. S&Tvar, Tvar and Svar denote the use of damage functions that include both spatial variation and temporal variability, and temporal variability and spatial variation alone, respectively. The bar labeled None shows the results produced with the damage function that does not include any type of variability in temperature change. Calculations are based on a 4% discount rate.

The SCC was decomposed into 15 sectors, which add up only approximately to the SCC totals described in the previous paragraph due to differences in warming between the various months and annual warming (Table S13). A sensitivity analysis of the discount rate is presented in Tables S14-S15. Figure 4b shows the results for the SSP5-8.5 scenario obtained using the four sets of monthly sectoral damage functions (results for all SSPs considered are shown in Figure S5). The sector with the highest SCC value is *Health (S&Tvar: US$15.96/*$tCO_2$ *; None: US$11.89/*$tCO_2$*)* which represents about 20% of the total SCC*,* followed by *Amenity (S&Tvar: US$13.92/*$tCO_2$*; None: US$10.37/*$tCO_2$*;* about 18% of total SCC*)*, *Catastrophe (S&Tvar: US$10.35/*$tCO_2$*; None: US$8.22/*$tCO_2$*;* about 13% of total SCC*)* and *Water (S&Tvar: US$8.50/*$tCO_2$*; None: US$7.42/*$tCO_2$*;* about 12% of total SCC*)*, while the only sector with negative SCC value is *Time use (S&Tvar: -US$12.06/*$tCO_2$*; None: -US$8.98/*$tCO_2$*)*. These estimates allow to identify that about 60% of the total SCC comes from impacts on four sectors (Health, Amenity, Water, and Ecosystems). Most of the damages occur in the winter and summer, and that damages in 3 of these sectors would exceed 0.1% of global GDP per month during this decade (Figure S4). This suggests prioritizing adaptation in these sectors and seasons is key for reducing climate change impacts. For some sectors, the SCC values obtained from the *Svar* damage functions are greater than those from the *S&Tvar* functions (*Agriculture*, *Forestry*, *Water* and *Air pollution*; Tables S13-S15). In these sectors, the SCC values from the Tvar damage functions are smaller than those obtained from the damage functions that consider no variability (*None*). Similar to what is discussed in the previous section, these results are produced in sectors that have strong seasonal dependence and in which at least some of the relevant months show lower warming than the annual mean warming.

Using modified versions of the damage functions from the RICE model (Methods; Table S16), the global SCC estimates were decomposed into 12 world regions. Figure 4c shows the SCC values for the SSP5-8.5 while Tables S17-S19 show the results for the SSP370, SSP245, SSP126 and SSP119, and three discount rates (4%, 3%, and 1.5%) The damages produced by an extra ton of CO2 are distributed in a highly heterogeneous way, concentrating in the most vulnerable regions. The highest SCC values occur for India (*S&Tvar: US$14.28/t*$CO_2$*; None: US$11.35/t*$CO_2$), Africa (*S&Tvar: US$14.25/t*$CO_2$*; None: US$11.32/t*$CO_2$), and China (*S&Tvar: US$11.55/t*$CO_2$*; None: US$9.17/t*$CO_2$). These three regions account for about 45% to 51% of the global SCC value, depending on the SSP scenario. In contrast, the SCC values are about one order of magnitude lower for Eurasia (*S&Tvar: US$0.99/t*$CO_2$*; None: US$0.79/t*$CO_2$), Russia (*S&Tvar: US$1.07/t*$CO_2$*; None: US$0.85/t*$CO_2$) and Japan (*S&Tvar: US$1.55/t*$CO_2$*; None: US$1.23/t*$CO_2$). The SCC for the most developed regions (US, WEU, OHI) represents 22% to 25% of the global SCC value.

## Conclusions

Most IAMs and economic assessments of the impacts of climate change rely on the assumption that changes in the mean of a chosen index (e.g., global mean annual temperature) provide an adequate representation of the effects of this phenomenon on natural and human systems. Although this assumption helps keeping IAMs as simple and tractable as possible, it has two important limitations. First, it fails to take advantage of the advancement that climate modelling has had over the last 30 years in producing spatially explicit projections with better representation of low- and high-frequency variability[37], and of the development of methods to emulate the output of complex climate models with much simpler and low computing costs[17,38]. Second, as is shown here, this simplification produces nontrivial biases in the estimated costs. Contrary to past studies which have focus on evaluating the effects temporal variability, here we show that omitting spatial variation produces large downward biases on the estimates of the costs of climate change, and that the interaction between temporal variability and spatial variation increases these biases. We show that current IAMs can be easily extended to account for these limitations and provide a more accurate representation of the effects of changes in climate both in the mean and its variability. We develop a set of damage functions that are able to account for temporal and spatial variability in temperature change. Moreover, annual, seasonal and monthly sectoral damage functions are proposed to fully take advantage of seasonal and spatially explicit temperature change projections.

Results show that accounting for variability in warming affects the expected level of losses, and not only the risk premium, as proposed in previous publications. Spatial variation and temporal variability together always lead to higher present value of losses and SCC estimates when aggregated across sectors. However, for sectors in which there is a strong seasonal dependence and in which the warming in months that a relevant for the sector is lower than the annual average, economic losses can be smaller than under the assumption of no seasonal variability in warming.

We show that traditional damage functions that rely on annual mean global temperature change may severely underestimate the expected losses from climate change. Under the SSP585 scenario, by 2100 global annual losses are projected to be about US$17-US$28 trillion dollars higher when spatial variation and temporal variability are accounted for, in comparison to commonly used damage functions that ignore warming variability. Similarly, the corresponding present value of losses during this century (2020-2100) is about US$47-US$66 trillion dollars higher using the S&Tvar damage functions. It is important to underline that the reported estimates are based on conservative calibrations of global damage functions and a relatively high discount rate is used (4%) for the main estimates. Comparing with previous estimates from Calel et al. there are differences to point out. Calel e al.[18] report that the expected damages do no change when accounting for this factor, but that the risk premium does by US$32 trillion. Here we show that spatial variation and temporal variability do affect the expected value of damages, increasing losses in the range of $47-$66 trillion or about ½ to ¾ of annual global GDP in 2020. These estimates are much larger than those in Calel et al.[18] although they use one of the damage functions that produce highest damages in the literature[39]. Another important difference is that Calel et al. conclude that accounting for temporal variability has no effect on the social cost of carbon. Our results show that ignoring spatial variation and temporal variability produces an important negative bias on SCC estimates.

The SCC estimates are also highly sensitive to the assumptions about warming variability embedded in damage functions. If spatial and temporal variability are included, the SCC values are about 25% higher than under the assumptions in traditional damage functions. Moreover, developing monthly sectoral damage functions allow to understand better how sectors with strong seasonal components can be affected by climate change and, in some cases, to avoid over estimating losses.

The damage functions developed in this paper are expected to be particularly useful when applied to climate scenarios in which warming spatial and temporal patterns are highly modified. This type of scenarios include climate catastrophes such as the shutdown of the thermohaline circulation, the Amazon dieback, and the disintegration of Wests Antarctica icesheet. Overly optimistic estimates of the costs and consequences of such events may be largely due to current IAMs ignoring large-scale and profound changes in climatic patterns would bring and concentrating only in the changes these events produce over global mean warming.

## Methods

### Data and methods

*Climate and economic data*

We use the global monthly gridded dataset of near surface temperature projection of the ensemble mean of the Coupled Model Intercomparison Project Phase 6 (CMIP6; one realization for each of the 37 climate models included). The sample period is 1850-2100. The SSP585 scenario was chosen to maximize the climate signal, although sensitivity analyses are presented for the SSP119, SSP126, SSP245 and SSP370 in the Supplementary Information. We used realizations from different climate models with various ensemble sizes: ACCESS-CM2 (ensemble mean of 3 realizations), CanESM (ensemble mean of 25 realizations), CESM2 (ensemble mean of 25 realizations), INM-CM5-0 (one realization), MPI-ESM1-2-HR (one realization), and ACCESS-ESM1-5 (one realization). All simulations are expressed as anomalies with respect to the 1850-1880 period. For each model, we used two different procedures to separate the climate signal from the weather (irregular) component. The first procedure (SP) is particularly convenient for IAMs. It involves calculating the scaling patterns from general circulation climate models' simulations, which allow to produce spatially explicit climate projections using a simple zero-dimensional climate model [14,16,40]. To generate these scaling patterns, we use ordinary least squares to estimate the slope parameters in regressions between local and global temperature for each grid cell with geographical coordinates $i, j$ [41]. To construct spatially explicit future scenarios, these scaling patterns are then scaled with global temperature change projections. All global temperature projections used here were generated with MAGICC7 [22,42]. The use of different models to obtain the corresponding spatial patterns of change, scales by a single realization of global temperature (per SSP investigated), allows to evaluate the effects of variability in a separate manner from the uncertainty in climate sensitivity. In the second procedure (RM), we approximate the climate signal using a rolling mean smoothing procedure with a 31-year window for each grid cell. This averaging minimizes the effects of high- to low-frequency oscillations and allows to approximate the underlying climate of the chosen region. This procedure also helps to alleviate some of the problems caused by nonstationarity in climate variables[23,43].

The effect of a pulse of 1GtC on global temperatures is approximated using the stylized response function proposed by Ricke and Caldeira[34], which provides an option that is physically sound and easy to implement in IAMs for estimating SCC values. It is composed of three-exponential function described by $\Delta T_t = -(a_1 + a_2 + a_3) + a_1 e^{-\frac{(t-t_0)}{\tau_1}} + a_2 e^{-\frac{(t-t_0)}{\tau_2}} + a_3 e^{-\frac{(t-t_0)}{\tau_3}}$, where the units of $\Delta T_t$ are mK/GtC. The SCC estimates presented here are US dollars per tCO2, which is converted from tC by multiplying by 12/44.

To represent economic growth for the remaining of this century (2020-2100), we use the estimates of global and for the RICE regions GDP produced by the OECD Env-Growth for the SSP5 narrative, included in the SSP Database[44]. These projections were linearly interpolated from 5-year to annual frequencies.

**Derivation of the proposed damage functions**

*Stochasticity*

The damage function originally proposed by William Nordhaus to transform global mean annual surface temperature increase from pre-industrial times ($\overline{\Delta T_t}$) into economic damages ($I_t$) has been updated over the years, but its main features have remained the same[24–27]. It consists of a power (=2) function of $\overline{\Delta T_t}$, scaled by some parameter values that represent the loss in global welfare for a given temperature change and a calibration temperature value ($C$). The various versions of this damage function can be succinctly described by the following equation:

$$I_t = a\left(\frac{\overline{\Delta T_t}}{C}\right)^2 = \alpha(\overline{\Delta T_t})^2 \qquad (1)$$

Where $a$ represents the losses estimated for an increase of $C$ in global temperature and $\alpha = \left(\frac{a}{C}\right)$. By means of a review of published estimates based on Tol (2009,2014; see Supplementary Information S1), the selected values for $a$ and $C$ we use here are 2.01% GDP and 2.5ºC, respectively.

It is important to note that $\overline{\Delta T_t}$ denotes the average (expected value) of temperature change over the entire planet and a sufficiently long period of time (e.g., 30 years). Making use of the derivation of the variance, $var(X) = E[X^2] - E[X]^2$, so that $E[X^2] = E[X]^2 + var(X)$, the damage function in equation (1) can be written as:

$$I_t^* = \alpha(\overline{\Delta T_t})^2 + \alpha[var(\Delta T_t)] \qquad (2)$$

Note that if $var(\Delta T_t) = 0$, equation (2) is reduced to equation (1). That is, Nordhaus' original formulation implicitly assumes that there is no temporal nor spatial variation in the temperature change. Alternatively, it assumes that the variance of the change in temperature changes in direct proportion to the change in temperature squared: $[var(\Delta T_t)] = \beta(\overline{\Delta T_t})^2$. To incorporate the effects of variability in temperature, the assumption of zero variance/proportionality is relaxed. The damages that include temperature variability are denoted $I_t^*$ and are computationally more efficient to calculate as follows:

$$I_t^* = \alpha(\overline{\Delta T_t})^2 + \alpha[var(\Delta T_t)] = \alpha E[(\Delta T_t^2)] = \alpha\overline{\Delta T_t^2} \qquad (3)$$

The difference between equations (1) and (3) is that the first calculates the damages associated with the average temperature change over space and time domains, while the second represents the average damages of temperature change over both dimensions. As a simple illustration consider the case of two numbers that are not the same, for example 1 and 3. If the average of these two numbers is taken first and then raised to the power of 2, the result is $[(1+3)/2]^2 = 2^2 = 4$. In contrast, if the numbers are raised to the power of two and then the average of the two is taken, the result is $(1^2+3^2)/2 = 10/2 = 5$. If there is variability in the data, the average of a sum of squares will be larger than the average of the same set of numbers risen to the power of 2. Since damage functions evaluate the losses caused by changes in climate, the value of $I_t^*$ in the reference year (i.e., preindustrial times) is subtracted from the $I_t^*$ sequence.

*Sectoral impacts*

The damage functions expressed by equations 1-3 assume that the impacts of climate change are uniform over the year. This is expected to be true for some of them such as sea level rise, which is driven by thermal expansion and ice melt, that is cumulative warming rather than warming in a specific month. However, such impacts as the demand for space heating or cooling, or cold- and heat-related mortality and morbidity are tied to the projected warming in winter and summer, respectively. We can thus improve on the naive calibration of Equation (3). We first compiled results from past studies in which the economic impacts of climate change have been estimated (Supplementary Information S1), particularly those in which sectoral decomposition is included. There are four studies that provide estimates for sectoral impacts[19,29,45], and they are summarized in Table S1. The total damage from all sectors is estimated to be 2.01% of global GDP for a 2.5ºC warming, where 40% and 46% of them are market and non-market damages, respectively, and 13% correspond to risk premium (catastrophes)[29].

*Seasonality*

As a first step to derive seasonal and monthly damage functions, for each sector seasonal weights are assigned to each sector depending on how active the sector is at that time of year (Table S2). The seasonal damage parameters become:

$$\alpha_{k,j}^s = \omega_{k,j}^s \alpha_j \qquad (4)$$

Where $\alpha_{k,j}^s$ and $\omega_{k,j}^s$ are the parameter of the seasonal damage function for season and the weight for season $k$ and sector $j$, and $\alpha_j$ is the original damage parameter obtained form the literature for sector $j$. Using these weights, the total damages per season for 2.5ºC warming are estimated to be 0.29%, 0.92%, 0.29% and 0.52% for spring, summer, fall and winter, respectively.

The seasonal weights and the differences in total GDP per hemisphere are used to derive monthly damage functions as described in the following equation:

$$\alpha_{l,j}^{m} = \omega_{h}^{GDP} \frac{1}{3} \alpha_{k,j}^{s} \quad (5)$$

Where $\alpha_{l,j}^{m}$ is the monthly damage coefficient for month $l$ and $l \in k$, $\omega_{h}^{GDP}$ is the ratio of GDP produced in hemisphere $h$ ($77,800 and $9,470 billion USD for the northern and southern hemispheres, respectively) and global GDP ($87,270 billion USD). The monthly damage coefficients imply economic losses (%GDP) for a 2.5ºC in warming of about 1.10% for December, January and February, 0.55% for March, April and May and, 1.69% for June, July and August, and 0.55% for September, October and November. Note that the sum of monthly (as well as seasonal) damages equals the annual total 2.01%. Figure S1 shows the seasonal and monthly aggregated damages as percent of GDP for 1ºC to 6ºC increase in annual mean global temperatures.

*Regional damage functions*

We adapt the regional damage functions included in the RICE model[19,20] to include temporal and spatial variability. These 12 regional functions project economic damages based on changes in annual mean global temperature. The regions included in the RICE model are the US, Western Europe (WEU), Japan, Russia, Eurasia, China, India, Middle East (MEAST), Africa, Latin America (LAM), other Asian countries (OASIA), and other high income countries (OHI)[19]. As first step, we standardized the specification of these regional functions, as half of them (China, India, EAST, Africa, LAM, and OASIA) include a linear term in addition to the quadratic function in equation (1). We use an optimization method to find the values of parameter $\alpha_r$ for each region *r* that minimizes the squared differences between the original functions and the one that only includes a quadratic term for a range of global temperature change from 0 to 6ºC. This approximation provides very good fit, with $R^2$ values larger than 0.99 in all cases. The original and fitted parameter values are provided in Table S16. For each region, the proposed damage functions that include temporal and spatial variability are obtained following equations (1) to (3).

Note that results from global damage functions such as the one in (1) and the aggregated total obtained from regional damage functions (like as those in RICE) do not exactly match[19]. To avoid discrepancies between results here we scale regional results to exactly match those based in equation (1). The scaling factor is the ratio of the results from the global damage function and the sum of those obtained for each region with RICE. This ensures that regional and global results are exactly consistent, while it maintains the regional differences generated by the region-specific damage functions.

References

1. World Meteorological Organization. *Guide to climatological practices*. (World Meteorological Organization Geneva, Switzerland, 2011).

2. Bathiany, S., Dakos, V., Scheffer, M. & Lenton, T. M. Climate models predict increasing temperature variability in poor countries. *Sci. Adv.* **4**, (2018).

3. Estrada, F., Wouter Botzen, W. J. & Tol, R. S. J. Economic losses from US hurricanes

consistent with an influence from climate change. *Nat. Geosci.* **8**, 880–885 (2015).

4. Nordhaus, W. D. The economics of hurricanes and implications of global warming. *Clim. Chang. Econ.* **01**, 1–20 (2010).

5. Tol, R. S. J. & Fankhauser, S. On the representation of impact in integrated assessment models of climate change. *Environ. Model. Assess.* **3**, 63–74 (1998).

6. Nordhaus, W. D. *The Climate casino : risk, uncertainty, and economics for a warming world*. (Yale University Press, 2013).

7. Estrada, F. & Tol, R. S. J. Toward Impact Functions for Stochastic Climate Change. *Clim. Chang. Econ.* **06**, 1550015 (2015).

8. Estrada, F., Tol, R. S. J. & Botzen, W. J. W. Global economic impacts of climate variability and change during the 20th century. *PLoS One* **12**, e0172201 (2017).

9. Dosio, A. & Fischer, E. M. Will Half a Degree Make a Difference? Robust Projections of Indices of Mean and Extreme Climate in Europe Under 1.5°C, 2°C, and 3°C Global Warming. *Geophys. Res. Lett.* **45**, 935–944 (2018).

10. Qian, C. & Zhang, X. Human influences on changes in the temperature seasonality in mid- to high-latitude land areas. *J. Clim.* **28**, 5908–5921 (2015).

11. Swanson, K. L., Sugihara, G. & Tsonis, A. A. Long-term natural variability and 20th century climate change. *Proc. Natl. Acad. Sci.* **106**, 16120–16123 (2009).

12. Estrada, F., Kim, D. & Perron, P. Spatial variations in the warming trend and the transition to more severe weather in midlatitudes. *Sci. Rep.* **11**, 145 (2021).

13. Cohen, J. *et al.* Divergent consensuses on Arctic amplification influence on midlatitude severe winter weather. *Nat. Clim. Chang.* **10**, 20–29 (2020).

14. Tebaldi, C. & Arblaster, J. M. Pattern scaling: Its strengths and limitations, and an update on the latest model simulations. *Clim. Change* **122**, 459–471 (2014).

15. Kravitz, B., Lynch, C., Hartin, C. & Bond-Lamberty, B. Exploring precipitation pattern scaling methodologies and robustness among CMIP5 models. *Geosci. Model Dev.* **10**, 1889–1902 (2017).

16. Santer, B. D., Wigley, T. M. L., Schlesinger, M. E. & Mitchell, J. F. B. Developing climate scenarios from equilibrium GCM results. *Report/Max-Planck-Institut für Meteorol.* **47**, (1990).

17. Estrada, F. & Botzen, W. J. W. Economic impacts and risks of climate change under failure and success of the Paris Agreement. *Ann. N. Y. Acad. Sci.* **1504**, 95–115 (2021).

18. Calel, R., Chapman, S. C., Stainforth, D. A. & Watkins, N. W. Temperature variability implies greater economic damages from climate change. *Nat. Commun. 2020 111* **11**, 1–5 (2020).

19. Nordhaus, W. D. & Boyer, J. *Warming the world: economic models of global warming*. (MIT press, 2003).

20. Nordhaus, W. D. Economic aspects of global warming in a post-Copenhagen environment. *Proc. Natl. Acad. Sci. U. S. A.* **107**, 11721–6 (2010).

21. Lynch, C., Hartin, C., Bond-Lamberty, B. & Kravitz, B. An open-access CMIP5 pattern library for temperature and precipitation: description and methodology. *Earth Syst. Sci. Data* **9**, 281–292 (2017).

22. Meinshausen, M. *et al.* The shared socio-economic pathway (SSP) greenhouse gas concentrations and their extensions to 2500. *Geosci. Model Dev.* **13**, 3571–3605 (2020).

23. Arguez, A. & Vose, R. S. The Definition of the Standard WMO Climate Normal: The Key to Deriving Alternative Climate Normals. *Bull. Am. Meteorol. Soc.* **92**, 699–704 (2010).

24. Tol, R. S. J. The Economic Effects of Climate Change. *J. Econ. Perspect.* **23**, 29–51 (2009).

25. Nordhaus, W. Projections and uncertainties about climate change in an era of minimal climate policies. *Am. Econ. J. Econ. Policy* **10**, 333–360 (2018).

26. de Bruin, K. C., Dellink, R. B. & Tol, R. S. J. AD-DICE: an implementation of adaptation in the DICE model. *Clim. Change* **95**, 63–81 (2009).

27. Estrada, F., Tol, R. S. J. & Botzen, W. J. W. Extending integrated assessment models' damage functions to include adaptation and dynamic sensitivity. *Environ. Model. Softw.* **121**, 104504 (2019).

28. Nordhaus, W. D. Roll the Dice again: the economics of global warming. *Draft Version* (1999).

29. Tol, R. S. J. The impact of climate change and the social cost of carbon. in *Routledge Handbook of Energy Economics* (eds. Soytas, U. & Sari, R.) 253–273 (Taylor and Francis, 2019).

30. Tol, R. S. J. Correction and Update: The Economic Effects of Climate Change †. *J. Econ. Perspect.* **28**, 221–226 (2014).

31. Chen, D. *et al.* Framing, Context, and Methods. in *Climate Change 2021: The Physical Science Basis. Contribution of Working Group I to the Sixth Assessment Report of the Intergovernmental Panel on Climate Change* (eds. Masson-Delmotte, V. et al.) 147–286 (Cambridge University Press, 2021).

32. Nordhaus, W. D. Revisiting the social cost of carbon. *Proc. Natl. Acad. Sci. U. S. A.* **114**, 1518–1523 (2017).

33. Tol, R. S. J. The Social Cost of Carbon: Trends, Outliers and Catastrophes. *Economics* **2**, (2008).

34. Ricke, K. L. & Caldeira, K. Maximum warming occurs about one decade after a carbon dioxide emission. *Environ. Res. Lett.* **9**, 124002 (2014).

35. Ricke, K., Drouet, L., Caldeira, K. & Tavoni, M. Country-level social cost of carbon. *Nat. Clim. Chang.* **8**, 895–900 (2018).

36. IAWG. *Technical Support Document: Social Cost of Carbon, Methane, and Nitrous Oxide: Interim Estimates under Executive Order 13990. US Inter Agency Working Group, Washington, DC* (2021).

37. IPCC. *Climate Change 2021: The Physical Science Basis. Contribution of Working Group I to*

*the Sixth Assessment Report of the Intergovernmental Panel on Climate Change*. (Cambridge University Press, 2021).

38. Ignjacevic, P. *et al.* CLIMRISK-RIVER: Accounting for local river flood risk in estimating the economic cost of climate change. *Environ. Model. Softw.* **132**, (2020).

39. Weitzman, M. L. On Modeling and Interpreting the Economics of Catastrophic Climate Change. *Rev. Econ. Stat.* **91**, 1–19 (2009).

40. Wigley, T. M. L. MAGICC/SCENGEN 5.3: User manual (version 2). *NCAR, Boulder, CO* **80**, (2008).

41. Lynch, C., Hartin, C., Bond-Lamberty, B. & Kravitz, B. An open-access CMIP5 pattern library for temperature and precipitation: description and methodology. *Earth Syst. Sci. Data* **9**, 281–292 (2017).

42. Meinshausen, M., Wigley, T. M. L. & Raper, S. C. B. Emulating atmosphere-ocean and carbon cycle models with a simpler model, MAGICC6 - Part 2: Applications. *Atmos. Chem. Phys.* (2011) doi:10.5194/acp-11-1457-2011.

43. Gay, C., Estrada, F. & Conde, C. Some implications of time series analysis for describing climatologic conditions and for forecasting. An illustrative case: Veracruz, México. *Atmosfera* (2007).

44. Riahi, K. *et al.* The Shared Socioeconomic Pathways and their energy, land use, and greenhouse gas emissions implications: An overview. *Glob. Environ. Chang.* **42**, 153–168 (2017).

45. Fankhauser, S. *Valuing Climate Change: The Economics of the Greenhouse*. (Routledge, 1995). doi:https://doi.org/10.4324/9781315070582.

## Supplementary Information

*Supplementary Tables*

Table S1. Summary of estimates of economic impacts (% of global GDP) per sector for a 2.5ºC increase in global temperature.

| Sector | Fankhauser | Berz | Tol | Nordhaus | Average |
|---|---|---|---|---|---|
| Agriculture | -0.20 | -0.19 | -0.13 | -0.13 | -0.16 |
| Forestry | -0.01 | -0.02 | -0.01 | -0.01 | -0.01 |
| Energy | -0.12 | -0.11 | -0.12 | -0.12 | -0.12 |
| Water | -0.24 | -0.23 | -0.24 | -0.24 | -0.24 |
| Coastal defense | 0.00 | -0.01 | -0.08 | -0.03 | -0.03 |
| Dryland | -0.07 | -0.07 | -0.09 | -0.08 | -0.08 |
| Wetland | -0.16 | -0.16 | -0.17 | -0.16 | -0.16 |
| Ecosystem | -0.21 | -0.20 | -0.19 | -0.20 | -0.20 |
| Health | -0.26 | -0.40 | -0.77 | -0.10 | -0.38 |
| Air pollution | -0.08 | -0.08 | -0.08 | -0.08 | -0.08 |
| Time use | 0.29 | 0.29 | 0.29 | 0.29 | 0.29 |
| Settlements | -0.17 | -0.17 | -0.17 | -0.17 | -0.17 |
| Catastrophe | -0.01 | -0.01 | -0.01 | -1.02 | -0.27 |
| Migration | -0.02 | -0.02 | -0.12 | -0.07 | -0.06 |
| Amenity | -0.33 | -0.33 | -0.33 | -0.33 | -0.33 |
| Total | -1.61 | -1.71 | -2.24 | -2.47 | -2.01 |

Table S2. Seasonal weights per sector used to calibrate the proposed damage functions.

| **Sector** | **Spring** | **Summer** | **Fall** | **Winter** |
|---|---|---|---|---|
| **Agriculture** | 0.25 | 0.5 | 0.25 | 0 |
| **Forestry** | 0.25 | 0.5 | 0.25 | 0 |
| **Energy** | 0 | 0.5 | 0 | 0.5 |
| **Water** | 0 | 1 | 0 | 0 |
| **Coastal defense** | 0.25 | 0.25 | 0.25 | 0.25 |
| **Dryland** | 0.25 | 0.25 | 0.25 | 0.25 |
| **Wetland** | 0.25 | 0.25 | 0.25 | 0.25 |
| **Ecosystem** | 0.25 | 0.25 | 0.25 | 0.25 |
| **Health** | 0 | 0.5 | 0 | 0.5 |
| **Air pollution** | 0 | 1 | 0 | 0 |
| **Time use** | 0 | 0.5 | 0 | 0.5 |
| **Settlements** | 0.25 | 0.25 | 0.25 | 0.25 |
| **Catastrophe** | 0.25 | 0.25 | 0.25 | 0.25 |
| **Migration** | 0.25 | 0.25 | 0.25 | 0.25 |
| **Amenity** | 0 | 0.5 | 0 | 0.5 |

Source: Assessment by the authors

Table S3. Differences between damage estimates from $I_t$ and $I_t^*$ incorporating spatial variation and temporal variability.

| SP 2050 | | | | | | |
|---|---|---|---|---|---|---|
| Climate model and damage function | S&Tvar | Tvar | Svar | S&Tvar (%) | Tvar (%) | Svar (%) |
| CMIP6 ensemble/annual damage function | $ 1,625.29 | $ 8.39 | $ 1,184.98 | 25.84% | 0.13% | 18.84% |
| CMIP6 ensemble/monthly damage function | $ 1,601.03 | $ 27.72 | $ 1,184.98 | 25.45% | 0.44% | 18.84% |
| ACCESS-ESM1-5/annual damage function | $ 2,225.79 | $ 6.37 | $ 1,344.89 | 35.36% | 0.10% | 21.37% |
| ACCESS-ESM1-5/monthly damage function | $ 2,244.18 | $ 25.30 | $ 1,344.89 | 35.66% | 0.40% | 21.37% |
| MPI-ESM1-2-HR/annual damage function | $ 2,130.07 | $ 14.19 | $ 1,416.10 | 33.93% | 0.23% | 22.56% |
| MPI-ESM1-2-HR/monthly damage function | $ 2,098.65 | $ (6.73) | $ 1,416.10 | 33.43% | -0.11% | 22.56% |
| INM-CM5/annual damage function | $ 2,260.94 | $ 15.64 | $ 1,377.93 | 35.98% | 0.25% | 21.93% |
| INM-CM5/monthly damage function | $ 2,129.90 | $ (12.05) | $ 1,377.93 | 33.90% | -0.19% | 21.93% |
| CESM2 ensemble/annual damage function | $ 1,435.53 | $ 7.01 | $ 914.44 | 22.84% | 0.11% | 14.55% |
| CESM2 ensemble/monthly damage function | $ 1,646.40 | $ 136.07 | $ 914.44 | 26.19% | 2.16% | 14.55% |
| CanESM5 ensemble/annual damage function | $ 1,917.96 | $ 7.88 | $ 1,352.10 | 30.51% | 0.13% | 21.51% |
| CanESM5 ensemble/monthly damage function | $ 1,813.72 | $ (20.71) | $ 1,352.10 | 28.85% | -0.33% | 21.51% |
| ACCESS-CM2 ensemble/annual damage function | $ 1,745.10 | $ 9.63 | $ 1,104.42 | 27.75% | 0.15% | 17.56% |
| ACCESS-CM2 ensemble/monthly damage function | $ 1,779.27 | $ 53.37 | $ 1,104.42 | 28.30% | 0.85% | 17.56% |
| SP 2100 | | | | | | |
| Climate model and damage function | S&Tvar | Tvar | Svar | S&Tvar (%) | Tvar (%) | Svar (%) |
| CMIP6 ensemble/annual damage function | $ 20,074.73 | $ 103.63 | $ 14,636.34 | 25.87% | 0.13% | 18.86% |
| CMIP6 ensemble/monthly damage function | $ 19,775.10 | $ 342.44 | $ 14,636.34 | 25.48% | 0.44% | 18.86% |
| ACCESS-ESM1-5/annual damage function | $ 27,491.86 | $ 78.73 | $ 16,611.39 | 35.40% | 0.10% | 21.39% |
| ACCESS-ESM1-5/monthly damage function | $ 27,719.03 | $ 312.44 | $ 16,611.39 | 35.70% | 0.40% | 21.39% |
| MPI-ESM1-2-HR/annual damage function | $ 26,309.53 | $ 175.24 | $ 17,490.93 | 33.97% | 0.23% | 22.58% |
| MPI-ESM1-2-HR/monthly damage function | $ 25,921.54 | $ (83.09) | $ 17,490.93 | 33.47% | -0.11% | 22.58% |
| INM-CM5/annual damage function | $ 27,926.04 | $ 193.18 | $ 17,019.55 | 36.02% | 0.25% | 21.95% |
| INM-CM5/monthly damage function | $ 26,307.43 | $ (148.82) | $ 17,019.55 | 33.93% | -0.19% | 21.95% |
| CESM2 ensemble/annual damage function | $ 17,731.00 | $ 86.63 | $ 11,294.77 | 22.86% | 0.11% | 14.56% |
| CESM2 ensemble/monthly damage function | $ 20,335.55 | $ 1,680.70 | $ 11,294.77 | 26.22% | 2.17% | 14.56% |
| CanESM5 ensemble/annual damage function | $ 23,689.68 | $ 97.34 | $ 16,700.43 | 30.54% | 0.13% | 21.53% |
| CanESM5 ensemble/monthly damage function | $ 22,402.13 | $ (255.85) | $ 16,700.43 | 28.88% | -0.33% | 21.53% |
| ACCESS-CM2 ensemble/annual damage function | $ 21,554.62 | $ 118.96 | $ 13,641.25 | 27.78% | 0.15% | 17.58% |
| ACCESS-CM2 ensemble/monthly damage function | $ 21,976.64 | $ 659.18 | $ 13,641.25 | 28.33% | 0.85% | 17.58% |
| RM 2050 | | | | | | |
| Climate model and damage function | S&Tvar | Tvar | Svar | S&Tvar (%) | Tvar (%) | Svar (%) |
| CMIP6 ensemble/annual damage function | $ 2,390.71 | $ 17.48 | $ 1,601.53 | 30.20% | 0.22% | 20.23% |
| CMIP6 ensemble/monthly damage function | $ 2,160.19 | $ (37.35) | $ 1,601.53 | 27.29% | -0.47% | 20.23% |
| ACCESS-ESM1-5/annual damage function | $ 3,161.66 | $ 18.27 | $ 1,904.61 | 44.40% | 0.26% | 26.75% |
| ACCESS-ESM1-5/monthly damage function | $ 2,931.09 | $ (35.89) | $ 1,904.61 | 41.16% | -0.50% | 26.75% |
| MPI-ESM1-2-HR/annual damage function | $ 1,877.47 | $ 14.11 | $ 1,191.16 | 38.47% | 0.29% | 24.40% |
| MPI-ESM1-2-HR/monthly damage function | $ 1,684.91 | $ (71.50) | $ 1,191.16 | 34.52% | -1.46% | 24.40% |
| INM-CM5/annual damage function | $ 1,965.67 | $ 14.62 | $ 1,217.40 | 37.42% | 0.28% | 23.18% |
| INM-CM5/monthly damage function | $ 1,726.14 | $ (50.16) | $ 1,217.40 | 32.86% | -0.95% | 23.18% |
| CESM2 ensemble/annual damage function | $ 3,483.14 | $ 18.42 | $ 2,052.91 | 37.67% | 0.20% | 22.20% |
| CESM2 ensemble/monthly damage function | $ 3,326.88 | $ 16.95 | $ 2,052.91 | 35.98% | 0.18% | 22.20% |
| CanESM5 ensemble/annual damage function | $ 7,175.67 | $ 47.22 | $ 4,610.83 | 45.29% | 0.30% | 29.10% |
| CanESM5 ensemble/monthly damage function | $ 6,514.98 | $ (160.22) | $ 4,610.83 | 41.12% | -1.01% | 29.10% |
| ACCESS-CM2 ensemble/annual damage function | $ 2,903.75 | $ 24.02 | $ 1,514.28 | 35.91% | 0.30% | 18.73% |
| ACCESS-CM2 ensemble/monthly damage function | $ 2,727.67 | $ 39.69 | $ 1,514.28 | 33.74% | 0.49% | 18.73% |
| RM 2085 | | | | | | |
| Climate model and damage function | S&Tvar | Tvar | Svar | S&Tvar (%) | Tvar (%) | Svar (%) |
| CMIP6 ensemble/annual damage function | $ 14,522.50 | $ 82.37 | $ 10,381.18 | 26.25% | 0.15% | 18.77% |
| CMIP6 ensemble/monthly damage function | $ 13,823.98 | $ 35.05 | $ 10,381.18 | 24.99% | 0.06% | 18.77% |
| ACCESS-ESM1-5/annual damage function | $ 17,201.24 | $ 56.68 | $ 10,887.57 | 35.77% | 0.12% | 22.64% |
| ACCESS-ESM1-5/monthly damage function | $ 16,830.96 | $ 71.93 | $ 10,887.57 | 35.00% | 0.15% | 22.64% |
| MPI-ESM1-2-HR/annual damage function | $ 10,849.32 | $ 66.38 | $ 7,321.49 | 32.65% | 0.20% | 22.03% |
| MPI-ESM1-2-HR/monthly damage function | $ 10,375.71 | $ (167.21) | $ 7,321.49 | 31.23% | -0.50% | 22.03% |
| INM-CM5/annual damage function | $ 10,220.96 | $ 69.01 | $ 6,519.94 | 34.61% | 0.23% | 22.08% |
| INM-CM5/monthly damage function | $ 9,311.31 | $ (147.70) | $ 6,519.94 | 31.53% | -0.50% | 22.08% |
| CESM2 ensemble/annual damage function | $ 18,815.14 | $ 85.40 | $ 12,071.08 | 27.55% | 0.13% | 17.68% |
| CESM2 ensemble/monthly damage function | $ 19,736.14 | $ 905.56 | $ 12,071.08 | 28.90% | 1.33% | 17.68% |
| CanESM5 ensemble/annual damage function | $ 34,650.57 | $ 177.01 | $ 23,886.33 | 31.07% | 0.16% | 21.42% |
| CanESM5 ensemble/monthly damage function | $ 32,085.69 | $ (665.32) | $ 23,886.33 | 28.77% | -0.60% | 21.42% |
| ACCESS-CM2 ensemble/annual damage function | $ 17,340.53 | $ 102.75 | $ 10,493.88 | 26.87% | 0.16% | 16.26% |
| ACCESS-CM2 ensemble/monthly damage function | $ 17,515.53 | $ 566.47 | $ 10,493.88 | 27.15% | 0.88% | 16.26% |

Figures are in billion US$2005. S&Tvar, Tvar and Svar denote the use of damage functions that include both spatial variation and temporal variability, and temporal variability and spatial variation alone, respectively. The column labeled None shows the results produced with the damage function that does not include any type of variability in temperature change. Columns with (%) denote the percent change with respect to None. Numbers in parenthesis denote benefits.

Table S4. Present value of the aggregated economic losses from climate change using different climate models, approaches to create climatological values, and ensemble sizes.

| SP | | | | | | | | | | |
|---|---|---|---|---|---|---|---|---|---|---|
| **Climate model and damage function** | **None** | **S&Tvar** | **Tvar** | **Svar** | **S&Tvar-None** | **Tvar-None** | **Svar-None** | **Sc** | **Tc** | **Int** |
| **CMIP6 ensemble/annual damage function** | $ 183,841.02 | $ 231,362.68 | $184,086.33 | $ 218,488.72 | $ 47,521.66 | $ 245.31 | $34,647.70 | 72.91% | 0.52% | 26.57% |
| **CMIP6 ensemble/monthly damage function** | $ 183,841.02 | $ 230,653.39 | $184,651.66 | $ 218,488.72 | $ 46,812.37 | $ 810.64 | $34,647.70 | 74.01% | 1.73% | 24.25% |
| **ACCESS-ESM1-5/annual damage function** | $ 183,943.06 | $ 249,022.83 | $184,129.42 | $ 223,266.16 | $ 65,079.77 | $ 186.36 | $39,323.10 | 60.42% | 0.29% | 39.29% |
| **ACCESS-ESM1-5/monthly damage function** | $ 183,943.06 | $ 249,560.58 | $184,682.68 | $ 223,266.16 | $ 65,617.52 | $ 739.62 | $39,323.10 | 59.93% | 1.13% | 38.95% |
| **MPI-ESM1-2-HR/annual damage function** | $ 183,464.79 | $ 245,745.69 | $183,879.62 | $ 224,869.98 | $ 62,280.90 | $ 414.82 | $41,405.19 | 66.48% | 0.67% | 32.85% |
| **MPI-ESM1-2-HR/monthly damage function** | $ 183,464.79 | $ 244,827.23 | $183,268.09 | $ 224,869.98 | $ 61,362.44 | $ (196.70) | $41,405.19 | 67.48% | -0.32% | 32.84% |
| **INM-CM5/annual damage function** | $ 183,642.39 | $ 249,749.95 | $184,099.70 | $ 223,931.71 | $ 66,107.56 | $ 457.31 | $40,289.32 | 60.95% | 0.69% | 38.36% |
| **INM-CM5/monthly damage function** | $ 183,642.39 | $ 245,918.33 | $183,290.09 | $ 223,931.71 | $ 62,275.95 | $ (352.30) | $40,289.32 | 64.69% | -0.57% | 35.87% |
| **CESM2 ensemble/annual damage function** | $ 183,705.72 | $ 225,679.22 | $183,910.79 | $ 210,443.11 | $ 41,973.50 | $ 205.07 | $26,737.40 | 63.70% | 0.49% | 35.81% |
| **CESM2 ensemble/monthly damage function** | $ 183,705.72 | $ 231,844.80 | $187,684.33 | $ 210,443.11 | $ 48,139.08 | $ 3,978.61 | $26,737.40 | 55.54% | 8.26% | 36.19% |
| **CanESM5 ensemble/annual damage function** | $ 183,727.68 | $ 239,806.77 | $183,958.10 | $ 223,261.56 | $ 56,079.09 | $ 230.42 | $39,533.89 | 70.50% | 0.41% | 29.09% |
| **CanESM5 ensemble/monthly damage function** | $ 183,727.68 | $ 236,758.83 | $183,122.01 | $ 223,261.56 | $ 53,031.15 | $ (605.67) | $39,533.89 | 74.55% | -1.14% | 26.59% |
| **ACCESS-CM2 ensemble/annual damage function** | $ 183,779.10 | $ 234,804.02 | $184,060.72 | $ 216,071.18 | $ 51,024.91 | $ 281.61 | $32,292.08 | 63.29% | 0.55% | 36.16% |
| **ACCESS-CM2 ensemble/monthly damage function** | $ 183,779.10 | $ 235,803.02 | $185,339.54 | $ 216,071.18 | $ 52,023.92 | $ 1,560.44 | $32,292.08 | 62.07% | 3.00% | 34.93% |
| **RM** | | | | | | | | | | |
| **Climate model and damage function** | **None** | **S&Tvar** | **Tvar** | **Svar** | **S&Tvar-None** | **Tvar-None** | **Svar-None** | **Sc** | **Tc** | **Int** |
| **CMIP6 ensemble/annual damage function** | $ 168,353.56 | $ 216,786.82 | $168,679.75 | $ 201,447.83 | $ 48,433.26 | $ 326.20 | $33,094.27 | 68.33% | 0.67% | 31.00% |
| **CMIP6 ensemble/monthly damage function** | $ 168,353.56 | $ 212,946.68 | $167,897.33 | $ 201,447.83 | $ 44,593.13 | $ (456.22) | $33,094.27 | 74.21% | -1.02% | 26.81% |
| **ACCESS-ESM1-5/annual damage function** | $ 149,910.73 | $ 211,656.90 | $150,201.71 | $ 187,504.03 | $ 61,746.16 | $ 290.98 | $37,593.30 | 60.88% | 0.47% | 38.65% |
| **ACCESS-ESM1-5/monthly damage function** | $ 149,910.73 | $ 208,697.81 | $149,692.36 | $ 187,504.03 | $ 58,787.07 | $ (218.37) | $37,593.30 | 63.95% | -0.37% | 36.42% |
| **MPI-ESM1-2-HR/annual damage function** | $ 102,973.27 | $ 140,038.48 | $103,220.25 | $ 126,758.31 | $ 37,065.20 | $ 246.98 | $23,785.04 | 64.17% | 0.67% | 35.16% |
| **MPI-ESM1-2-HR/monthly damage function** | $ 102,973.27 | $ 137,214.87 | $101,912.12 | $ 126,758.31 | $ 34,241.60 | $ (1,061.15) | $23,785.04 | 69.46% | -3.10% | 33.64% |
| **INM-CM5/annual damage function** | $ 103,618.45 | $ 141,361.24 | $103,877.03 | $ 127,193.35 | $ 37,742.79 | $ 258.59 | $23,574.90 | 62.46% | 0.69% | 36.85% |
| **INM-CM5/monthly damage function** | $ 103,618.45 | $ 137,319.84 | $102,853.47 | $ 127,193.35 | $ 33,701.39 | $ (764.97) | $23,574.90 | 69.95% | -2.27% | 32.32% |
| **CESM2 ensemble/annual damage function** | $ 199,298.91 | $ 268,267.17 | $199,642.74 | $ 240,678.97 | $ 68,968.26 | $ 343.83 | $41,380.06 | 60.00% | 0.50% | 39.50% |
| **CESM2 ensemble/monthly damage function** | $ 199,298.91 | $ 267,253.50 | $200,555.91 | $ 240,678.97 | $ 67,954.59 | $ 1,257.01 | $41,380.06 | 60.89% | 1.85% | 37.26% |
| **CanESM5 ensemble/annual damage function** | $ 340,641.79 | $ 478,498.88 | $341,523.22 | $ 430,796.27 | $ 137,857.10 | $ 881.43 | $90,154.49 | 65.40% | 0.64% | 33.96% |
| **CanESM5 ensemble/monthly damage function** | $ 340,641.79 | $ 466,231.76 | $337,505.20 | $ 430,796.27 | $ 125,589.98 | $ (3,136.59) | $90,154.49 | 71.78% | -2.50% | 30.71% |
| **ACCESS-CM2 ensemble/annual damage function** | $ 180,397.50 | $ 237,669.31 | $180,826.82 | $ 211,327.34 | $ 57,271.81 | $ 429.32 | $30,929.84 | 54.01% | 0.75% | 45.25% |
| **ACCESS-CM2 ensemble/monthly damage function** | $ 180,397.50 | $ 235,964.08 | $181,668.23 | $ 211,327.34 | $ 55,566.58 | $ 1,270.73 | $30,929.84 | 55.66% | 2.29% | 42.05% |

Figures are in billion US$2005. S&Tvar, Tvar and Svar denote the use of damage functions that include both spatial variation and temporal variability, and temporal variability and spatial variation alone, respectively. The column labeled None shows the results produced with the damage function that does not include any type of variability in temperature change. Columns S&Tvar-None, Tvar-None and Svar-None represent the differences between the present value between S&Tvar, Tvar and Svar, and None, respectively. Columns Sc, Tc and Int show the contributions of spatial variation, temporal variability and their interaction, respectively, to the difference between S&Tvar-None. Calculations are based on a 4% discount rate. Numbers in parenthesis denote benefits.

Table S5. Present value of the aggregated economic losses from climate change for different SSP trajectories and the benefits of mitigation.

| A) Scenario and damage function | None | S&Tvar | Tvar | Svar | S&Tvar-None |
|---|---|---|---|---|---|
| **SSP585/annual damage function** | $183,841.02 | $ 231,362.68 | $184,086.33 | $218,488.72 | $ 47,521.66 |
| **SSP585/monthly damage function** | $183,841.02 | $ 230,653.39 | $184,651.66 | $218,488.72 | $ 46,812.37 |
| **SSP370/annual damage function** | $ 54,568.66 | $ 68,663.40 | $ 54,641.42 | $ 64,845.03 | $ 14,094.74 |
| **SSP370/monthly damage function** | $ 54,568.66 | $ 68,453.03 | $ 54,809.09 | $ 64,845.03 | $ 13,884.37 |
| **SSP245/annual damage function** | $ 59,659.38 | $ 75,064.53 | $ 59,738.90 | $ 70,891.16 | $ 15,405.15 |
| **SSP245/monthly damage function** | $ 59,659.38 | $ 74,834.60 | $ 59,922.16 | $ 70,891.16 | $ 15,175.22 |
| **SSP126/annual damage function** | $ 50,198.79 | $ 63,151.06 | $ 50,265.65 | $ 59,642.20 | $ 12,952.27 |
| **SSP126/monthly damage function** | $ 50,198.79 | $ 62,957.74 | $ 50,419.74 | $ 59,642.20 | $ 12,758.94 |
| **SSP119/annual damage function** | $ 41,237.49 | $ 51,871.03 | $ 41,292.39 | $ 48,990.33 | $ 10,633.53 |
| **SSP119/monthly damage function** | $ 41,237.49 | $ 51,712.32 | $ 41,418.88 | $ 48,990.33 | $ 10,474.82 |
| **B) SSP scenario (annual damage function)** | **None** | **S&Tvar** | **Tvar** | **Svar** | |
| **SSP585** | $183,841.02 | $ 231,362.68 | $184,086.33 | $218,488.72 | |
| **SSP370** | $131,435.61 | $ 165,397.40 | $131,610.92 | $156,196.91 | |
| **SSP245** | $ 96,260.98 | $ 121,121.36 | $ 96,389.31 | $114,386.50 | |
| **SSP126** | $ 66,407.10 | $ 83,542.81 | $ 66,495.56 | $ 78,900.62 | |
| **C) Avoided losses mitigation (SSP5)** | **None** | **S&Tvar** | **Tvar** | **Svar** | |
| **RCP7.0** | $ 52,405.41 | $ 65,965.28 | $ 52,475.41 | $ 62,291.81 | |
| **RCP4.5** | $ 87,580.05 | $ 110,241.33 | $ 87,697.02 | $104,102.22 | |
| **RCP2.6** | $117,433.92 | $ 147,819.87 | $117,590.77 | $139,588.10 | |

Section A) show the present value estimates for the SSP585, SSP370, SSP245, SSP126 and SSP119, and for both annual and monthly damage functions. Section B) presents the present value estimates for different emissions trajectories, but maintaining the SSP5 socioeconomic scenario, and section C) shows the avoided damages from mitigation with respect to the RCP8.5 trajectory. Figures are in billion US$2005. S&T variability, T variability and S variability denote the use of damage functions that include both spatial and temporal variability, and temporal and spatial variability alone, respectively. The column labeled None shows the results produced with the damage function that does not include any type of variability in temperature change. Calculations are based on a 4% discount rate.

Table S6. Present value of the economic losses from climate change for 15 sectors and different SSP trajectories.

| SSP585 | Agriculture | Forestry | Energy | Water | Coastal defence | Dryland | Wetland | Ecosystem | Health | Air pollution | Time use | Settlements | Catastrophe | Migration | Amenity | Total |
|---|---|---|---|---|---|---|---|---|---|---|---|---|---|---|---|---|
| None | $14,952.84 | $1,242.51 | $10,842.14 | $21,918.96 | $2,859.53 | $7,036.21 | $14,878.61 | $18,436.13 | $35,111.65 | $7,228.09 | $(26,538.81) | $15,557.23 | $24,277.18 | $5,421.28 | $30,617.46 | $183,841.02 |
| S&Tvar | $17,346.36 | $1,441.41 | $14,551.53 | $25,103.42 | $3,598.70 | $8,855.03 | $18,724.63 | $23,201.75 | $47,124.31 | $8,278.22 | $(35,618.46) | $19,578.67 | $30,552.67 | $6,822.64 | $41,092.53 | $230,653.39 |
| Tvar | $14,606.49 | $1,213.73 | $11,183.89 | $21,535.94 | $2,863.35 | $7,045.60 | $14,898.47 | $18,460.73 | $36,218.38 | $7,101.79 | $(27,375.32) | $15,577.99 | $24,309.57 | $5,428.51 | $31,582.54 | $184,651.66 |
| Svar | $17,770.93 | $1,476.69 | $12,885.51 | $26,049.93 | $3,398.45 | $8,362.30 | $17,682.72 | $21,910.71 | $41,728.98 | $8,590.34 | $(31,540.46) | $18,489.24 | $28,852.59 | $6,443.00 | $36,387.80 | $218,488.72 |
| **SSP370** | | | | | | | | | | | | | | | | |
| None | $4,438.38 | $368.81 | $3,218.22 | $6,506.10 | $848.78 | $2,088.53 | $4,416.35 | $5,472.31 | $10,422.02 | $2,145.48 | $(7,877.39) | $4,617.78 | $7,206.08 | $1,609.17 | $9,088.04 | $54,568.66 |
| S&Tvar | $5,148.29 | $427.80 | $4,318.41 | $7,450.60 | $1,068.02 | $2,627.98 | $5,557.06 | $6,885.77 | $13,984.93 | $2,456.94 | $(10,570.38) | $5,810.52 | $9,067.37 | $2,024.81 | $12,194.90 | $68,453.03 |
| Tvar | $4,335.65 | $360.27 | $3,319.58 | $6,392.50 | $849.91 | $2,091.31 | $4,422.24 | $5,479.61 | $10,750.28 | $2,108.02 | $(8,125.50) | $4,623.94 | $7,215.69 | $1,611.32 | $9,374.27 | $54,809.09 |
| Svar | $5,274.22 | $438.26 | $3,824.28 | $7,731.33 | $1,008.62 | $2,481.84 | $5,248.03 | $6,502.85 | $12,384.70 | $2,549.52 | $(9,360.86) | $5,487.40 | $8,563.13 | $1,912.21 | $10,799.50 | $64,845.03 |
| **SSP245** | | | | | | | | | | | | | | | | |
| None | $4,852.44 | $403.22 | $3,518.45 | $7,113.06 | $927.96 | $2,283.36 | $4,828.35 | $5,982.82 | $11,394.30 | $2,345.63 | $(8,612.27) | $5,048.57 | $7,878.34 | $1,759.29 | $9,935.86 | $59,659.38 |
| S&Tvar | $5,628.35 | $467.69 | $4,720.93 | $8,145.37 | $1,167.58 | $2,872.97 | $6,075.12 | $7,527.70 | $15,288.45 | $2,686.05 | $(11,555.63) | $6,352.21 | $9,912.67 | $2,213.57 | $13,331.58 | $74,834.60 |
| Tvar | $4,740.16 | $393.89 | $3,629.23 | $6,988.89 | $929.20 | $2,286.41 | $4,834.79 | $5,990.80 | $11,753.07 | $2,304.69 | $(8,883.45) | $5,055.30 | $7,888.84 | $1,761.64 | $10,248.71 | $59,922.16 |
| Svar | $5,765.98 | $479.13 | $4,180.85 | $8,452.20 | $1,102.67 | $2,713.24 | $5,737.36 | $7,109.18 | $13,539.45 | $2,787.23 | $(10,233.66) | $5,999.04 | $9,361.55 | $2,090.51 | $11,806.44 | $70,891.16 |
| **SSP126** | | | | | | | | | | | | | | | | |
| None | $4,082.95 | $339.28 | $2,960.51 | $5,985.09 | $780.81 | $1,921.28 | $4,062.69 | $5,034.09 | $9,587.43 | $1,973.67 | $(7,246.57) | $4,247.99 | $6,629.02 | $1,480.31 | $8,360.26 | $50,198.79 |
| S&Tvar | $4,735.32 | $393.48 | $3,971.52 | $6,853.03 | $982.27 | $2,417.00 | $5,110.94 | $6,332.98 | $12,861.54 | $2,259.89 | $(9,721.27) | $5,344.05 | $8,339.43 | $1,862.26 | $11,215.30 | $62,957.74 |
| Tvar | $3,988.55 | $331.43 | $3,053.65 | $5,880.70 | $781.85 | $1,923.84 | $4,068.10 | $5,040.79 | $9,889.07 | $1,939.25 | $(7,474.56) | $4,253.65 | $6,637.85 | $1,482.28 | $8,623.30 | $50,419.74 |
| Svar | $4,851.04 | $403.10 | $3,517.44 | $7,111.01 | $927.70 | $2,282.71 | $4,826.96 | $5,981.10 | $11,391.01 | $2,344.96 | $(8,609.79) | $5,047.12 | $7,876.07 | $1,758.79 | $9,933.00 | $59,642.20 |
| **SSP119** | | | | | | | | | | | | | | | | |
| None | $3,354.08 | $278.71 | $2,432.01 | $4,916.66 | $641.42 | $1,578.30 | $3,337.43 | $4,135.42 | $7,875.92 | $1,621.34 | $(5,952.94) | $3,489.65 | $5,445.63 | $1,216.05 | $6,867.82 | $41,237.49 |
| S&Tvar | $3,889.66 | $323.21 | $3,262.03 | $5,629.22 | $806.82 | $1,985.28 | $4,198.02 | $5,201.78 | $10,563.89 | $1,856.32 | $(7,984.62) | $4,389.50 | $6,849.84 | $1,529.62 | $9,211.74 | $51,712.32 |
| Tvar | $3,276.58 | $272.27 | $2,508.48 | $4,830.95 | $642.28 | $1,580.40 | $3,341.87 | $4,140.92 | $8,123.56 | $1,593.08 | $(6,140.12) | $3,494.30 | $5,452.88 | $1,217.67 | $7,083.77 | $41,418.88 |
| Svar | $3,984.66 | $331.11 | $2,889.24 | $5,841.01 | $762.01 | $1,875.02 | $3,964.88 | $4,912.90 | $9,356.62 | $1,926.16 | $(7,072.12) | $4,145.72 | $6,469.43 | $1,444.67 | $8,159.00 | $48,990.33 |

Figures are in billion US$2005. S&Tvar, Tvar and Svar denote the use of damage functions that include both spatial variation and temporal variability, and temporal variability and spatial variation alone, respectively. The column labeled None shows the results produced with the damage function that does not include any type of variability in temperature change. Results are based on the CMIP6 ensemble and the SP approach. Numbers in parenthesis denote benefits. Calculations are based on a 4% discount rate.

Table S7. Present value of the economic losses from climate change for 15 sectors under the SSP585 and different climate models.

| ACCESS-ESM1-5 | Agriculture | Forestry | Energy | Water | Coastal defence | Dryland | Wetland | Ecosystem | Health | Air pollution | Time use | Settlements | Catastrophe | Migration | Amenity | Total |
|---|---|---|---|---|---|---|---|---|---|---|---|---|---|---|---|---|
| **None** | $14,961.14 | $1,243.20 | $10,848.16 | $21,931.13 | $ 2,861.12 | $7,040.12 | $14,886.87 | $18,446.36 | $35,131.14 | $ 7,232.11 | $(26,553.54) | $15,565.87 | $24,290.65 | $5,424.29 | $30,634.46 | $183,943.06 |
| **S&Tvar** | $18,480.27 | $1,535.63 | $15,970.15 | $27,084.53 | $ 3,873.39 | $9,530.94 | $20,153.90 | $24,972.76 | $51,718.42 | $ 8,931.52 | $(39,090.88) | $21,073.13 | $32,884.78 | $7,343.42 | $45,098.62 | $249,560.58 |
| **Tvar** | $14,604.25 | $1,213.55 | $11,188.69 | $21,528.83 | $ 2,864.02 | $7,047.25 | $14,901.95 | $18,465.05 | $36,233.94 | $ 7,099.44 | $(27,387.08) | $15,581.64 | $24,315.26 | $5,429.78 | $31,596.10 | $184,682.68 |
| **Svar** | $18,159.51 | $1,508.98 | $13,167.26 | $26,619.53 | $ 3,472.76 | $8,545.14 | $18,069.36 | $22,389.80 | $42,641.42 | $ 8,778.18 | $(32,230.12) | $18,893.52 | $29,483.48 | $6,583.88 | $37,183.45 | $223,266.16 |
| **MPI-ESM1-2-HR** | | | | | | | | | | | | | | | | |
| **None** | $14,922.24 | $1,239.97 | $10,819.95 | $21,874.10 | $ 2,853.68 | $7,021.81 | $14,848.16 | $18,398.40 | $35,039.79 | $ 7,213.30 | $(26,484.50) | $15,525.40 | $24,227.50 | $5,410.18 | $30,554.80 | $183,464.79 |
| **S&Tvar** | $18,095.68 | $1,503.67 | $15,658.00 | $26,107.41 | $ 3,822.42 | $9,405.51 | $19,888.68 | $24,644.12 | $50,707.54 | $ 8,609.29 | $(38,326.82) | $20,795.81 | $32,452.02 | $7,246.78 | $44,217.13 | $244,827.23 |
| **Tvar** | $14,421.73 | $1,198.38 | $11,125.11 | $20,924.75 | $ 2,860.13 | $7,037.69 | $14,881.74 | $18,440.00 | $36,028.03 | $ 6,900.24 | $(27,231.45) | $15,560.50 | $24,282.28 | $5,422.42 | $31,416.55 | $183,268.09 |
| **Svar** | $18,289.96 | $1,519.81 | $13,261.85 | $26,810.75 | $ 3,497.71 | $8,606.53 | $18,199.17 | $22,550.64 | $42,947.74 | $ 8,841.23 | $(32,461.64) | $19,029.24 | $29,695.27 | $6,631.18 | $37,450.55 | $224,869.98 |
| **INM-CM5** | | | | | | | | | | | | | | | | |
| **None** | $14,936.68 | $1,241.17 | $10,830.43 | $21,895.28 | $ 2,856.44 | $7,028.61 | $14,862.54 | $18,416.21 | $35,073.71 | $ 7,220.28 | $(26,510.13) | $15,540.42 | $24,250.95 | $5,415.42 | $30,584.38 | $183,642.39 |
| **S&Tvar** | $18,507.00 | $1,537.85 | $15,436.84 | $25,911.36 | $ 3,884.70 | $9,558.77 | $20,212.75 | $25,045.68 | $49,991.31 | $ 8,544.64 | $(37,785.46) | $21,134.66 | $32,980.80 | $7,364.86 | $43,592.57 | $245,918.33 |
| **Tvar** | $14,382.66 | $1,195.14 | $11,149.83 | $20,807.70 | $ 2,863.55 | $7,046.11 | $14,899.55 | $18,462.07 | $36,108.07 | $ 6,861.64 | $(27,291.95) | $15,579.12 | $24,311.34 | $5,428.91 | $31,486.35 | $183,290.09 |
| **Svar** | $18,213.64 | $1,513.47 | $13,206.51 | $26,698.88 | $ 3,483.12 | $8,570.62 | $18,123.23 | $22,456.54 | $42,768.54 | $ 8,804.34 | $(32,326.20) | $18,949.84 | $29,571.37 | $6,603.51 | $37,294.29 | $223,931.71 |
| **CESM2 ensemble** | | | | | | | | | | | | | | | | |
| **None** | $14,941.83 | $1,241.60 | $10,834.16 | $21,902.83 | $ 2,857.43 | $7,031.03 | $14,867.66 | $18,422.56 | $35,085.81 | $ 7,222.77 | $(26,519.28) | $15,545.78 | $24,259.31 | $5,417.29 | $30,594.93 | $183,705.72 |
| **S&Tvar** | $17,530.78 | $1,456.73 | $14,716.60 | $27,333.05 | $ 3,510.30 | $8,637.50 | $18,264.66 | $22,631.79 | $47,658.86 | $ 9,013.47 | $(36,022.50) | $19,097.72 | $29,802.13 | $6,655.04 | $41,558.67 | $231,844.80 |
| **Tvar** | $14,840.22 | $1,233.16 | $11,443.66 | $22,788.22 | $ 2,860.62 | $7,038.88 | $14,884.26 | $18,443.13 | $37,059.63 | $ 7,514.75 | $(28,011.18) | $15,563.14 | $24,286.39 | $5,423.34 | $32,316.11 | $187,684.33 |
| **Svar** | $17,116.54 | $1,422.31 | $12,411.02 | $25,090.67 | $ 3,273.31 | $8,054.36 | $17,031.57 | $21,103.87 | $40,192.36 | $ 8,274.01 | $(30,379.02) | $17,808.39 | $27,790.13 | $6,205.75 | $35,047.86 | $210,443.11 |
| **CanESM5 ensemble** | | | | | | | | | | | | | | | | |
| **None** | $14,943.62 | $1,241.75 | $10,835.46 | $21,905.45 | $ 2,857.77 | $7,031.87 | $14,869.44 | $18,424.76 | $35,090.00 | $ 7,223.64 | $(26,522.45) | $15,547.64 | $24,262.21 | $5,417.94 | $30,598.59 | $183,727.68 |
| **S&Tvar** | $17,909.86 | $1,488.23 | $14,813.40 | $25,271.34 | $ 3,730.04 | $9,178.21 | $19,408.03 | $24,048.54 | $47,972.35 | $ 8,333.59 | $(36,259.45) | $20,293.24 | $31,667.75 | $7,071.65 | $41,832.04 | $236,758.83 |
| **Tvar** | $14,544.90 | $1,208.62 | $11,019.32 | $21,053.28 | $ 2,861.35 | $7,040.69 | $14,888.09 | $18,447.87 | $35,685.44 | $ 6,942.62 | $(26,972.51) | $15,567.14 | $24,292.64 | $5,424.73 | $31,117.81 | $183,122.01 |
| **Svar** | $18,159.14 | $1,508.94 | $13,166.99 | $26,618.98 | $ 3,472.69 | $8,544.97 | $18,068.99 | $22,389.34 | $42,640.55 | $ 8,777.99 | $(32,229.46) | $18,893.13 | $29,482.87 | $6,583.75 | $37,182.68 | $223,261.56 |
| **ACCESS-CM2 ensemble** | | | | | | | | | | | | | | | | |
| **None** | $14,947.80 | $1,242.10 | $10,838.49 | $21,911.58 | $ 2,858.57 | $7,033.84 | $14,873.60 | $18,429.92 | $35,099.82 | $ 7,225.66 | $(26,529.87) | $15,551.99 | $24,269.00 | $5,419.45 | $30,607.15 | $183,779.10 |
| **S&Tvar** | $17,504.13 | $1,454.52 | $15,071.72 | $25,796.99 | $ 3,652.23 | $8,986.74 | $19,003.15 | $23,546.85 | $48,808.89 | $ 8,506.93 | $(36,891.74) | $19,869.89 | $31,007.11 | $6,924.12 | $42,561.49 | $235,803.02 |
| **Tvar** | $14,603.42 | $1,213.48 | $11,280.75 | $21,728.88 | $ 2,862.95 | $7,044.62 | $14,896.39 | $18,458.16 | $36,532.06 | $ 7,165.41 | $(27,612.41) | $15,575.82 | $24,306.19 | $5,427.76 | $31,856.06 | $185,339.54 |
| **Svar** | $17,574.30 | $1,460.35 | $12,742.93 | $25,761.69 | $ 3,360.85 | $8,269.77 | $17,487.06 | $21,668.27 | $41,267.26 | $ 8,495.29 | $(31,191.47) | $18,284.66 | $28,533.34 | $6,371.71 | $35,985.17 | $216,071.18 |

Figures are in billion US$2005. S&Tvar, Tvar and Svar denote the use of damage functions that include both spatial variation and temporal variability, and temporal variability and spatial variation alone, respectively. The column labeled None shows the results produced with the damage function that does not include any type of variability in temperature change. Results are based on the SP approach. Numbers in parenthesis denote benefits. Calculations are based on a 4% discount rate.

Table S8. Present value of the economic losses from climate change for 15 sectors under the SSP370 and different climate models.

| ACCESS-ESM1-5 | Agriculture | Forestry | Energy | Water | Coastal defence | Dryland | Wetland | Ecosystem | Health | Air pollution | Time use | Settlements | Catastrophe | Migration | Amenity | Total |
|---|---|---|---|---|---|---|---|---|---|---|---|---|---|---|---|---|
| **None** | $ 4,440.84 | $ 369.01 | $ 3,220.01 | $ 6,509.71 | $ 849.25 | $2,089.68 | $ 4,418.80 | $ 5,475.35 | $10,427.81 | $ 2,146.67 | $ (7,881.76) | $ 4,620.34 | $ 7,210.08 | $1,610.07 | $ 9,093.08 | $ 54,598.95 |
| **S&Tvar** | $ 5,484.60 | $ 455.75 | $ 4,739.17 | $ 8,038.19 | $ 1,149.49 | $2,828.45 | $ 5,980.98 | $ 7,411.05 | $15,347.53 | $ 2,650.71 | $(11,600.29) | $ 6,253.78 | $ 9,759.06 | $2,179.27 | $13,383.09 | $ 74,060.85 |
| **Tvar** | $ 4,334.99 | $ 360.22 | $ 3,321.01 | $ 6,390.39 | $ 850.11 | $2,091.80 | $ 4,423.27 | $ 5,480.89 | $10,754.90 | $ 2,107.32 | $ (8,128.99) | $ 4,625.02 | $ 7,217.38 | $1,611.70 | $ 9,378.30 | $ 54,818.31 |
| **Svar** | $ 5,389.47 | $ 447.84 | $ 3,907.84 | $ 7,900.28 | $ 1,030.66 | $2,536.07 | $ 5,362.71 | $ 6,644.96 | $12,655.33 | $ 2,605.23 | $ (9,565.41) | $ 5,607.31 | $ 8,750.25 | $1,954.00 | $11,035.49 | $ 66,262.03 |
| **MPI-ESM1-2-HR** | | | | | | | | | | | | | | | | |
| **None** | $ 4,429.30 | $ 368.06 | $ 3,211.64 | $ 6,492.79 | $ 847.04 | $2,084.25 | $ 4,407.31 | $ 5,461.11 | $10,400.70 | $ 2,141.09 | $ (7,861.27) | $ 4,608.33 | $ 7,191.33 | $1,605.88 | $ 9,069.44 | $ 54,456.98 |
| **S&Tvar** | $ 5,370.53 | $ 446.27 | $ 4,646.58 | $ 7,748.37 | $ 1,134.37 | $2,791.25 | $ 5,902.31 | $ 7,313.57 | $15,047.69 | $ 2,555.14 | $(11,373.65) | $ 6,171.51 | $ 9,630.69 | $2,150.61 | $13,121.63 | $ 72,656.85 |
| **Tvar** | $ 4,280.85 | $ 355.72 | $ 3,302.14 | $ 6,211.21 | $ 848.96 | $2,088.96 | $ 4,417.27 | $ 5,473.45 | $10,693.80 | $ 2,048.24 | $ (8,082.81) | $ 4,618.74 | $ 7,207.58 | $1,609.51 | $ 9,325.03 | $ 54,398.64 |
| **Svar** | $ 5,428.15 | $ 451.06 | $ 3,935.89 | $ 7,956.98 | $ 1,038.06 | $2,554.27 | $ 5,401.20 | $ 6,692.65 | $12,746.16 | $ 2,623.93 | $ (9,634.07) | $ 5,647.56 | $ 8,813.05 | $1,968.02 | $11,114.69 | $ 66,737.60 |
| **INM-CM5** | | | | | | | | | | | | | | | | |
| **None** | $ 4,433.58 | $ 368.41 | $ 3,214.74 | $ 6,499.07 | $ 847.86 | $2,086.27 | $ 4,411.58 | $ 5,466.40 | $10,410.76 | $ 2,143.16 | $ (7,868.88) | $ 4,612.79 | $ 7,198.29 | $1,607.43 | $ 9,078.22 | $ 54,509.70 |
| **S&Tvar** | $ 5,492.53 | $ 456.40 | $ 4,580.99 | $ 7,690.22 | $ 1,152.84 | $2,836.70 | $ 5,998.43 | $ 7,432.67 | $14,835.27 | $ 2,535.96 | $(11,213.10) | $ 6,272.02 | $ 9,787.53 | $2,185.63 | $12,936.40 | $ 72,980.51 |
| **Tvar** | $ 4,269.26 | $ 354.76 | $ 3,309.48 | $ 6,176.50 | $ 849.97 | $2,091.46 | $ 4,422.55 | $ 5,480.00 | $10,717.55 | $ 2,036.79 | $ (8,100.76) | $ 4,624.27 | $ 7,216.21 | $1,611.43 | $ 9,345.74 | $ 54,405.21 |
| **Svar** | $ 5,405.52 | $ 449.17 | $ 3,919.48 | $ 7,923.80 | $ 1,033.73 | $2,543.62 | $ 5,378.69 | $ 6,664.74 | $12,693.02 | $ 2,612.99 | $ (9,593.90) | $ 5,624.01 | $ 8,776.31 | $1,959.82 | $11,068.35 | $ 66,459.36 |
| **CESM2 ensemble** | | | | | | | | | | | | | | | | |
| **None** | $ 4,435.11 | $ 368.54 | $ 3,215.85 | $ 6,501.31 | $ 848.16 | $2,086.99 | $ 4,413.10 | $ 5,468.28 | $10,414.35 | $ 2,143.90 | $ (7,871.59) | $ 4,614.38 | $ 7,200.78 | $1,607.99 | $ 9,081.35 | $ 54,528.50 |
| **S&Tvar** | $ 5,202.98 | $ 432.35 | $ 4,367.37 | $ 8,111.90 | $ 1,041.80 | $2,563.46 | $ 5,420.63 | $ 6,716.72 | $14,143.47 | $ 2,675.01 | $(10,690.21) | $ 5,667.87 | $ 8,844.76 | $1,975.10 | $12,333.15 | $ 68,806.36 |
| **Tvar** | $ 4,404.98 | $ 366.03 | $ 3,396.63 | $ 6,763.92 | $ 849.10 | $2,089.32 | $ 4,418.02 | $ 5,474.38 | $10,999.78 | $ 2,230.50 | $ (8,314.08) | $ 4,619.53 | $ 7,208.81 | $1,609.78 | $ 9,591.84 | $ 55,708.54 |
| **Svar** | $ 5,080.12 | $ 422.14 | $ 3,683.54 | $ 7,446.81 | $ 971.51 | $2,390.50 | $ 5,054.90 | $ 6,263.55 | $11,928.94 | $ 2,455.69 | $ (9,016.38) | $ 5,285.46 | $ 8,248.00 | $1,841.84 | $10,402.07 | $ 62,458.71 |
| **CanESM5 ensemble** | | | | | | | | | | | | | | | | |
| **None** | $ 4,435.64 | $ 368.58 | $ 3,216.24 | $ 6,502.09 | $ 848.26 | $2,087.24 | $ 4,413.62 | $ 5,468.94 | $10,415.60 | $ 2,144.16 | $ (7,872.53) | $ 4,614.93 | $ 7,201.64 | $1,608.18 | $ 9,082.43 | $ 54,535.01 |
| **S&Tvar** | $ 5,315.42 | $ 441.69 | $ 4,396.08 | $ 7,500.40 | $ 1,106.97 | $2,723.83 | $ 5,759.75 | $ 7,136.93 | $14,236.46 | $ 2,473.37 | $(10,760.49) | $ 6,022.46 | $ 9,398.09 | $2,098.67 | $12,414.23 | $ 70,263.85 |
| **Tvar** | $ 4,317.39 | $ 358.76 | $ 3,270.77 | $ 6,249.34 | $ 849.32 | $2,089.85 | $ 4,419.16 | $ 5,475.79 | $10,592.20 | $ 2,060.81 | $ (8,006.02) | $ 4,620.72 | $ 7,210.66 | $1,610.20 | $ 9,236.43 | $ 54,355.38 |
| **Svar** | $ 5,389.35 | $ 447.83 | $ 3,907.76 | $ 7,900.11 | $ 1,030.64 | $2,536.02 | $ 5,362.60 | $ 6,644.81 | $12,655.06 | $ 2,605.17 | $ (9,565.21) | $ 5,607.19 | $ 8,750.06 | $1,953.96 | $11,035.25 | $ 66,260.61 |
| **ACCESS-CM2 ensemble** | | | | | | | | | | | | | | | | |
| **None** | $ 4,436.89 | $ 368.69 | $ 3,217.14 | $ 6,503.91 | $ 848.50 | $2,087.82 | $ 4,414.86 | $ 5,470.47 | $10,418.51 | $ 2,144.76 | $ (7,874.74) | $ 4,616.22 | $ 7,203.65 | $1,608.63 | $ 9,084.97 | $ 54,550.28 |
| **S&Tvar** | $ 5,195.08 | $ 431.69 | $ 4,472.70 | $ 7,656.31 | $ 1,083.89 | $2,667.04 | $ 5,639.67 | $ 6,988.13 | $14,484.57 | $ 2,524.78 | $(10,948.03) | $ 5,896.90 | $ 9,202.15 | $2,054.91 | $12,630.59 | $ 69,980.37 |
| **Tvar** | $ 4,334.74 | $ 360.20 | $ 3,348.31 | $ 6,449.72 | $ 849.79 | $2,091.02 | $ 4,421.62 | $ 5,478.84 | $10,843.31 | $ 2,126.89 | $ (8,195.81) | $ 4,623.29 | $ 7,214.68 | $1,611.09 | $ 9,455.40 | $ 55,013.10 |
| **Svar** | $ 5,215.89 | $ 433.42 | $ 3,781.99 | $ 7,645.84 | $ 997.47 | $2,454.39 | $ 5,190.00 | $ 6,430.95 | $12,247.75 | $ 2,521.33 | $ (9,257.35) | $ 5,426.72 | $ 8,468.44 | $1,891.07 | $10,680.08 | $ 64,127.99 |

Figures are in billion US$2005. S&Tvar, Tvar and Svar denote the use of damage functions that include both spatial variation and temporal variability, and temporal variability and spatial variation alone, respectively. The column labeled None shows the results produced with the damage function that does not include any type of variability in temperature change. Results are based on the SP approach. Numbers in parenthesis denote benefits. Calculations are based on a 4% discount rate.

Table S9. Present value of the economic losses from climate change for 15 sectors under the SSP245 and different climate models.

| ACCESS-ESM1-5 | Agriculture | Forestry | Energy | Water | Coastal defence | Dryland | Wetland | Ecosystem | Health | Air pollution | Time use | Settlements | Catastrophe | Migration | Amenity | Total |
|---|---|---|---|---|---|---|---|---|---|---|---|---|---|---|---|---|
| **None** | $ 4,855.13 | $ 403.44 | $ 3,520.40 | $ 7,117.00 | $ 928.48 | $ 2,284.63 | $ 4,831.03 | $ 5,986.14 | $ 11,400.62 | $ 2,346.93 | $ (8,617.05) | $ 5,051.38 | $ 7,882.71 | $ 1,760.27 | $ 9,941.38 | $ 59,692.49 |
| **S&Tvar** | $ 5,995.93 | $ 498.24 | $ 5,180.81 | $ 8,787.59 | $ 1,256.63 | $ 3,092.08 | $ 6,538.45 | $ 8,101.81 | $ 16,777.74 | $ 2,897.83 | $ (12,681.30) | $ 6,836.67 | $ 10,668.68 | $ 2,382.40 | $ 14,630.24 | $ 80,963.80 |
| **Tvar** | $ 4,739.44 | $ 393.83 | $ 3,630.79 | $ 6,986.59 | $ 929.42 | $ 2,286.94 | $ 4,835.92 | $ 5,992.20 | $ 11,758.12 | $ 2,303.93 | $ (8,887.26) | $ 5,056.49 | $ 7,890.69 | $ 1,762.05 | $ 10,253.11 | $ 59,932.26 |
| **Svar** | $ 5,891.95 | $ 489.60 | $ 4,272.19 | $ 8,636.85 | $ 1,126.76 | $ 2,772.52 | $ 5,862.70 | $ 7,264.49 | $ 13,835.24 | $ 2,848.13 | $ (10,457.24) | $ 6,130.10 | $ 9,566.07 | $ 2,136.18 | $ 12,064.37 | $ 72,439.91 |
| **MPI-ESM1-2-HR** | | | | | | | | | | | | | | | | |
| **None** | $ 4,842.51 | $ 402.39 | $ 3,511.25 | $ 7,098.50 | $ 926.06 | $ 2,278.69 | $ 4,818.47 | $ 5,970.58 | $ 11,370.98 | $ 2,340.83 | $ (8,594.65) | $ 5,038.24 | $ 7,862.21 | $ 1,755.69 | $ 9,915.53 | $ 59,537.29 |
| **S&Tvar** | $ 5,871.24 | $ 487.87 | $ 5,079.61 | $ 8,470.81 | $ 1,240.10 | $ 3,051.42 | $ 6,452.46 | $ 7,995.26 | $ 16,450.01 | $ 2,793.37 | $ (12,433.59) | $ 6,746.76 | $ 10,528.37 | $ 2,351.06 | $ 14,344.46 | $ 79,429.22 |
| **Tvar** | $ 4,680.26 | $ 388.91 | $ 3,610.17 | $ 6,790.75 | $ 928.16 | $ 2,283.84 | $ 4,829.35 | $ 5,984.06 | $ 11,691.34 | $ 2,239.35 | $ (8,836.79) | $ 5,049.62 | $ 7,879.97 | $ 1,759.66 | $ 10,194.88 | $ 59,473.52 |
| **Svar** | $ 5,934.22 | $ 493.11 | $ 4,302.84 | $ 8,698.82 | $ 1,134.84 | $ 2,792.41 | $ 5,904.77 | $ 7,316.61 | $ 13,934.51 | $ 2,868.56 | $ (10,532.26) | $ 6,174.09 | $ 9,634.71 | $ 2,151.50 | $ 12,150.93 | $ 72,959.65 |
| **INM-CM5** | | | | | | | | | | | | | | | | |
| **None** | $ 4,847.19 | $ 402.78 | $ 3,514.65 | $ 7,105.37 | $ 926.96 | $ 2,280.90 | $ 4,823.13 | $ 5,976.36 | $ 11,381.99 | $ 2,343.10 | $ (8,602.97) | $ 5,043.12 | $ 7,869.82 | $ 1,757.39 | $ 9,925.13 | $ 59,594.92 |
| **S&Tvar** | $ 6,004.59 | $ 498.95 | $ 5,007.91 | $ 8,407.27 | $ 1,260.29 | $ 3,101.10 | $ 6,557.52 | $ 8,125.44 | $ 16,217.84 | $ 2,772.42 | $ (12,258.10) | $ 6,856.61 | $ 10,699.79 | $ 2,389.34 | $ 14,142.01 | $ 79,782.98 |
| **Tvar** | $ 4,667.60 | $ 387.86 | $ 3,618.19 | $ 6,752.81 | $ 929.27 | $ 2,286.57 | $ 4,835.13 | $ 5,991.22 | $ 11,717.30 | $ 2,226.84 | $ (8,856.41) | $ 5,055.66 | $ 7,889.40 | $ 1,761.76 | $ 10,217.52 | $ 59,480.71 |
| **Svar** | $ 5,909.49 | $ 491.05 | $ 4,284.91 | $ 8,662.56 | $ 1,130.11 | $ 2,780.77 | $ 5,880.16 | $ 7,286.12 | $ 13,876.43 | $ 2,856.60 | $ (10,488.37) | $ 6,148.35 | $ 9,594.55 | $ 2,142.54 | $ 12,100.28 | $ 72,655.55 |
| **CESM2 ensemble** | | | | | | | | | | | | | | | | |
| **None** | $ 4,848.87 | $ 402.92 | $ 3,515.86 | $ 7,107.82 | $ 927.28 | $ 2,281.68 | $ 4,824.80 | $ 5,978.42 | $ 11,385.91 | $ 2,343.91 | $ (8,605.93) | $ 5,044.86 | $ 7,872.54 | $ 1,758.00 | $ 9,928.55 | $ 59,615.47 |
| **S&Tvar** | $ 5,688.13 | $ 472.66 | $ 4,774.43 | $ 8,868.14 | $ 1,138.92 | $ 2,802.45 | $ 5,926.00 | $ 7,342.93 | $ 15,461.73 | $ 2,924.40 | $ (11,686.61) | $ 6,196.29 | $ 9,669.36 | $ 2,159.24 | $ 13,482.68 | $ 75,220.77 |
| **Tvar** | $ 4,815.93 | $ 400.18 | $ 3,713.44 | $ 7,394.84 | $ 928.31 | $ 2,284.23 | $ 4,830.18 | $ 5,985.09 | $ 12,025.77 | $ 2,438.56 | $ (9,089.56) | $ 5,050.48 | $ 7,881.32 | $ 1,759.96 | $ 10,486.51 | $ 60,905.22 |
| **Svar** | $ 5,553.84 | $ 461.50 | $ 4,027.03 | $ 8,141.23 | $ 1,062.10 | $ 2,613.42 | $ 5,526.27 | $ 6,847.62 | $ 13,041.31 | $ 2,684.69 | $ (9,857.15) | $ 5,778.33 | $ 9,017.13 | $ 2,013.59 | $ 11,372.06 | $ 68,282.96 |
| **CanESM5 ensemble** | | | | | | | | | | | | | | | | |
| **None** | $ 4,849.45 | $ 402.97 | $ 3,516.28 | $ 7,108.67 | $ 927.39 | $ 2,281.96 | $ 4,825.37 | $ 5,979.13 | $ 11,387.27 | $ 2,344.19 | $ (8,606.96) | $ 5,045.46 | $ 7,873.48 | $ 1,758.21 | $ 9,929.73 | $ 59,622.60 |
| **S&Tvar** | $ 5,811.02 | $ 482.87 | $ 4,805.82 | $ 8,199.80 | $ 1,210.16 | $ 2,977.74 | $ 6,296.65 | $ 7,802.20 | $ 15,563.36 | $ 2,704.00 | $ (11,763.42) | $ 6,583.85 | $ 10,274.14 | $ 2,294.29 | $ 13,571.30 | $ 76,813.77 |
| **Tvar** | $ 4,720.19 | $ 392.23 | $ 3,575.88 | $ 6,832.42 | $ 928.55 | $ 2,284.82 | $ 4,831.42 | $ 5,986.62 | $ 11,580.30 | $ 2,253.09 | $ (8,752.86) | $ 5,051.78 | $ 7,883.34 | $ 1,760.41 | $ 10,098.05 | $ 59,426.26 |
| **Svar** | $ 5,891.82 | $ 489.58 | $ 4,272.10 | $ 8,636.66 | $ 1,126.73 | $ 2,772.46 | $ 5,862.58 | $ 7,264.33 | $ 13,834.94 | $ 2,848.06 | $ (10,457.01) | $ 6,129.97 | $ 9,565.87 | $ 2,136.13 | $ 12,064.11 | $ 72,438.34 |
| **ACCESS-CM2 ensemble** | | | | | | | | | | | | | | | | |
| **None** | $ 4,850.80 | $ 403.08 | $ 3,517.26 | $ 7,110.66 | $ 927.65 | $ 2,282.60 | $ 4,826.72 | $ 5,980.81 | $ 11,390.46 | $ 2,344.84 | $ (8,609.37) | $ 5,046.87 | $ 7,875.68 | $ 1,758.70 | $ 9,932.51 | $ 59,639.29 |
| **S&Tvar** | $ 5,679.49 | $ 471.94 | $ 4,889.55 | $ 8,370.20 | $ 1,184.93 | $ 2,915.67 | $ 6,165.40 | $ 7,639.57 | $ 15,834.54 | $ 2,760.19 | $ (11,968.39) | $ 6,446.61 | $ 10,059.98 | $ 2,246.47 | $ 13,807.77 | $ 76,503.94 |
| **Tvar** | $ 4,739.16 | $ 393.80 | $ 3,660.63 | $ 7,051.43 | $ 929.07 | $ 2,286.09 | $ 4,834.11 | $ 5,989.96 | $ 11,854.75 | $ 2,325.31 | $ (8,960.30) | $ 5,054.60 | $ 7,887.74 | $ 1,761.39 | $ 10,337.38 | $ 60,145.13 |
| **Svar** | $ 5,702.24 | $ 473.83 | $ 4,134.63 | $ 8,358.76 | $ 1,090.48 | $ 2,683.25 | $ 5,673.93 | $ 7,030.59 | $ 13,389.76 | $ 2,756.42 | $ (10,120.53) | $ 5,932.72 | $ 9,258.06 | $ 2,067.39 | $ 11,675.91 | $ 70,107.45 |

Figures are in billion US$2005. S&Tvar, Tvar and Svar denote the use of damage functions that include both spatial variation and temporal variability, and temporal variability and spatial variation alone, respectively. The column labeled None shows the results produced with the damage function that does not include any type of variability in temperature change. Results are based on the SP approach. Numbers in parenthesis denote benefits. Calculations are based on a 4% discount rate.

Table S10. Present value of the economic losses from climate change for 15 sectors under the SSP126 and different climate models.

| ACCESS-ESM1-5 | Agriculture | Forestry | Energy | Water | Coastal defence | Dryland | Wetland | Ecosystem | Health | Air pollution | Time use | Settlements | Catastrophe | Migration | Amenity | Total |
|---|---|---|---|---|---|---|---|---|---|---|---|---|---|---|---|---|
| None | $ 4,085.22 | $ 339.46 | $ 2,962.15 | $ 5,988.41 | $ 781.24 | $1,922.34 | $ 4,064.94 | $ 5,036.88 | $ 9,592.75 | $ 1,974.77 | $ (7,250.59) | $ 4,250.34 | $ 6,632.70 | $1,481.13 | $ 8,364.90 | $ 50,226.65 |
| S&Tvar | $ 5,044.38 | $ 419.17 | $ 4,358.17 | $ 7,393.00 | $ 1,057.14 | $2,601.23 | $ 5,500.50 | $ 6,815.68 | $14,113.69 | $ 2,437.95 | $(10,667.70) | $ 5,751.38 | $ 8,975.07 | $2,004.20 | $12,307.18 | $ 68,111.03 |
| Tvar | $ 3,987.95 | $ 331.38 | $ 3,054.96 | $ 5,878.77 | $ 782.03 | $1,924.29 | $ 4,069.05 | $ 5,041.97 | $ 9,893.32 | $ 1,938.61 | $ (7,477.77) | $ 4,254.64 | $ 6,639.40 | $1,482.63 | $ 8,627.01 | $ 50,428.24 |
| Svar | $ 4,956.95 | $ 411.90 | $ 3,594.23 | $ 7,266.26 | $ 947.95 | $2,332.55 | $ 4,932.35 | $ 6,111.68 | $11,639.71 | $ 2,396.15 | $ (8,797.77) | $ 5,157.31 | $ 8,048.02 | $1,797.18 | $10,149.87 | $ 60,944.36 |
| MPI-ESM1-2-HR | | | | | | | | | | | | | | | | |
| None | $ 4,074.60 | $ 338.58 | $ 2,954.45 | $ 5,972.84 | $ 779.21 | $1,917.34 | $ 4,054.37 | $ 5,023.78 | $ 9,567.81 | $ 1,969.63 | $ (7,231.74) | $ 4,239.29 | $ 6,615.45 | $1,477.28 | $ 8,343.16 | $ 50,096.06 |
| S&Tvar | $ 4,939.53 | $ 410.45 | $ 4,273.08 | $ 7,126.65 | $ 1,043.25 | $2,567.03 | $ 5,428.19 | $ 6,726.08 | $13,838.13 | $ 2,350.12 | $(10,459.42) | $ 5,675.77 | $ 8,857.08 | $1,977.85 | $12,066.89 | $ 66,820.70 |
| Tvar | $ 3,938.18 | $ 327.25 | $ 3,037.62 | $ 5,714.09 | $ 780.97 | $1,921.67 | $ 4,063.52 | $ 5,035.12 | $ 9,837.15 | $ 1,884.30 | $ (7,435.32) | $ 4,248.86 | $ 6,630.38 | $1,480.61 | $ 8,578.03 | $ 50,042.45 |
| Svar | $ 4,992.49 | $ 414.85 | $ 3,620.00 | $ 7,318.35 | $ 954.75 | $2,349.27 | $ 4,967.70 | $ 6,155.50 | $11,723.16 | $ 2,413.33 | $ (8,860.84) | $ 5,194.28 | $ 8,105.72 | $1,810.07 | $10,222.63 | $ 61,381.25 |
| INM-CM5 | | | | | | | | | | | | | | | | |
| None | $ 4,078.54 | $ 338.91 | $ 2,957.31 | $ 5,978.62 | $ 779.97 | $1,919.20 | $ 4,058.30 | $ 5,028.65 | $ 9,577.07 | $ 1,971.54 | $ (7,238.74) | $ 4,243.40 | $ 6,621.85 | $1,478.71 | $ 8,351.23 | $ 50,144.55 |
| S&Tvar | $ 5,051.65 | $ 419.77 | $ 4,212.81 | $ 7,073.23 | $ 1,060.22 | $2,608.81 | $ 5,516.52 | $ 6,835.54 | $13,642.93 | $ 2,332.50 | $(10,311.88) | $ 5,768.14 | $ 9,001.22 | $2,010.04 | $11,896.68 | $ 67,118.17 |
| Tvar | $ 3,927.54 | $ 326.36 | $ 3,044.36 | $ 5,682.20 | $ 781.91 | $1,923.97 | $ 4,068.38 | $ 5,041.15 | $ 9,858.99 | $ 1,873.79 | $ (7,451.82) | $ 4,253.94 | $ 6,638.31 | $1,482.39 | $ 8,597.07 | $ 50,048.53 |
| Svar | $ 4,971.69 | $ 413.13 | $ 3,604.92 | $ 7,287.87 | $ 950.77 | $2,339.48 | $ 4,947.01 | $ 6,129.86 | $11,674.33 | $ 2,403.28 | $ (8,823.93) | $ 5,172.65 | $ 8,071.96 | $1,802.53 | $10,180.05 | $ 61,125.61 |
| CESM2 ensemble | | | | | | | | | | | | | | | | |
| None | $ 4,079.95 | $ 339.03 | $ 2,958.33 | $ 5,980.69 | $ 780.24 | $1,919.86 | $ 4,059.70 | $ 5,030.38 | $ 9,580.37 | $ 1,972.22 | $ (7,241.23) | $ 4,244.86 | $ 6,624.14 | $1,479.22 | $ 8,354.11 | $ 50,161.85 |
| S&Tvar | $ 4,785.58 | $ 397.66 | $ 4,016.50 | $ 7,460.72 | $ 958.18 | $2,357.71 | $ 4,985.56 | $ 6,177.63 | $13,007.22 | $ 2,460.28 | $ (9,831.38) | $ 5,212.96 | $ 8,134.86 | $1,816.58 | $11,342.33 | $ 63,282.39 |
| Tvar | $ 4,052.25 | $ 336.72 | $ 3,124.45 | $ 6,222.01 | $ 781.11 | $1,922.00 | $ 4,064.22 | $ 5,035.99 | $10,118.35 | $ 2,051.80 | $ (7,647.86) | $ 4,249.59 | $ 6,631.52 | $1,480.87 | $ 8,823.23 | $ 51,246.24 |
| Svar | $ 4,672.68 | $ 388.28 | $ 3,388.11 | $ 6,849.55 | $ 893.59 | $2,198.78 | $ 4,649.48 | $ 5,761.18 | $10,972.19 | $ 2,258.74 | $ (8,293.22) | $ 4,861.55 | $ 7,586.48 | $1,694.12 | $ 9,567.78 | $ 57,449.26 |
| CanESM5 ensemble | | | | | | | | | | | | | | | | |
| None | $ 4,080.44 | $ 339.07 | $ 2,958.68 | $ 5,981.40 | $ 780.33 | $1,920.09 | $ 4,060.18 | $ 5,030.98 | $ 9,581.52 | $ 1,972.45 | $ (7,242.10) | $ 4,245.37 | $ 6,624.93 | $1,479.40 | $ 8,355.11 | $ 50,167.84 |
| S&Tvar | $ 4,888.90 | $ 406.25 | $ 4,042.89 | $ 6,898.79 | $ 1,018.07 | $2,505.09 | $ 5,297.20 | $ 6,563.77 | $13,092.67 | $ 2,274.98 | $ (9,895.97) | $ 5,538.80 | $ 8,643.35 | $1,930.12 | $11,416.84 | $ 64,621.75 |
| Tvar | $ 3,971.76 | $ 330.04 | $ 3,008.79 | $ 5,749.14 | $ 781.31 | $1,922.50 | $ 4,065.26 | $ 5,037.28 | $ 9,743.81 | $ 1,895.86 | $ (7,364.76) | $ 4,250.68 | $ 6,633.22 | $1,481.25 | $ 8,496.63 | $ 50,002.77 |
| Svar | $ 4,956.84 | $ 411.89 | $ 3,594.15 | $ 7,266.10 | $ 947.93 | $2,332.49 | $ 4,932.24 | $ 6,111.55 | $11,639.45 | $ 2,396.10 | $ (8,797.57) | $ 5,157.20 | $ 8,047.85 | $1,797.14 | $10,149.64 | $ 60,943.00 |
| ACCESS-CM2 ensemble | | | | | | | | | | | | | | | | |
| None | $ 4,081.58 | $ 339.16 | $ 2,959.51 | $ 5,983.08 | $ 780.55 | $1,920.63 | $ 4,061.32 | $ 5,032.39 | $ 9,584.20 | $ 1,973.01 | $ (7,244.13) | $ 4,246.56 | $ 6,626.78 | $1,479.81 | $ 8,357.45 | $ 50,181.88 |
| S&Tvar | $ 4,778.32 | $ 397.06 | $ 4,113.30 | $ 7,042.07 | $ 996.86 | $2,452.90 | $ 5,186.85 | $ 6,427.04 | $13,320.67 | $ 2,322.22 | $(10,068.30) | $ 5,423.42 | $ 8,463.29 | $1,889.92 | $11,615.66 | $ 64,361.26 |
| Tvar | $ 3,987.72 | $ 331.36 | $ 3,080.05 | $ 5,933.28 | $ 781.74 | $1,923.57 | $ 4,067.53 | $ 5,040.09 | $ 9,974.56 | $ 1,956.58 | $ (7,539.18) | $ 4,253.05 | $ 6,636.92 | $1,482.07 | $ 8,697.85 | $ 50,607.19 |
| Svar | $ 4,797.44 | $ 398.65 | $ 3,478.57 | $ 7,032.44 | $ 917.45 | $2,257.49 | $ 4,773.63 | $ 5,915.02 | $11,265.16 | $ 2,319.05 | $ (8,514.67) | $ 4,991.36 | $ 7,789.05 | $1,739.35 | $ 9,823.26 | $ 58,983.25 |

Figures are in billion US$2005. S&Tvar, Tvar and Svar denote the use of damage functions that include both spatial variation and temporal variability, and temporal variability and spatial variation alone, respectively. The column labeled None shows the results produced with the damage function that does not include any type of variability in temperature change. Results are based on the SP approach. Numbers in parenthesis denote benefits. Calculations are based on a 4% discount rate.

Table S11. Present value of the economic losses from climate change for 15 sectors under the SSP119 and different climate models.

| ACCESS-ESM1-5 | Agriculture | Forestry | Energy | Water | Coastal defence | Dryland | Wetland | Ecosystem | Health | Air pollution | Time use | Settlements | Catastrophe | Migration | Amenity | Total |
|---|---|---|---|---|---|---|---|---|---|---|---|---|---|---|---|---|
| **None** | $ 3,355.94 | $ 278.86 | $ 2,433.36 | $ 4,919.38 | $ 641.78 | $1,579.17 | $ 3,339.28 | $ 4,137.72 | $ 7,880.29 | $ 1,622.24 | $ (5,956.24) | $ 3,491.59 | $ 5,448.65 | $1,216.73 | $ 6,871.63 | $ 41,260.38 |
| **S&Tvar** | $ 4,143.39 | $ 344.30 | $ 3,579.46 | $ 6,072.52 | $ 868.29 | $2,136.52 | $ 4,517.84 | $ 5,598.07 | $11,591.89 | $ 2,002.50 | $ (8,761.62) | $ 4,723.91 | $ 7,371.69 | $1,646.15 | $10,108.16 | $ 55,943.08 |
| **Tvar** | $ 3,276.08 | $ 272.23 | $ 2,509.56 | $ 4,829.37 | $ 642.43 | $1,580.77 | $ 3,342.66 | $ 4,141.90 | $ 8,127.05 | $ 1,592.55 | $ (6,142.76) | $ 3,495.12 | $ 5,454.16 | $1,217.95 | $ 7,086.81 | $ 41,425.88 |
| **Svar** | $ 4,071.62 | $ 338.33 | $ 2,952.28 | $ 5,968.47 | $ 778.64 | $1,915.94 | $ 4,051.40 | $ 5,020.11 | $ 9,560.80 | $ 1,968.19 | $ (7,226.44) | $ 4,236.19 | $ 6,610.61 | $1,476.20 | $ 8,337.05 | $ 50,059.39 |
| **MPI-ESM1-2-HR** | | | | | | | | | | | | | | | | |
| **None** | $ 3,347.22 | $ 278.14 | $ 2,427.03 | $ 4,906.59 | $ 640.11 | $1,575.07 | $ 3,330.60 | $ 4,126.96 | $ 7,859.80 | $ 1,618.02 | $ (5,940.75) | $ 3,482.51 | $ 5,434.48 | $1,213.56 | $ 6,853.77 | $ 41,153.10 |
| **S&Tvar** | $ 4,057.31 | $ 337.14 | $ 3,509.60 | $ 5,853.85 | $ 856.88 | $2,108.45 | $ 4,458.48 | $ 5,524.51 | $11,365.64 | $ 1,930.39 | $ (8,590.61) | $ 4,661.83 | $ 7,274.82 | $1,624.52 | $ 9,910.87 | $ 54,883.67 |
| **Tvar** | $ 3,235.22 | $ 268.83 | $ 2,495.31 | $ 4,694.16 | $ 641.55 | $1,578.62 | $ 3,338.11 | $ 4,136.27 | $ 8,080.93 | $ 1,547.97 | $ (6,107.89) | $ 3,490.37 | $ 5,446.74 | $1,216.30 | $ 7,046.59 | $ 41,109.09 |
| **Svar** | $ 4,100.78 | $ 340.76 | $ 2,973.43 | $ 6,011.23 | $ 784.22 | $1,929.67 | $ 4,080.43 | $ 5,056.07 | $ 9,629.29 | $ 1,982.29 | $ (7,278.21) | $ 4,266.54 | $ 6,657.96 | $1,486.77 | $ 8,396.77 | $ 50,418.00 |
| **INM-CM5** | | | | | | | | | | | | | | | | |
| **None** | $ 3,350.46 | $ 278.41 | $ 2,429.38 | $ 4,911.34 | $ 640.73 | $1,576.59 | $ 3,333.82 | $ 4,130.95 | $ 7,867.41 | $ 1,619.59 | $ (5,946.50) | $ 3,485.88 | $ 5,439.75 | $1,214.74 | $ 6,860.40 | $ 41,192.94 |
| **S&Tvar** | $ 4,149.36 | $ 344.79 | $ 3,460.12 | $ 5,809.99 | $ 870.82 | $2,142.74 | $ 4,531.00 | $ 5,614.37 | $11,205.40 | $ 1,915.93 | $ (8,469.49) | $ 4,737.66 | $ 7,393.15 | $1,650.95 | $ 9,771.14 | $ 55,127.92 |
| **Tvar** | $ 3,226.49 | $ 268.11 | $ 2,500.85 | $ 4,667.98 | $ 642.32 | $1,580.51 | $ 3,342.11 | $ 4,141.21 | $ 8,098.86 | $ 1,539.34 | $ (6,121.44) | $ 3,494.54 | $ 5,453.26 | $1,217.75 | $ 7,062.23 | $ 41,114.11 |
| **Svar** | $ 4,083.72 | $ 339.34 | $ 2,961.06 | $ 5,986.21 | $ 780.96 | $1,921.63 | $ 4,063.44 | $ 5,035.02 | $ 9,589.21 | $ 1,974.04 | $ (7,247.92) | $ 4,248.78 | $ 6,630.25 | $1,480.59 | $ 8,361.82 | $ 50,208.15 |
| **CESM2 ensemble** | | | | | | | | | | | | | | | | |
| **None** | $ 3,351.61 | $ 278.50 | $ 2,430.22 | $ 4,913.04 | $ 640.95 | $1,577.14 | $ 3,334.97 | $ 4,132.38 | $ 7,870.12 | $ 1,620.14 | $ (5,948.56) | $ 3,487.08 | $ 5,441.62 | $1,215.16 | $ 6,862.77 | $ 41,207.14 |
| **S&Tvar** | $ 3,930.92 | $ 326.64 | $ 3,298.96 | $ 6,128.11 | $ 787.04 | $1,936.60 | $ 4,095.09 | $ 5,074.24 | $10,683.49 | $ 2,020.83 | $ (8,075.01) | $ 4,281.87 | $ 6,681.89 | $1,492.12 | $ 9,316.03 | $ 51,978.83 |
| **Tvar** | $ 3,328.88 | $ 276.61 | $ 2,566.60 | $ 5,111.15 | $ 641.66 | $1,578.89 | $ 3,338.69 | $ 4,136.98 | $ 8,311.79 | $ 1,685.48 | $ (6,282.39) | $ 3,490.97 | $ 5,447.68 | $1,216.51 | $ 7,247.90 | $ 42,097.40 |
| **Svar** | $ 3,838.23 | $ 318.94 | $ 2,783.06 | $ 5,626.35 | $ 734.01 | $1,806.12 | $ 3,819.17 | $ 4,732.35 | $ 9,012.77 | $ 1,855.37 | $ (6,812.22) | $ 3,993.37 | $ 6,231.68 | $1,391.58 | $ 7,859.16 | $ 47,189.95 |
| **CanESM5 ensemble** | | | | | | | | | | | | | | | | |
| **None** | $ 3,352.01 | $ 278.54 | $ 2,430.51 | $ 4,913.62 | $ 641.03 | $1,577.32 | $ 3,335.37 | $ 4,132.87 | $ 7,871.06 | $ 1,620.34 | $ (5,949.27) | $ 3,487.50 | $ 5,442.27 | $1,215.30 | $ 6,863.59 | $ 41,212.07 |
| **S&Tvar** | $ 4,015.74 | $ 333.69 | $ 3,320.62 | $ 5,666.78 | $ 836.21 | $2,057.59 | $ 4,350.94 | $ 5,391.26 | $10,753.64 | $ 1,868.70 | $ (8,128.04) | $ 4,549.38 | $ 7,099.35 | $1,585.34 | $ 9,377.21 | $ 53,078.42 |
| **Tvar** | $ 3,262.80 | $ 271.12 | $ 2,471.65 | $ 4,722.94 | $ 641.83 | $1,579.30 | $ 3,339.55 | $ 4,138.04 | $ 8,004.30 | $ 1,557.46 | $ (6,049.97) | $ 3,491.86 | $ 5,449.08 | $1,216.82 | $ 6,979.77 | $ 41,076.54 |
| **Svar** | $ 4,071.52 | $ 338.33 | $ 2,952.22 | $ 5,968.33 | $ 778.62 | $1,915.90 | $ 4,051.31 | $ 5,019.99 | $ 9,560.58 | $ 1,968.14 | $ (7,226.28) | $ 4,236.09 | $ 6,610.46 | $1,476.16 | $ 8,336.86 | $ 50,058.25 |
| **ACCESS-CM2 ensemble** | | | | | | | | | | | | | | | | |
| **None** | $ 3,352.95 | $ 278.62 | $ 2,431.19 | $ 4,915.00 | $ 641.21 | $1,577.77 | $ 3,336.31 | $ 4,134.03 | $ 7,873.26 | $ 1,620.79 | $ (5,950.93) | $ 3,488.48 | $ 5,443.79 | $1,215.64 | $ 6,865.51 | $ 41,223.60 |
| **S&Tvar** | $ 3,924.96 | $ 326.15 | $ 3,378.42 | $ 5,784.41 | $ 818.80 | $2,014.75 | $ 4,260.34 | $ 5,279.00 | $10,940.83 | $ 1,907.49 | $ (8,269.52) | $ 4,454.66 | $ 6,951.53 | $1,552.33 | $ 9,540.43 | $ 52,864.57 |
| **Tvar** | $ 3,275.89 | $ 272.21 | $ 2,530.15 | $ 4,874.12 | $ 642.19 | $1,580.18 | $ 3,341.41 | $ 4,140.35 | $ 8,193.74 | $ 1,607.31 | $ (6,193.16) | $ 3,493.81 | $ 5,452.12 | $1,217.50 | $ 7,144.97 | $ 41,572.77 |
| **Svar** | $ 3,940.66 | $ 327.45 | $ 2,857.33 | $ 5,776.51 | $ 753.60 | $1,854.32 | $ 3,921.10 | $ 4,858.65 | $ 9,253.30 | $ 1,904.89 | $ (6,994.02) | $ 4,099.94 | $ 6,397.99 | $1,428.72 | $ 8,068.91 | $ 48,449.34 |

Figures are in billion US$2005. S&Tvar, Tvar and Svar denote the use of damage functions that include both spatial variation and temporal variability, and temporal variability and spatial variation alone, respectively. The column labeled None shows the results produced with the damage function that does not include any type of variability in temperature change. Results are based on the SP approach. Numbers in parenthesis denote benefits. Calculations are based on a 4% discount rate.

Table S12. Global social cost of carbon for five SSP scenarios and three discount rates.

| Annual damage function and scenario | SSP585 | SSP370 | SSP245 | SSP126 | SSP119 |
|---|---|---|---|---|---|
| **Discount rate: 4%** | | | | | |
| **None** | $ 62.24 | $ 25.08 | $ 30.42 | $ 30.79 | $ 28.05 |
| **S&Tvar** | $ 78.34 | $ 31.57 | $ 38.29 | $ 38.76 | $ 35.30 |
| **Tvar** | $ 62.32 | $ 25.11 | $ 30.46 | $ 30.83 | $ 28.08 |
| **Svar** | $ 73.98 | $ 29.81 | $ 36.16 | $ 36.60 | $ 33.34 |
| **Discount rate: 3%** | | | | | |
| **None** | $ 95.41 | $ 35.32 | $ 43.69 | $ 43.25 | $ 38.91 |
| **S&Tvar** | $ 120.10 | $ 44.46 | $ 54.99 | $ 54.44 | $ 48.98 |
| **Tvar** | $ 95.54 | $ 35.37 | $ 43.75 | $ 43.31 | $ 38.96 |
| **Svar** | $ 113.41 | $ 41.98 | $ 51.93 | $ 51.41 | $ 46.25 |
| **Discount rate: 1.5%** | | | | | |
| **None** | $ 197.32 | $ 64.74 | $ 82.40 | $ 78.34 | $ 69.11 |
| **S&Tvar** | $ 248.37 | $ 81.49 | $ 103.72 | $ 98.62 | $ 86.99 |
| **Tvar** | $ 197.58 | $ 64.83 | $ 82.51 | $ 78.45 | $ 69.20 |
| **Svar** | $ 234.54 | $ 76.96 | $ 97.95 | $ 93.12 | $ 82.14 |

Figures are in dollars US$2005. S&Tvar, Tvar and Svar denote the use of damage functions that include both spatial variation and temporal variability, and temporal variability and spatial variation alone, respectively. The column labeled None shows the results produced with the damage function that does not include any type of variability in temperature change.

Table S13. Social cost of carbon for 15 sectors and different SSP trajectories (4% discount rate).

| | Agriculture | Forestry | Energy | Water | Coastal defence | Dryland | Wetland | Ecosystem | Health | Air pollution | Time use | Settlements | Catastrophe | Migration | Amenity | Total |
|---|---|---|---|---|---|---|---|---|---|---|---|---|---|---|---|---|
| **SSP585** | | | | | | | | | | | | | | | | |
| **None** | $ 5.06 | $ 0.42 | $ 3.67 | $ 7.42 | $ 0.97 | $ 2.38 | $ 5.04 | $ 6.24 | $ 11.89 | $ 2.45 | $ (8.98) | $ 5.27 | $ 8.22 | $ 1.84 | $ 10.37 | $ 62.24 |
| **S&Tvar** | $ 5.87 | $ 0.49 | $ 4.93 | $ 8.50 | $ 1.22 | $ 3.00 | $ 6.34 | $ 7.86 | $ 15.96 | $ 2.80 | $ (12.06) | $ 6.63 | $ 10.35 | $ 2.31 | $ 13.92 | $ 78.10 |
| **Tvar** | $ 4.94 | $ 0.41 | $ 3.79 | $ 7.29 | $ 0.97 | $ 2.39 | $ 5.04 | $ 6.25 | $ 12.26 | $ 2.40 | $ (9.27) | $ 5.27 | $ 8.23 | $ 1.84 | $ 10.69 | $ 62.51 |
| **Svar** | $ 6.02 | $ 0.50 | $ 4.36 | $ 8.82 | $ 1.15 | $ 2.83 | $ 5.99 | $ 7.42 | $ 14.13 | $ 2.91 | $ (10.68) | $ 6.26 | $ 9.77 | $ 2.18 | $ 12.32 | $ 73.98 |
| **SSP370** | | | | | | | | | | | | | | | | |
| **None** | $ 2.04 | $ 0.17 | $ 1.48 | $ 2.99 | $ 0.39 | $ 0.96 | $ 2.03 | $ 2.51 | $ 4.79 | $ 0.99 | $ (3.62) | $ 2.12 | $ 3.31 | $ 0.74 | $ 4.18 | $ 25.08 |
| **S&Tvar** | $ 2.37 | $ 0.20 | $ 1.99 | $ 3.42 | $ 0.49 | $ 1.21 | $ 2.55 | $ 3.17 | $ 6.43 | $ 1.13 | $ (4.86) | $ 2.67 | $ 4.17 | $ 0.93 | $ 5.61 | $ 31.47 |
| **Tvar** | $ 1.99 | $ 0.17 | $ 1.53 | $ 2.94 | $ 0.39 | $ 0.96 | $ 2.03 | $ 2.52 | $ 4.94 | $ 0.97 | $ (3.73) | $ 2.12 | $ 3.32 | $ 0.74 | $ 4.31 | $ 25.19 |
| **Svar** | $ 2.42 | $ 0.20 | $ 1.76 | $ 3.55 | $ 0.46 | $ 1.14 | $ 2.41 | $ 2.99 | $ 5.69 | $ 1.17 | $ (4.30) | $ 2.52 | $ 3.94 | $ 0.88 | $ 4.96 | $ 29.81 |
| **SSP245** | | | | | | | | | | | | | | | | |
| **None** | $ 2.47 | $ 0.21 | $ 1.79 | $ 3.63 | $ 0.47 | $ 1.16 | $ 2.46 | $ 3.05 | $ 5.81 | $ 1.20 | $ (4.39) | $ 2.57 | $ 4.02 | $ 0.90 | $ 5.07 | $ 30.42 |
| **S&Tvar** | $ 2.87 | $ 0.24 | $ 2.41 | $ 4.15 | $ 0.60 | $ 1.47 | $ 3.10 | $ 3.84 | $ 7.80 | $ 1.37 | $ (5.89) | $ 3.24 | $ 5.06 | $ 1.13 | $ 6.80 | $ 38.17 |
| **Tvar** | $ 2.42 | $ 0.20 | $ 1.85 | $ 3.56 | $ 0.47 | $ 1.17 | $ 2.47 | $ 3.05 | $ 5.99 | $ 1.18 | $ (4.53) | $ 2.58 | $ 4.02 | $ 0.90 | $ 5.23 | $ 30.55 |
| **Svar** | $ 2.94 | $ 0.24 | $ 2.13 | $ 4.31 | $ 0.56 | $ 1.38 | $ 2.93 | $ 3.63 | $ 6.91 | $ 1.42 | $ (5.22) | $ 3.06 | $ 4.77 | $ 1.07 | $ 6.02 | $ 36.16 |
| **SSP126** | | | | | | | | | | | | | | | | |
| **None** | $ 2.50 | $ 0.21 | $ 1.82 | $ 3.67 | $ 0.48 | $ 1.18 | $ 2.49 | $ 3.09 | $ 5.88 | $ 1.21 | $ (4.44) | $ 2.61 | $ 4.07 | $ 0.91 | $ 5.13 | $ 30.79 |
| **S&Tvar** | $ 2.91 | $ 0.24 | $ 2.44 | $ 4.20 | $ 0.60 | $ 1.48 | $ 3.14 | $ 3.89 | $ 7.89 | $ 1.39 | $ (5.97) | $ 3.28 | $ 5.12 | $ 1.14 | $ 6.88 | $ 38.64 |
| **Tvar** | $ 2.45 | $ 0.20 | $ 1.87 | $ 3.61 | $ 0.48 | $ 1.18 | $ 2.50 | $ 3.09 | $ 6.07 | $ 1.19 | $ (4.59) | $ 2.61 | $ 4.07 | $ 0.91 | $ 5.29 | $ 30.93 |
| **Svar** | $ 2.98 | $ 0.25 | $ 2.16 | $ 4.36 | $ 0.57 | $ 1.40 | $ 2.96 | $ 3.67 | $ 6.99 | $ 1.44 | $ (5.28) | $ 3.10 | $ 4.83 | $ 1.08 | $ 6.10 | $ 36.60 |
| **SSP119** | | | | | | | | | | | | | | | | |
| **None** | $ 2.28 | $ 0.19 | $ 1.65 | $ 3.34 | $ 0.44 | $ 1.07 | $ 2.27 | $ 2.81 | $ 5.36 | $ 1.10 | $ (4.05) | $ 2.37 | $ 3.70 | $ 0.83 | $ 4.67 | $ 28.05 |
| **S&Tvar** | $ 2.65 | $ 0.22 | $ 2.22 | $ 3.83 | $ 0.55 | $ 1.35 | $ 2.86 | $ 3.54 | $ 7.19 | $ 1.26 | $ (5.44) | $ 2.99 | $ 4.66 | $ 1.04 | $ 6.27 | $ 35.20 |
| **Tvar** | $ 2.23 | $ 0.19 | $ 1.71 | $ 3.29 | $ 0.44 | $ 1.07 | $ 2.27 | $ 2.82 | $ 5.53 | $ 1.08 | $ (4.18) | $ 2.38 | $ 3.71 | $ 0.83 | $ 4.82 | $ 28.17 |
| **Svar** | $ 2.71 | $ 0.23 | $ 1.97 | $ 3.97 | $ 0.52 | $ 1.28 | $ 2.70 | $ 3.34 | $ 6.37 | $ 1.31 | $ (4.81) | $ 2.82 | $ 4.40 | $ 0.98 | $ 5.55 | $ 33.34 |

Figures are in dollars US$2005. Negative quantities are in parenthesis. S&Tvar, Tvar and Svar denote the use of damage functions that include both spatial variation and temporal variability, and temporal variability and spatial variation alone, respectively. The column labeled None shows the results produced with the damage function that does not include any type of variability in temperature change.

Table S14. Social cost of carbon for 15 sectors and different SSP trajectories (3% discount rate).

| | Agriculture | Forestry | Energy | Water | Coastal defence | Dryland | Wetland | Ecosystem | Health | Air pollution | Time use | Settlements | Catastrophe | Migration | Amenity | Total |
|---|---|---|---|---|---|---|---|---|---|---|---|---|---|---|---|---|
| **SSP585** | | | | | | | | | | | | | | | | |
| **None** | $ 7.76 | $ 0.64 | $ 5.63 | $ 11.38 | $ 1.48 | $ 3.65 | $ 7.72 | $ 9.57 | $ 18.22 | $ 3.75 | $ (13.77) | $ 8.07 | $ 12.60 | $ 2.81 | $ 15.89 | $ 95.41 |
| **S&Tvar** | $ 9.00 | $ 0.75 | $ 7.55 | $ 13.03 | $ 1.87 | $ 4.60 | $ 9.72 | $ 12.04 | $ 24.46 | $ 4.30 | $ (18.49) | $ 10.16 | $ 15.86 | $ 3.54 | $ 21.33 | $ 119.73 |
| **Tvar** | $ 7.58 | $ 0.63 | $ 5.80 | $ 11.18 | $ 1.49 | $ 3.66 | $ 7.73 | $ 9.58 | $ 18.80 | $ 3.69 | $ (14.21) | $ 8.09 | $ 12.62 | $ 2.82 | $ 16.39 | $ 95.83 |
| **Svar** | $ 9.22 | $ 0.77 | $ 6.69 | $ 13.52 | $ 1.76 | $ 4.34 | $ 9.18 | $ 11.37 | $ 21.66 | $ 4.46 | $ (16.37) | $ 9.60 | $ 14.98 | $ 3.34 | $ 18.89 | $ 113.41 |
| **SSP370** | | | | | | | | | | | | | | | | |
| **None** | $ 2.87 | $ 0.24 | $ 2.08 | $ 4.21 | $ 0.55 | $ 1.35 | $ 2.86 | $ 3.54 | $ 6.75 | $ 1.39 | $ (5.10) | $ 2.99 | $ 4.66 | $ 1.04 | $ 5.88 | $ 35.32 |
| **S&Tvar** | $ 3.33 | $ 0.28 | $ 2.80 | $ 4.82 | $ 0.69 | $ 1.70 | $ 3.60 | $ 4.46 | $ 9.06 | $ 1.59 | $ (6.84) | $ 3.76 | $ 5.87 | $ 1.31 | $ 7.90 | $ 44.32 |
| **Tvar** | $ 2.81 | $ 0.23 | $ 2.15 | $ 4.14 | $ 0.55 | $ 1.35 | $ 2.86 | $ 3.55 | $ 6.96 | $ 1.36 | $ (5.26) | $ 2.99 | $ 4.67 | $ 1.04 | $ 6.07 | $ 35.47 |
| **Svar** | $ 3.41 | $ 0.28 | $ 2.48 | $ 5.01 | $ 0.65 | $ 1.61 | $ 3.40 | $ 4.21 | $ 8.02 | $ 1.65 | $ (6.06) | $ 3.55 | $ 5.54 | $ 1.24 | $ 6.99 | $ 41.98 |
| **SSP245** | | | | | | | | | | | | | | | | |
| **None** | $ 3.55 | $ 0.30 | $ 2.58 | $ 5.21 | $ 0.68 | $ 1.67 | $ 3.54 | $ 4.38 | $ 8.34 | $ 1.72 | $ (6.31) | $ 3.70 | $ 5.77 | $ 1.29 | $ 7.28 | $ 43.69 |
| **S&Tvar** | $ 4.12 | $ 0.34 | $ 3.46 | $ 5.97 | $ 0.86 | $ 2.10 | $ 4.45 | $ 5.51 | $ 11.20 | $ 1.97 | $ (8.47) | $ 4.65 | $ 7.26 | $ 1.62 | $ 9.77 | $ 54.82 |
| **Tvar** | $ 3.47 | $ 0.29 | $ 2.66 | $ 5.12 | $ 0.68 | $ 1.67 | $ 3.54 | $ 4.39 | $ 8.61 | $ 1.69 | $ (6.51) | $ 3.70 | $ 5.78 | $ 1.29 | $ 7.51 | $ 43.88 |
| **Svar** | $ 4.22 | $ 0.35 | $ 3.06 | $ 6.19 | $ 0.81 | $ 1.99 | $ 4.20 | $ 5.21 | $ 9.92 | $ 2.04 | $ (7.50) | $ 4.39 | $ 6.86 | $ 1.53 | $ 8.65 | $ 51.93 |
| **SSP126** | | | | | | | | | | | | | | | | |
| **None** | $ 3.52 | $ 0.29 | $ 2.55 | $ 5.16 | $ 0.67 | $ 1.66 | $ 3.50 | $ 4.34 | $ 8.26 | $ 1.70 | $ (6.24) | $ 3.66 | $ 5.71 | $ 1.28 | $ 7.20 | $ 43.25 |
| **S&Tvar** | $ 4.08 | $ 0.34 | $ 3.42 | $ 5.91 | $ 0.85 | $ 2.08 | $ 4.41 | $ 5.46 | $ 11.09 | $ 1.95 | $ (8.38) | $ 4.61 | $ 7.19 | $ 1.61 | $ 9.67 | $ 54.27 |
| **Tvar** | $ 3.44 | $ 0.29 | $ 2.63 | $ 5.07 | $ 0.67 | $ 1.66 | $ 3.50 | $ 4.34 | $ 8.52 | $ 1.67 | $ (6.44) | $ 3.66 | $ 5.72 | $ 1.28 | $ 7.43 | $ 43.44 |
| **Svar** | $ 4.18 | $ 0.35 | $ 3.03 | $ 6.13 | $ 0.80 | $ 1.97 | $ 4.16 | $ 5.16 | $ 9.82 | $ 2.02 | $ (7.42) | $ 4.35 | $ 6.79 | $ 1.52 | $ 8.56 | $ 51.41 |
| **SSP119** | | | | | | | | | | | | | | | | |
| **None** | $ 3.16 | $ 0.26 | $ 2.29 | $ 4.64 | $ 0.61 | $ 1.49 | $ 3.15 | $ 3.90 | $ 7.43 | $ 1.53 | $ (5.62) | $ 3.29 | $ 5.14 | $ 1.15 | $ 6.48 | $ 38.91 |
| **S&Tvar** | $ 3.67 | $ 0.31 | $ 3.08 | $ 5.31 | $ 0.76 | $ 1.87 | $ 3.96 | $ 4.91 | $ 9.98 | $ 1.75 | $ (7.54) | $ 4.14 | $ 6.47 | $ 1.44 | $ 8.70 | $ 48.83 |
| **Tvar** | $ 3.09 | $ 0.26 | $ 2.37 | $ 4.56 | $ 0.61 | $ 1.49 | $ 3.15 | $ 3.91 | $ 7.67 | $ 1.50 | $ (5.79) | $ 3.30 | $ 5.14 | $ 1.15 | $ 6.68 | $ 39.08 |
| **Svar** | $ 3.76 | $ 0.31 | $ 2.73 | $ 5.51 | $ 0.72 | $ 1.77 | $ 3.74 | $ 4.64 | $ 8.83 | $ 1.82 | $ (6.68) | $ 3.91 | $ 6.11 | $ 1.36 | $ 7.70 | $ 46.25 |

Figures are in dollars US$2005. Negative quantities are in parenthesis. S&Tvar, Tvar and Svar denote the use of damage functions that include both spatial variation and temporal variability, and temporal variability and spatial variation alone, respectively. The column labeled None shows the results produced with the damage function that does not include any type of variability in temperature change.

Table S15. Social cost of carbon for 15 sectors and different SSP trajectories (1.5% discount rate).

| | Agriculture | Forestry | Energy | Water | Coastal defence | Dryland | Wetland | Ecosystem | Health | Air pollution | Time use | Settlements | Catastrophe | Migration | Amenity | Total |
|---|---|---|---|---|---|---|---|---|---|---|---|---|---|---|---|---|
| **SSP585** | | | | | | | | | | | | | | | | |
| **None** | $ 16.05 | $ 1.33 | $11.64 | $ 23.53 | $ 3.07 | $ 7.55 | $ 15.97 | $ 19.79 | $ 37.69 | $ 7.76 | $ (28.48) | $ 16.70 | $ 26.06 | $ 5.82 | $ 32.86 | $ 197.32 |
| **S&Tvar** | $ 18.62 | $ 1.55 | $15.62 | $ 26.95 | $ 3.86 | $ 9.51 | $ 20.10 | $ 24.91 | $ 50.59 | $ 8.89 | $ (38.24) | $ 21.02 | $ 32.80 | $ 7.32 | $ 44.12 | $ 247.61 |
| **Tvar** | $ 15.68 | $ 1.30 | $12.00 | $ 23.11 | $ 3.07 | $ 7.56 | $ 15.99 | $ 19.81 | $ 38.87 | $ 7.62 | $ (29.38) | $ 16.72 | $ 26.09 | $ 5.83 | $ 33.90 | $ 198.19 |
| **Svar** | $ 19.08 | $ 1.59 | $13.83 | $ 27.96 | $ 3.65 | $ 8.98 | $ 18.98 | $ 23.52 | $ 44.79 | $ 9.22 | $ (33.86) | $ 19.85 | $ 30.97 | $ 6.92 | $ 39.06 | $ 234.54 |
| **SSP370** | | | | | | | | | | | | | | | | |
| **None** | $ 5.27 | $ 0.44 | $ 3.82 | $ 7.72 | $ 1.01 | $ 2.48 | $ 5.24 | $ 6.49 | $ 12.37 | $ 2.55 | $ (9.35) | $ 5.48 | $ 8.55 | $ 1.91 | $ 10.78 | $ 64.74 |
| **S&Tvar** | $ 6.11 | $ 0.51 | $ 5.13 | $ 8.84 | $ 1.27 | $ 3.12 | $ 6.60 | $ 8.17 | $ 16.60 | $ 2.92 | $ (12.55) | $ 6.90 | $ 10.76 | $ 2.40 | $ 14.48 | $ 81.24 |
| **Tvar** | $ 5.14 | $ 0.43 | $ 3.94 | $ 7.58 | $ 1.01 | $ 2.48 | $ 5.25 | $ 6.50 | $ 12.76 | $ 2.50 | $ (9.64) | $ 5.49 | $ 8.56 | $ 1.91 | $ 11.12 | $ 65.03 |
| **Svar** | $ 6.26 | $ 0.52 | $ 4.54 | $ 9.18 | $ 1.20 | $ 2.95 | $ 6.23 | $ 7.72 | $ 14.70 | $ 3.03 | $ (11.11) | $ 6.51 | $ 10.16 | $ 2.27 | $ 12.82 | $ 76.96 |
| **SSP245** | | | | | | | | | | | | | | | | |
| **None** | $ 6.70 | $ 0.56 | $ 4.86 | $ 9.82 | $ 1.28 | $ 3.15 | $ 6.67 | $ 8.26 | $ 15.74 | $ 3.24 | $ (11.90) | $ 6.97 | $ 10.88 | $ 2.43 | $ 13.72 | $ 82.40 |
| **S&Tvar** | $ 7.78 | $ 0.65 | $ 6.52 | $ 11.25 | $ 1.61 | $ 3.97 | $ 8.39 | $ 10.40 | $ 21.13 | $ 3.71 | $ (15.97) | $ 8.78 | $ 13.70 | $ 3.06 | $ 18.42 | $ 103.40 |
| **Tvar** | $ 6.55 | $ 0.54 | $ 5.01 | $ 9.65 | $ 1.28 | $ 3.16 | $ 6.68 | $ 8.27 | $ 16.23 | $ 3.18 | $ (12.27) | $ 6.98 | $ 10.90 | $ 2.43 | $ 14.16 | $ 82.76 |
| **Svar** | $ 7.97 | $ 0.66 | $ 5.78 | $ 11.68 | $ 1.52 | $ 3.75 | $ 7.93 | $ 9.82 | $ 18.71 | $ 3.85 | $ (14.14) | $ 8.29 | $ 12.93 | $ 2.89 | $ 16.31 | $ 97.95 |
| **SSP126** | | | | | | | | | | | | | | | | |
| **None** | $ 6.37 | $ 0.53 | $ 4.62 | $ 9.34 | $ 1.22 | $ 3.00 | $ 6.34 | $ 7.86 | $ 14.96 | $ 3.08 | $ (11.31) | $ 6.63 | $ 10.35 | $ 2.31 | $ 13.05 | $ 78.34 |
| **S&Tvar** | $ 7.39 | $ 0.61 | $ 6.20 | $ 10.70 | $ 1.53 | $ 3.77 | $ 7.98 | $ 9.89 | $ 20.09 | $ 3.53 | $ (15.18) | $ 8.35 | $ 13.02 | $ 2.91 | $ 17.52 | $ 98.31 |
| **Tvar** | $ 6.22 | $ 0.52 | $ 4.77 | $ 9.18 | $ 1.22 | $ 3.00 | $ 6.35 | $ 7.87 | $ 15.43 | $ 3.03 | $ (11.67) | $ 6.64 | $ 10.36 | $ 2.31 | $ 13.46 | $ 78.69 |
| **Svar** | $ 7.57 | $ 0.63 | $ 5.49 | $ 11.10 | $ 1.45 | $ 3.56 | $ 7.54 | $ 9.34 | $ 17.79 | $ 3.66 | $ (13.44) | $ 7.88 | $ 12.30 | $ 2.75 | $ 15.51 | $ 93.12 |
| **SSP119** | | | | | | | | | | | | | | | | |
| **None** | $ 5.62 | $ 0.47 | $ 4.08 | $ 8.24 | $ 1.07 | $ 2.64 | $ 5.59 | $ 6.93 | $ 13.20 | $ 2.72 | $ (9.98) | $ 5.85 | $ 9.13 | $ 2.04 | $ 11.51 | $ 69.11 |
| **S&Tvar** | $ 6.52 | $ 0.54 | $ 5.47 | $ 9.44 | $ 1.35 | $ 3.33 | $ 7.04 | $ 8.72 | $ 17.72 | $ 3.11 | $ (13.39) | $ 7.36 | $ 11.49 | $ 2.57 | $ 15.45 | $ 86.72 |
| **Tvar** | $ 5.49 | $ 0.46 | $ 4.20 | $ 8.10 | $ 1.08 | $ 2.65 | $ 5.60 | $ 6.94 | $ 13.62 | $ 2.67 | $ (10.29) | $ 5.86 | $ 9.14 | $ 2.04 | $ 11.87 | $ 69.41 |
| **Svar** | $ 6.68 | $ 0.56 | $ 4.84 | $ 9.79 | $ 1.28 | $ 3.14 | $ 6.65 | $ 8.24 | $ 15.69 | $ 3.23 | $ (11.86) | $ 6.95 | $ 10.85 | $ 2.42 | $ 13.68 | $ 82.14 |

Figures are in dollars US$2005. Negative quantities are in parenthesis. S&Tvar, Tvar and Svar denote the use of damage functions that include both spatial variation and temporal variability, and temporal variability and spatial variation alone, respectively. The column labeled None shows the results produced with the damage function that does not include any type of variability in temperature change.

Table S16. Original parameter values of the RICE model and optimized parameter values for a quadratic term specification.

| | US | EU | JAPAN | RUSSIA | EURASIA | CHINA | INDIA | MEAST | AFRICA | LAM | OHI | OTHASIA | |
|---|---|---|---|---|---|---|---|---|---|---|---|---|---|
| Original parameter values | | | | | | | | | | | | | |
| $\gamma_r$ | 0 | 0 | 0 | 0 | 0 | 0.0785 | 0.4385 | 0.278 | 0.341 | 0.0609 | 0 | 0.1755 | |
| $\alpha_r$ | 0.1414 | 0.1591 | 0.1617 | 0.1151 | 0.1305 | 0.1259 | 0.1689 | 0.1586 | 0.1983 | 0.1345 | 0.1564 | 0.1734 | |
| Optimized parameter values for a quadratic term specification | | | | | | | | | | | | | |
| | US | EU | JAPAN | RUSSIA | EURASIA | CHINA | INDIA | MEAST | AFRICA | LAM | OHI | OTHASIA | |
| $\gamma_r$ | 0 | 0 | 0 | 0 | 0 | 0 | 0 | 0 | 0 | 0 | 0 | 0 | |
| $\alpha_r$ | 0.1414 | 0.1591 | 0.1617 | 0.1151 | 0.1305 | 0.1411 | 0.2539 | 0.2125 | 0.2644 | 0.1463 | 0.1564 | 0.2074 | |

Table S17. Social cost of carbon for 12 regions and different SSP trajectories (4% discount rate).

| | USA | WEU | JAPAN | RUSSIA | EURASIA | CHINA | INDIA | MEAST | AFRICA | LAM | OHI | OASIA | Total |
|---|---|---|---|---|---|---|---|---|---|---|---|---|---|
| **SSP585** | | | | | | | | | | | | | |
| **None** | $ 5.59 | $ 6.49 | $ 1.23 | $ 0.85 | $ 0.79 | $ 9.17 | $ 11.35 | $ 2.94 | $ 11.32 | $ 3.65 | $ 1.89 | $ 6.97 | $ 62.24 |
| **S&Tvar** | $ 7.04 | $ 8.17 | $ 1.55 | $ 1.07 | $ 0.99 | $ 11.55 | $ 14.28 | $ 3.71 | $ 14.25 | $ 4.59 | $ 2.37 | $ 8.77 | $ 78.34 |
| **Tvar** | $ 5.60 | $ 6.50 | $ 1.23 | $ 0.85 | $ 0.79 | $ 9.18 | $ 11.36 | $ 2.95 | $ 11.34 | $ 3.65 | $ 1.89 | $ 6.98 | $ 62.32 |
| **Svar** | $ 6.65 | $ 7.71 | $ 1.46 | $ 1.01 | $ 0.94 | $ 10.90 | $ 13.49 | $ 3.50 | $ 13.46 | $ 4.34 | $ 2.24 | $ 8.28 | $ 73.98 |
| **SSP370** | | | | | | | | | | | | | |
| **None** | $ 2.51 | $ 2.91 | $ 0.56 | $ 0.44 | $ 0.42 | $ 4.15 | $ 3.82 | $ 1.52 | $ 3.41 | $ 1.90 | $ 0.82 | $ 2.63 | $ 25.08 |
| **S&Tvar** | $ 3.15 | $ 3.67 | $ 0.70 | $ 0.55 | $ 0.53 | $ 5.22 | $ 4.81 | $ 1.91 | $ 4.30 | $ 2.39 | $ 1.03 | $ 3.30 | $ 31.57 |
| **Tvar** | $ 2.51 | $ 2.92 | $ 0.56 | $ 0.44 | $ 0.42 | $ 4.15 | $ 3.83 | $ 1.52 | $ 3.42 | $ 1.90 | $ 0.82 | $ 2.63 | $ 25.11 |
| **Svar** | $ 2.98 | $ 3.46 | $ 0.66 | $ 0.52 | $ 0.50 | $ 4.93 | $ 4.54 | $ 1.81 | $ 4.06 | $ 2.25 | $ 0.97 | $ 3.12 | $ 29.81 |
| **SSP245** | | | | | | | | | | | | | |
| **None** | $ 2.82 | $ 3.49 | $ 0.64 | $ 0.47 | $ 0.45 | $ 4.82 | $ 5.19 | $ 1.57 | $ 4.68 | $ 2.07 | $ 0.95 | $ 3.27 | $ 30.42 |
| **S&Tvar** | $ 3.55 | $ 4.39 | $ 0.81 | $ 0.59 | $ 0.56 | $ 6.06 | $ 6.53 | $ 1.97 | $ 5.90 | $ 2.60 | $ 1.20 | $ 4.12 | $ 38.29 |
| **Tvar** | $ 2.82 | $ 3.49 | $ 0.65 | $ 0.47 | $ 0.45 | $ 4.82 | $ 5.20 | $ 1.57 | $ 4.69 | $ 2.07 | $ 0.96 | $ 3.28 | $ 30.46 |
| **Svar** | $ 3.35 | $ 4.15 | $ 0.77 | $ 0.56 | $ 0.53 | $ 5.73 | $ 6.17 | $ 1.86 | $ 5.57 | $ 2.46 | $ 1.13 | $ 3.89 | $ 36.16 |
| **SSP126** | | | | | | | | | | | | | |
| **None** | $ 2.78 | $ 3.25 | $ 0.67 | $ 0.45 | $ 0.44 | $ 5.25 | $ 5.44 | $ 1.38 | $ 4.85 | $ 1.99 | $ 0.89 | $ 3.41 | $ 30.79 |
| **S&Tvar** | $ 3.50 | $ 4.10 | $ 0.84 | $ 0.56 | $ 0.55 | $ 6.61 | $ 6.85 | $ 1.74 | $ 6.10 | $ 2.51 | $ 1.12 | $ 4.29 | $ 38.76 |
| **Tvar** | $ 2.78 | $ 3.26 | $ 0.67 | $ 0.45 | $ 0.44 | $ 5.26 | $ 5.45 | $ 1.38 | $ 4.85 | $ 2.00 | $ 0.89 | $ 3.41 | $ 30.83 |
| **Svar** | $ 3.30 | $ 3.87 | $ 0.79 | $ 0.53 | $ 0.52 | $ 6.24 | $ 6.47 | $ 1.64 | $ 5.76 | $ 2.37 | $ 1.06 | $ 4.05 | $ 36.60 |
| **SSP119** | | | | | | | | | | | | | |
| **None** | $ 2.57 | $ 3.00 | $ 0.62 | $ 0.41 | $ 0.40 | $ 4.85 | $ 4.91 | $ 1.26 | $ 4.30 | $ 1.82 | $ 0.82 | $ 3.08 | $ 28.05 |
| **S&Tvar** | $ 3.23 | $ 3.78 | $ 0.78 | $ 0.52 | $ 0.50 | $ 6.11 | $ 6.18 | $ 1.59 | $ 5.41 | $ 2.30 | $ 1.03 | $ 3.88 | $ 35.30 |
| **Tvar** | $ 2.57 | $ 3.00 | $ 0.62 | $ 0.41 | $ 0.40 | $ 4.86 | $ 4.91 | $ 1.26 | $ 4.31 | $ 1.83 | $ 0.82 | $ 3.08 | $ 28.08 |
| **Svar** | $ 3.05 | $ 3.57 | $ 0.74 | $ 0.49 | $ 0.48 | $ 5.77 | $ 5.83 | $ 1.50 | $ 5.11 | $ 2.17 | $ 0.98 | $ 3.66 | $ 33.34 |

Figures are in dollars US$2005. S&Tvar, Tvar and Svar denote the use of damage functions that include both spatial variation and temporal variability, and temporal variability and spatial variation alone, respectively. The column labeled None shows the results produced with the damage function that does not include any type of variability in temperature change.

Table S18. Social cost of carbon for 12 regions and different SSP trajectories (3% discount rate).

| | USA | WEU | JAPAN | RUSSIA | EURASIA | CHINA | INDIA | MEAST | AFRICA | LAM | OHI | OASIA | Total |
|---|---|---|---|---|---|---|---|---|---|---|---|---|---|
| **SSP585** | | | | | | | | | | | | | |
| **None** | $ 8.37 | $ 9.75 | $ 1.80 | $ 1.24 | $ 1.16 | $ 13.10 | $ 17.71 | $ 4.47 | $ 18.65 | $ 5.48 | $ 2.84 | $ 10.84 | $ 95.41 |
| **S&Tvar** | $ 10.54 | $ 12.28 | $ 2.26 | $ 1.56 | $ 1.46 | $ 16.50 | $ 22.30 | $ 5.62 | $ 23.47 | $ 6.90 | $ 3.58 | $ 13.65 | $ 120.10 |
| **Tvar** | $ 8.38 | $ 9.77 | $ 1.80 | $ 1.24 | $ 1.16 | $ 13.12 | $ 17.74 | $ 4.47 | $ 18.67 | $ 5.49 | $ 2.85 | $ 10.86 | $ 95.54 |
| **Svar** | $ 9.95 | $ 11.59 | $ 2.14 | $ 1.47 | $ 1.38 | $ 15.58 | $ 21.05 | $ 5.31 | $ 22.17 | $ 6.51 | $ 3.38 | $ 12.89 | $ 113.41 |
| **SSP370** | | | | | | | | | | | | | |
| **None** | $ 3.37 | $ 3.94 | $ 0.73 | $ 0.61 | $ 0.59 | $ 5.61 | $ 5.52 | $ 2.17 | $ 5.20 | $ 2.68 | $ 1.11 | $ 3.81 | $ 35.32 |
| **S&Tvar** | $ 4.24 | $ 4.96 | $ 0.91 | $ 0.77 | $ 0.74 | $ 7.06 | $ 6.94 | $ 2.73 | $ 6.55 | $ 3.37 | $ 1.39 | $ 4.79 | $ 44.46 |
| **Tvar** | $ 3.37 | $ 3.94 | $ 0.73 | $ 0.61 | $ 0.59 | $ 5.61 | $ 5.52 | $ 2.17 | $ 5.21 | $ 2.68 | $ 1.11 | $ 3.81 | $ 35.37 |
| **Svar** | $ 4.01 | $ 4.68 | $ 0.86 | $ 0.72 | $ 0.70 | $ 6.66 | $ 6.56 | $ 2.58 | $ 6.18 | $ 3.18 | $ 1.32 | $ 4.53 | $ 41.98 |
| **SSP245** | | | | | | | | | | | | | |
| **None** | $ 3.83 | $ 4.83 | $ 0.86 | $ 0.64 | $ 0.62 | $ 6.56 | $ 7.68 | $ 2.25 | $ 7.33 | $ 2.94 | $ 1.31 | $ 4.83 | $ 43.69 |
| **S&Tvar** | $ 4.82 | $ 6.08 | $ 1.08 | $ 0.81 | $ 0.78 | $ 8.26 | $ 9.67 | $ 2.83 | $ 9.23 | $ 3.70 | $ 1.65 | $ 6.08 | $ 54.99 |
| **Tvar** | $ 3.84 | $ 4.83 | $ 0.86 | $ 0.65 | $ 0.62 | $ 6.57 | $ 7.69 | $ 2.25 | $ 7.34 | $ 2.94 | $ 1.32 | $ 4.84 | $ 43.75 |
| **Svar** | $ 4.56 | $ 5.74 | $ 1.02 | $ 0.77 | $ 0.74 | $ 7.80 | $ 9.13 | $ 2.67 | $ 8.71 | $ 3.49 | $ 1.56 | $ 5.74 | $ 51.93 |
| **SSP126** | | | | | | | | | | | | | |
| **None** | $ 3.73 | $ 4.39 | $ 0.88 | $ 0.60 | $ 0.59 | $ 7.00 | $ 7.86 | $ 1.92 | $ 7.41 | $ 2.76 | $ 1.20 | $ 4.91 | $ 43.25 |
| **S&Tvar** | $ 4.69 | $ 5.53 | $ 1.11 | $ 0.75 | $ 0.75 | $ 8.81 | $ 9.90 | $ 2.42 | $ 9.32 | $ 3.48 | $ 1.51 | $ 6.18 | $ 54.44 |
| **Tvar** | $ 3.73 | $ 4.40 | $ 0.88 | $ 0.60 | $ 0.59 | $ 7.01 | $ 7.87 | $ 1.92 | $ 7.42 | $ 2.77 | $ 1.20 | $ 4.92 | $ 43.31 |
| **Svar** | $ 4.43 | $ 5.22 | $ 1.04 | $ 0.71 | $ 0.70 | $ 8.32 | $ 9.34 | $ 2.28 | $ 8.80 | $ 3.28 | $ 1.43 | $ 5.83 | $ 51.41 |
| **SSP119** | | | | | | | | | | | | | |
| **None** | $ 3.40 | $ 4.00 | $ 0.81 | $ 0.55 | $ 0.54 | $ 6.41 | $ 7.01 | $ 1.73 | $ 6.49 | $ 2.50 | $ 1.09 | $ 4.38 | $ 38.91 |
| **S&Tvar** | $ 4.29 | $ 5.04 | $ 1.02 | $ 0.69 | $ 0.68 | $ 8.06 | $ 8.82 | $ 2.18 | $ 8.17 | $ 3.14 | $ 1.38 | $ 5.51 | $ 48.98 |
| **Tvar** | $ 3.41 | $ 4.01 | $ 0.81 | $ 0.55 | $ 0.54 | $ 6.41 | $ 7.02 | $ 1.73 | $ 6.50 | $ 2.50 | $ 1.10 | $ 4.39 | $ 38.96 |
| **Svar** | $ 4.05 | $ 4.76 | $ 0.96 | $ 0.65 | $ 0.64 | $ 7.61 | $ 8.33 | $ 2.06 | $ 7.71 | $ 2.97 | $ 1.30 | $ 5.21 | $ 46.25 |

Figures are in dollars US$2005. S&Tvar, Tvar and Svar denote the use of damage functions that include both spatial variation and temporal variability, and temporal variability and spatial variation alone, respectively. The column labeled None shows the results produced with the damage function that does not include any type of variability in temperature change.

Table S19. Social cost of carbon for 12 regions and different SSP trajectories (1.5% discount rate).

| | USA | WEU | JAPAN | RUSSIA | EURASIA | CHINA | INDIA | MEAST | AFRICA | LAM | OHI | OASIA | Total |
|---|---|---|---|---|---|---|---|---|---|---|---|---|---|
| **SSP585** | | | | | | | | | | | | | |
| **None** | $ 16.83 | $ 19.72 | $ 3.48 | $ 2.37 | $ 2.26 | $ 24.32 | $ 37.34 | $ 9.10 | $ 42.32 | $ 11.00 | $ 5.76 | $ 22.81 | $ 197.32 |
| **S&Tvar** | $ 21.19 | $ 24.83 | $ 4.38 | $ 2.99 | $ 2.84 | $ 30.61 | $ 47.01 | $ 11.45 | $ 53.27 | $ 13.85 | $ 7.25 | $ 28.71 | $ 248.37 |
| **Tvar** | $ 16.85 | $ 19.75 | $ 3.49 | $ 2.38 | $ 2.26 | $ 24.35 | $ 37.39 | $ 9.11 | $ 42.38 | $ 11.02 | $ 5.77 | $ 22.84 | $ 197.58 |
| **Svar** | $ 20.01 | $ 23.44 | $ 4.14 | $ 2.82 | $ 2.68 | $ 28.91 | $ 44.39 | $ 10.81 | $ 50.31 | $ 13.08 | $ 6.84 | $ 27.11 | $ 234.54 |
| **SSP370** | | | | | | | | | | | | | |
| **None** | $ 5.72 | $ 6.76 | $ 1.16 | $ 1.08 | $ 1.07 | $ 9.59 | $ 10.46 | $ 4.06 | $ 10.71 | $ 4.94 | $ 1.90 | $ 7.29 | $ 64.74 |
| **S&Tvar** | $ 7.20 | $ 8.51 | $ 1.47 | $ 1.36 | $ 1.35 | $ 12.07 | $ 13.16 | $ 5.11 | $ 13.48 | $ 6.21 | $ 2.39 | $ 9.18 | $ 81.49 |
| **Tvar** | $ 5.73 | $ 6.77 | $ 1.17 | $ 1.08 | $ 1.08 | $ 9.60 | $ 10.47 | $ 4.06 | $ 10.72 | $ 4.94 | $ 1.90 | $ 7.30 | $ 64.83 |
| **Svar** | $ 6.80 | $ 8.03 | $ 1.38 | $ 1.29 | $ 1.28 | $ 11.40 | $ 12.43 | $ 4.82 | $ 12.73 | $ 5.87 | $ 2.26 | $ 8.67 | $ 76.96 |
| **SSP245** | | | | | | | | | | | | | |
| **None** | $ 6.65 | $ 8.61 | $ 1.44 | $ 1.14 | $ 1.12 | $ 11.31 | $ 15.07 | $ 4.22 | $ 15.61 | $ 5.46 | $ 2.32 | $ 9.46 | $ 82.40 |
| **S&Tvar** | $ 8.37 | $ 10.84 | $ 1.81 | $ 1.43 | $ 1.41 | $ 14.24 | $ 18.97 | $ 5.31 | $ 19.65 | $ 6.87 | $ 2.93 | $ 11.91 | $ 103.72 |
| **Tvar** | $ 6.66 | $ 8.62 | $ 1.44 | $ 1.14 | $ 1.13 | $ 11.33 | $ 15.09 | $ 4.22 | $ 15.63 | $ 5.47 | $ 2.33 | $ 9.47 | $ 82.51 |
| **Svar** | $ 7.90 | $ 10.24 | $ 1.71 | $ 1.35 | $ 1.34 | $ 13.44 | $ 17.91 | $ 5.01 | $ 18.55 | $ 6.49 | $ 2.76 | $ 11.24 | $ 97.95 |
| **SSP126** | | | | | | | | | | | | | |
| **None** | $ 6.31 | $ 7.51 | $ 1.44 | $ 0.99 | $ 1.02 | $ 11.58 | $ 14.78 | $ 3.42 | $ 15.15 | $ 4.90 | $ 2.05 | $ 9.20 | $ 78.34 |
| **S&Tvar** | $ 7.94 | $ 9.45 | $ 1.82 | $ 1.25 | $ 1.28 | $ 14.58 | $ 18.60 | $ 4.30 | $ 19.07 | $ 6.16 | $ 2.58 | $ 11.58 | $ 98.62 |
| **Tvar** | $ 6.32 | $ 7.52 | $ 1.44 | $ 0.99 | $ 1.02 | $ 11.60 | $ 14.80 | $ 3.42 | $ 15.17 | $ 4.90 | $ 2.05 | $ 9.21 | $ 78.45 |
| **Svar** | $ 7.50 | $ 8.93 | $ 1.71 | $ 1.18 | $ 1.21 | $ 13.77 | $ 17.57 | $ 4.06 | $ 18.01 | $ 5.82 | $ 2.44 | $ 10.93 | $ 93.12 |
| **SSP119** | | | | | | | | | | | | | |
| **None** | $ 5.64 | $ 6.70 | $ 1.30 | $ 0.89 | $ 0.91 | $ 10.42 | $ 12.94 | $ 3.03 | $ 13.05 | $ 4.34 | $ 1.83 | $ 8.06 | $ 69.11 |
| **S&Tvar** | $ 7.10 | $ 8.44 | $ 1.63 | $ 1.12 | $ 1.14 | $ 13.11 | $ 16.29 | $ 3.81 | $ 16.43 | $ 5.46 | $ 2.30 | $ 10.15 | $ 86.99 |
| **Tvar** | $ 5.65 | $ 6.71 | $ 1.30 | $ 0.89 | $ 0.91 | $ 10.43 | $ 12.96 | $ 3.03 | $ 13.07 | $ 4.34 | $ 1.83 | $ 8.07 | $ 69.20 |
| **Svar** | $ 6.71 | $ 7.97 | $ 1.54 | $ 1.06 | $ 1.08 | $ 12.38 | $ 15.38 | $ 3.60 | $ 15.51 | $ 5.16 | $ 2.18 | $ 9.58 | $ 82.14 |

Figures are in dollars US$2005. S&Tvar, Tvar and Svar denote the use of damage functions that include both spatial variation and temporal variability, and temporal variability and spatial variation alone, respectively. The column labeled None shows the results produced with the damage function that does not include any type of variability in temperature change.

*Supplementary Figures*

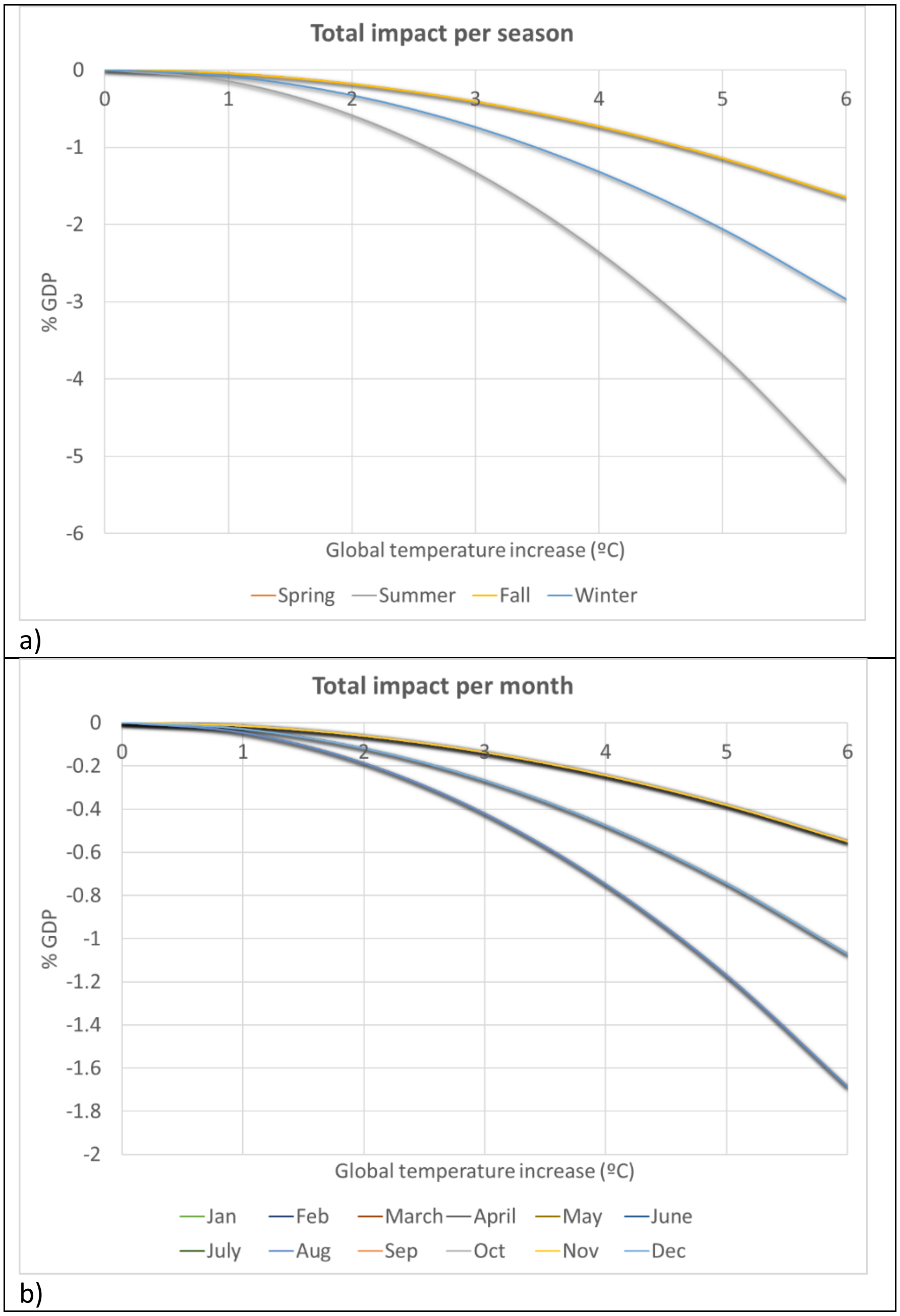

Figure S1. Seasonal and monthly damage function projections for different levels of warming. Panel a) shows the percent loss in welfare (%GDP) per season from increases in annual mean global surface temperature (ºC). Panel b) shows the percent loss in welfare (%GDP) per month from increases in annual mean global surface temperature (ºC).

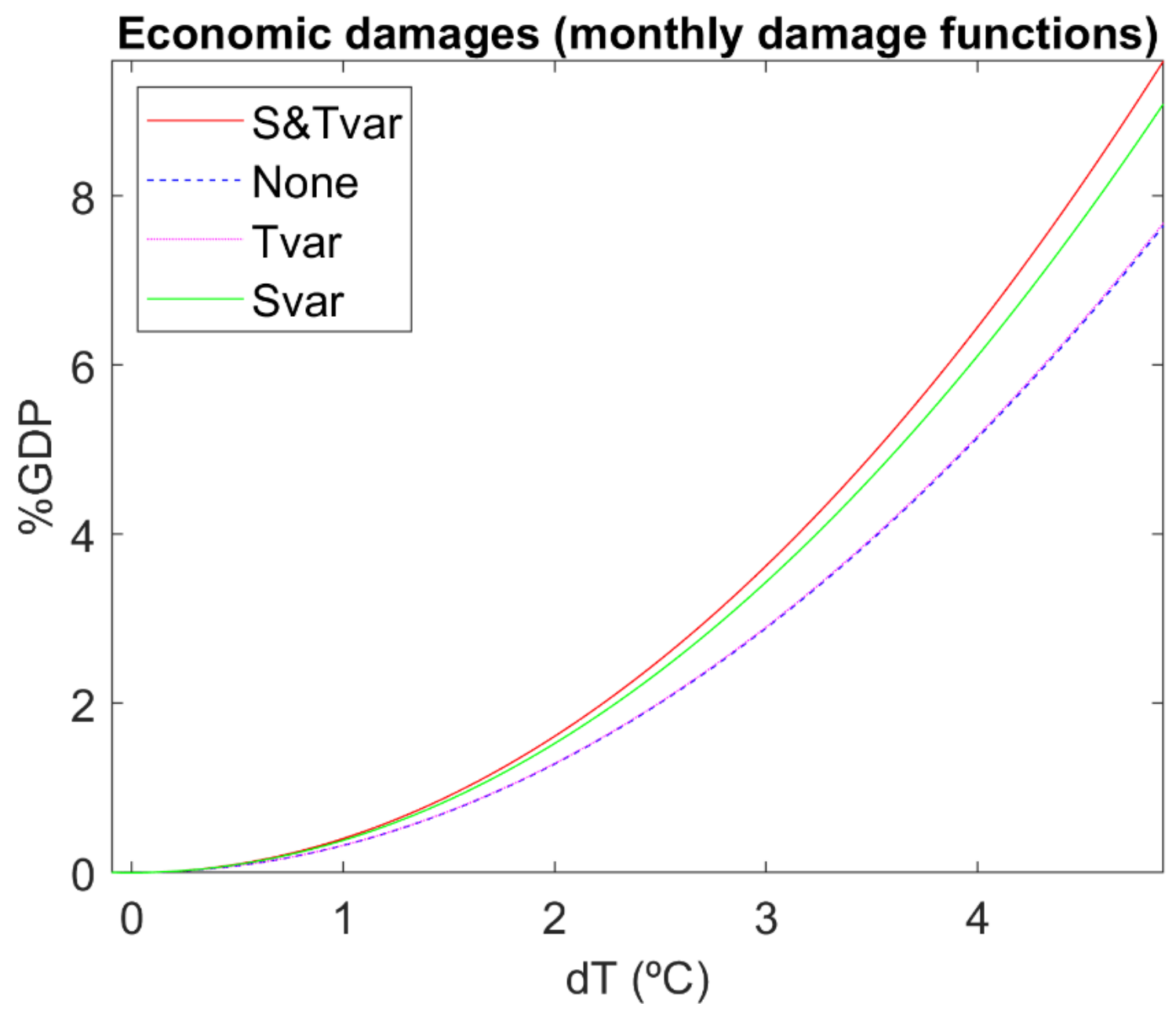

Figure S2. Projected economic damages (% GDP) as a function of changes in global temperature according to the aggregated monthly damage function and the SSP585 scenario. S&Tvar (red), Tvar (dotted red) and Svar (green) denote the use of damage functions that include both spatial variation and temporal variability, and temporal variability and spatial variation alone, respectively. The slashed blue line labeled None shows the results produced with the damage function that does not include any type of variability in temperature change.

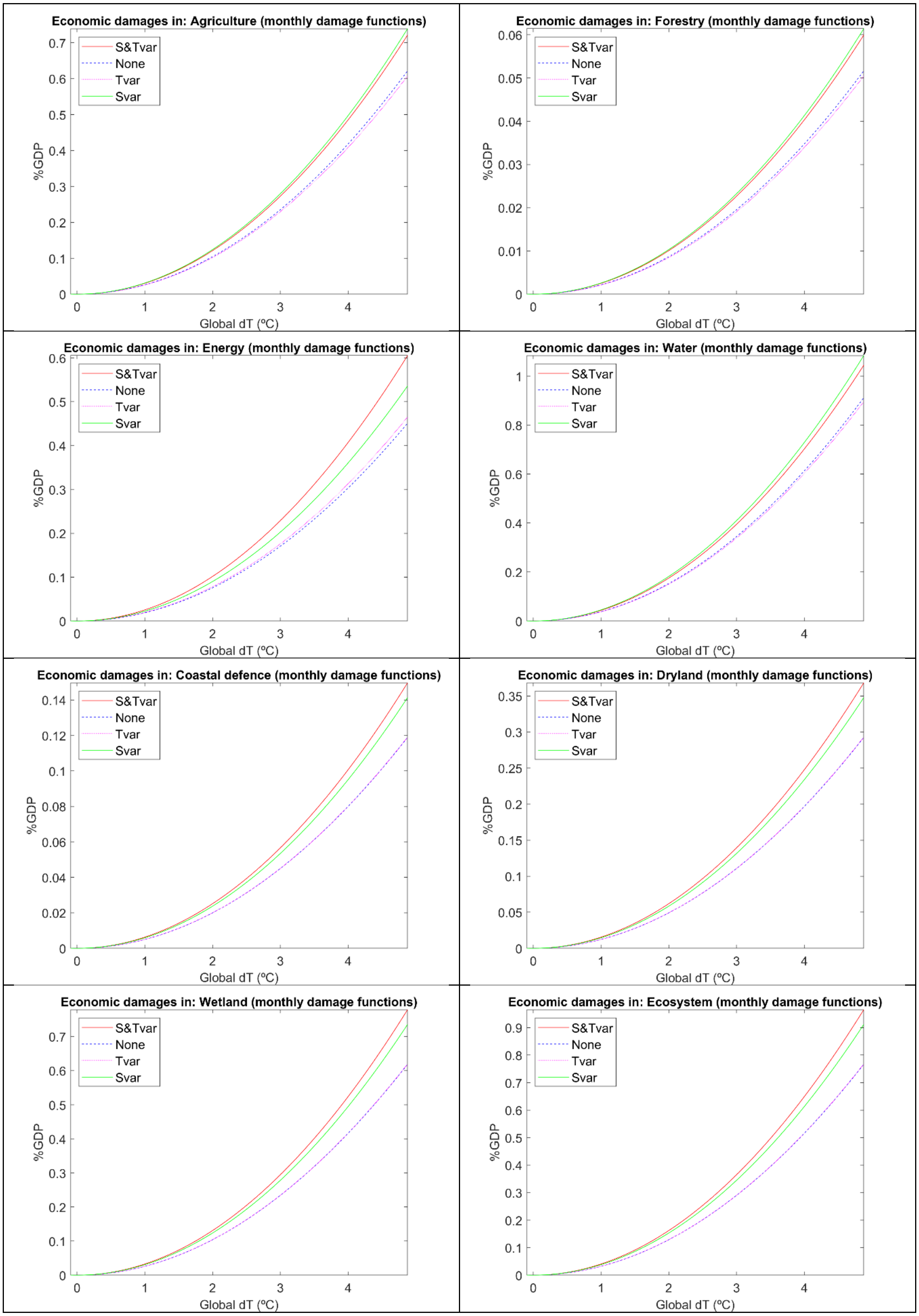
Economic damages in: Agriculture (monthly damage functions)
Economic damages in: Forestry (monthly damage functions)
Economic damages in: Energy (monthly damage functions)
Economic damages in: Water (monthly damage functions)
Economic damages in: Coastal defence (monthly damage functions)
Economic damages in: Dryland (monthly damage functions)
Economic damages in: Wetland (monthly damage functions)
Economic damages in: Ecosystem (monthly damage functions)
S&Tvar
None
Tvar
Svar
%GDP
Global dT (ºC)

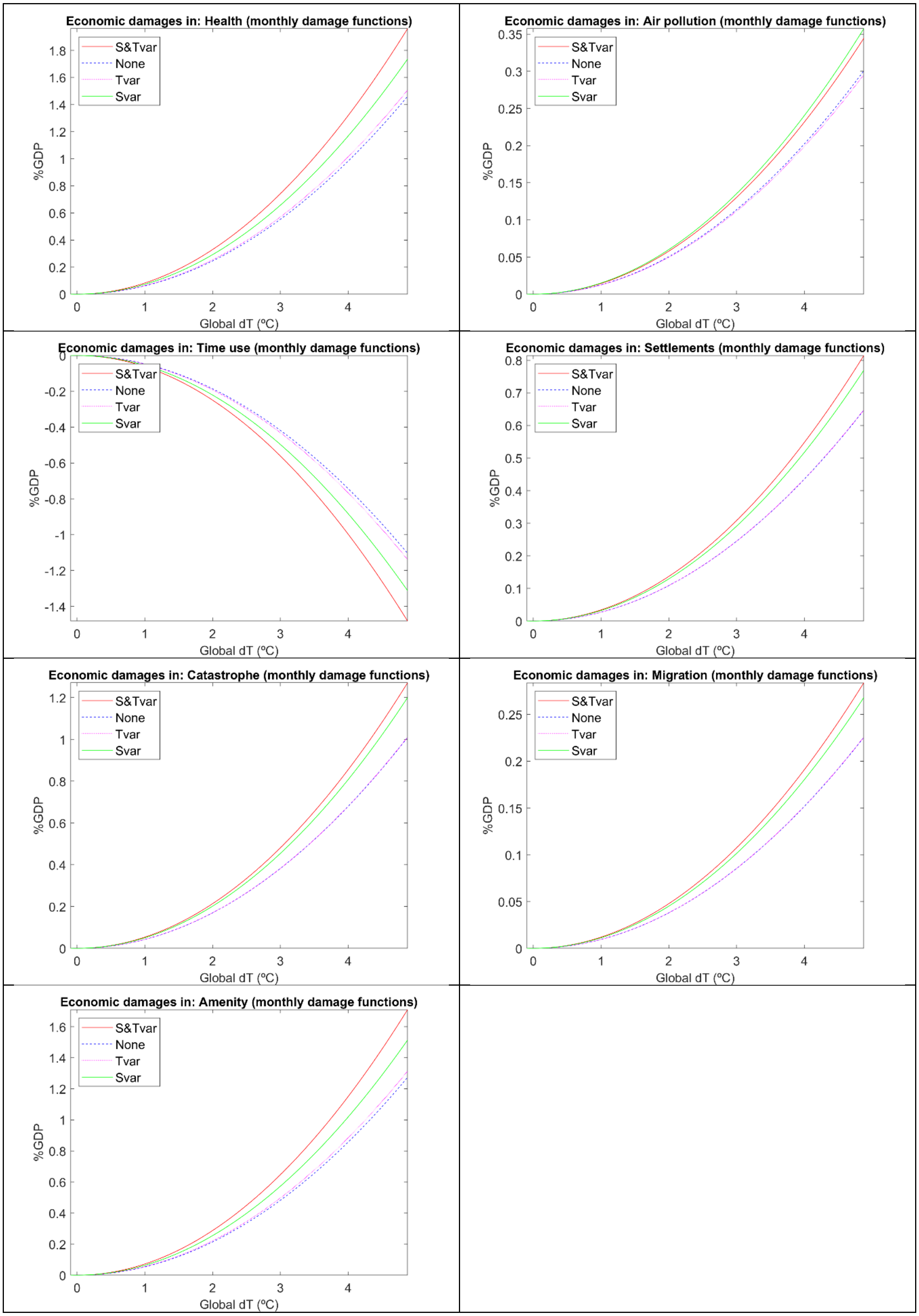
Economic damages in: Health (monthly damage functions)
S&Tvar
None
Tvar
Svar
%GDP
Global dT (ºC)
Economic damages in: Air pollution (monthly damage functions)
S&Tvar
None
Tvar
Svar
%GDP
Global dT (ºC)
Economic damages in: Time use (monthly damage functions)
S&Tvar
None
Tvar
Svar
%GDP
Global dT (ºC)
Economic damages in: Settlements (monthly damage functions)
S&Tvar
None
Tvar
Svar
%GDP
Global dT (ºC)
Economic damages in: Catastrophe (monthly damage functions)
S&Tvar
None
Tvar
Svar
%GDP
Global dT (ºC)
Economic damages in: Migration (monthly damage functions)
S&Tvar
None
Tvar
Svar
%GDP
Global dT (ºC)
Economic damages in: Amenity (monthly damage functions)
S&Tvar
None
Tvar
Svar
%GDP
Global dT (ºC)

Figure S3. Projected economic damages (% GDP) as a function of changes in global temperature according to the sector level monthly damage functions and the SSP585 scenario. S&Tvar (red), Tvar (dotted red) and Svar (green) denote the use of damage functions that include both spatial and temporal variability, and temporal and spatial variability alone, respectively. The slashed blue line labeled None shows the results produced with the damage function that does not include any type of variability in temperature change.

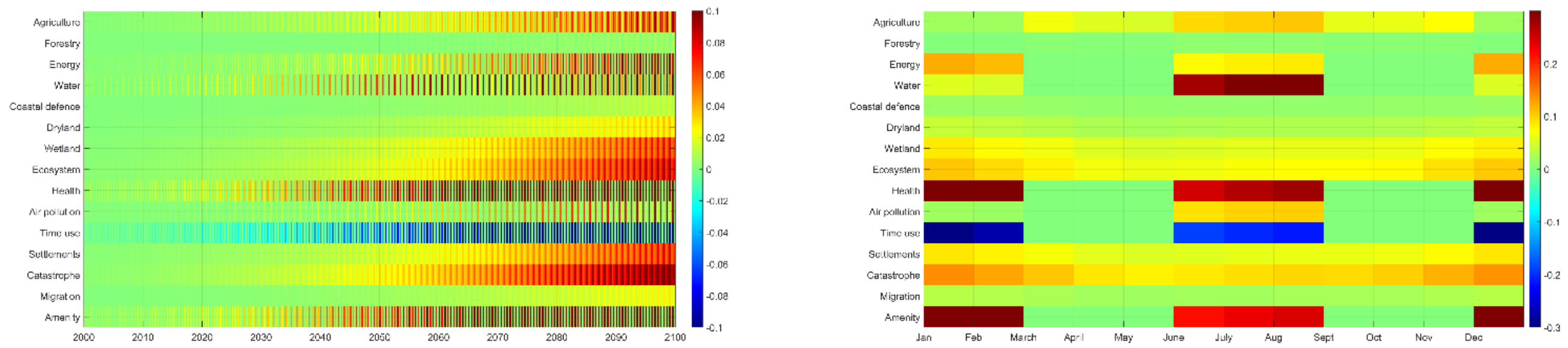

A

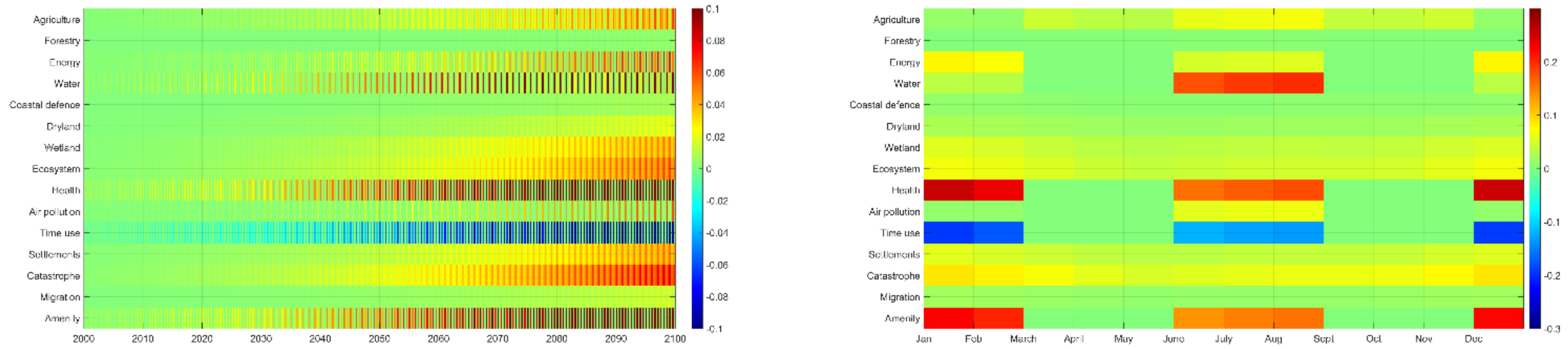

B

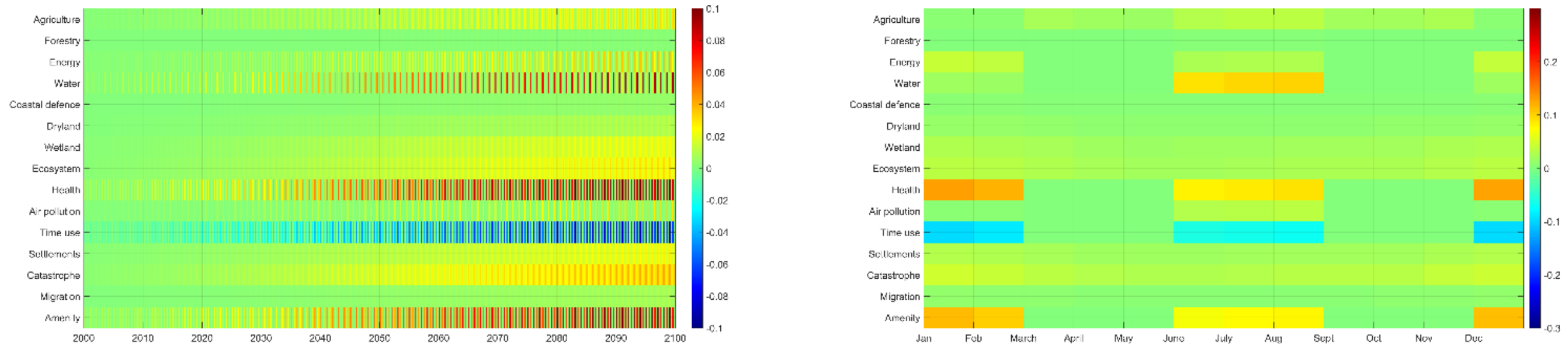

C

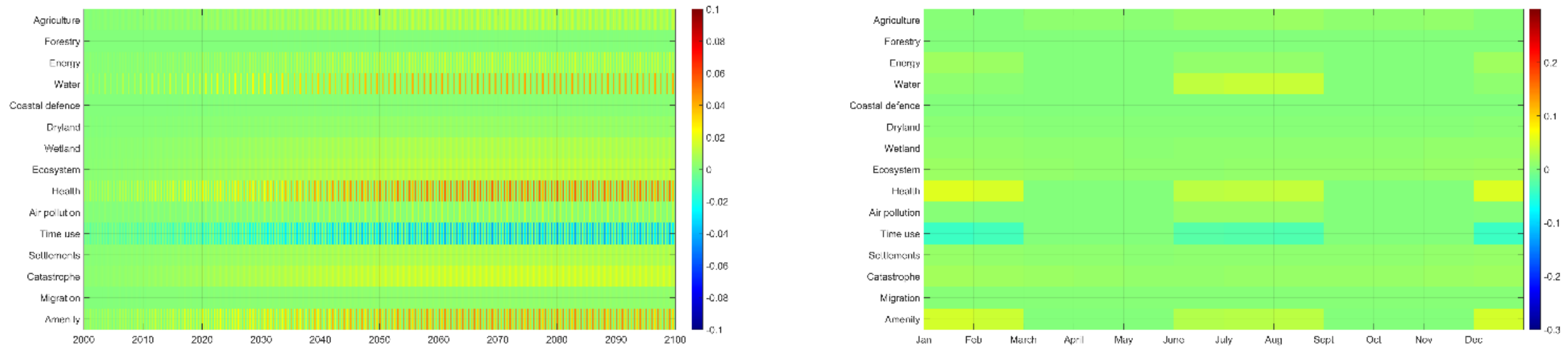

D

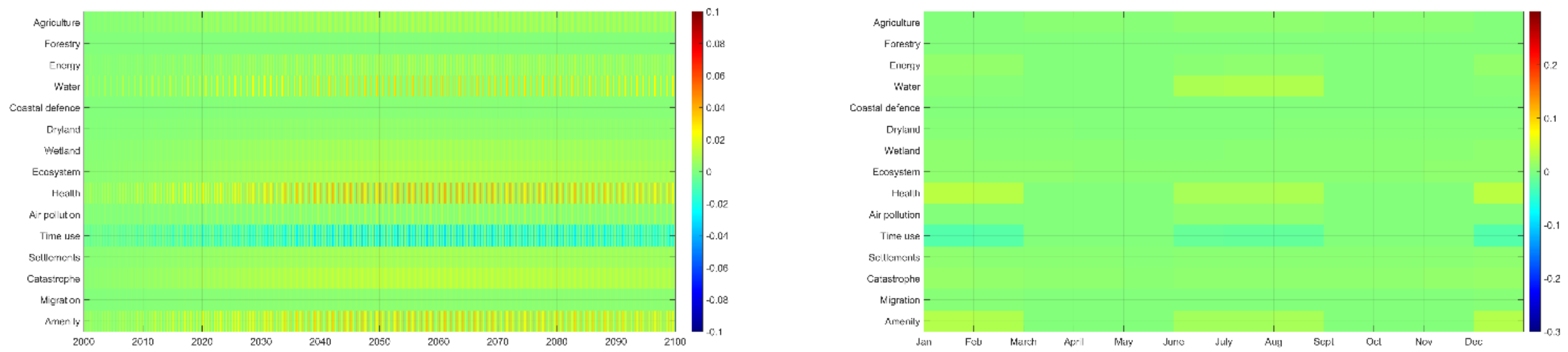

**E**

Figure S4. Projected economic losses Projected economic damages (% GDP) per sector and month for different SSP trajectories. Th left panel shows the percent of GDP lost during the period 2000-2100 and the right panel zooms into the monthly impacts per sector in 2100. Rows A-E show results for the SSP585, SSP370, SSP245, SSP126 and SSP119, respectively.

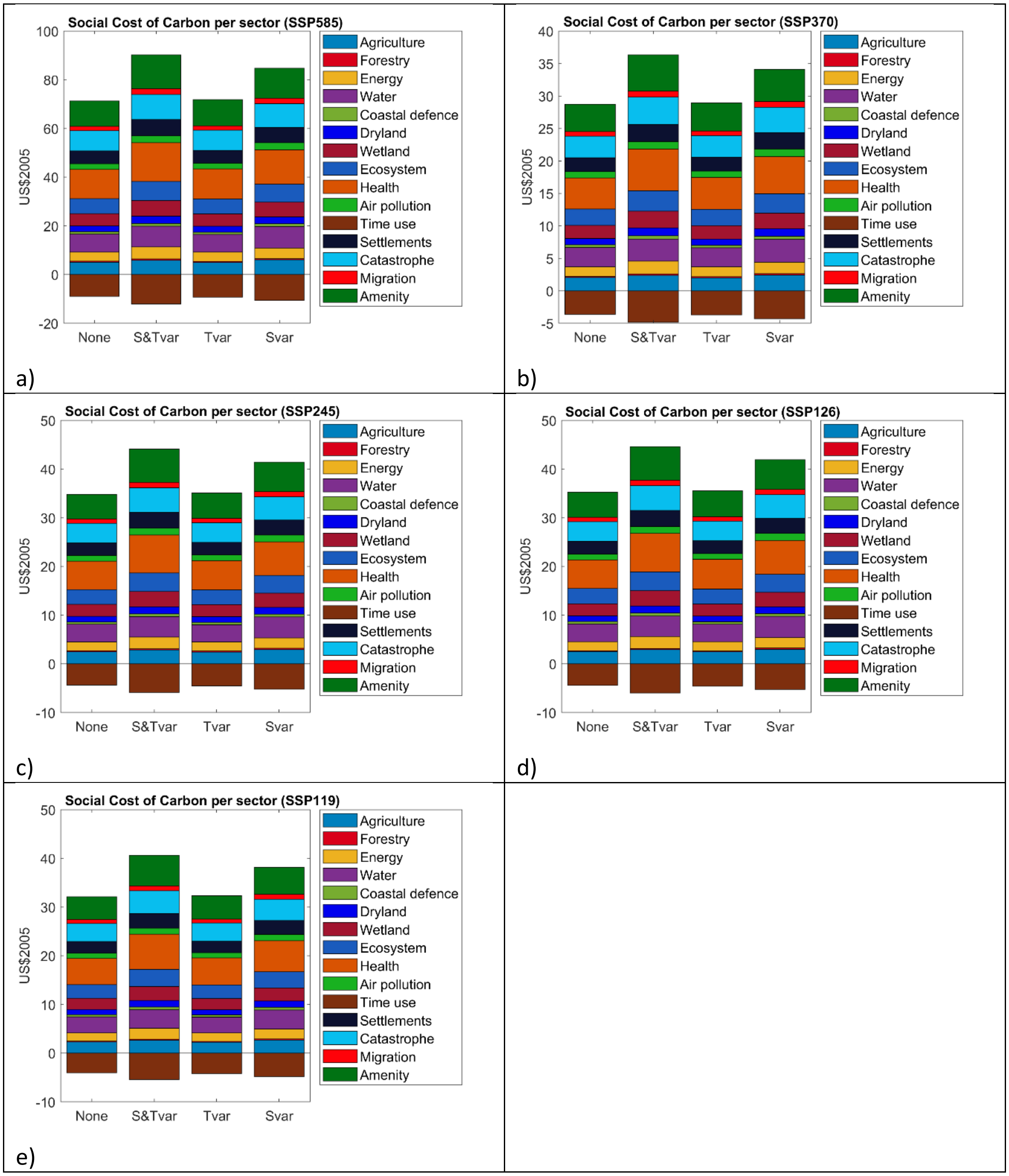

Figure S5. Sector level SCC values for different SSP trajectories using four damage functions. Panels a) to e) show results for the SSP585, SSP370, SSP245, SSP126 and SSP119, respectively. S&Tvar, Tvar and Svar denote the use of damage functions that include both spatial variation and temporal variability, and temporal variability and spatial variation alone, respectively. The bar labeled None shows the results produced with the damage function that does not include any type of variability in temperature change. Calculations are based on a 4% discount rate.